\documentclass[12pt]{article}
\usepackage{epsfig}
\usepackage{amsmath}
\usepackage{amssymb}
\usepackage{axodraw}

\def\lsim{\mathrel{\rlap{\lower4pt\hbox{\hskip1pt$\sim$}}
    \raise1pt\hbox{$<$}}}                
\def\gsim{\mathrel{\rlap{\lower4pt\hbox{\hskip1pt$\sim$}}
    \raise1pt\hbox{$>$}}}                

\newcommand{\alphas}{\alpha_{\mathrm{s}}}

\newcommand{\pT}{p_{\perp}}
\newcommand{\ECM}{E_{\mathrm{CM}}}

\renewcommand{\b}{\mathrm{b}}
\renewcommand{\c}{\mathrm{c}}
\renewcommand{\d}{\mathrm{d}}
\newcommand{\e}{\mathrm{e}}

\newcommand{\g}{\mathrm{g}}
\renewcommand{\j}{\mathrm{j}}
\newcommand{\J}{\mathrm{J}}
\newcommand{\hrm}{\mathrm{h}}
\newcommand{\p}{\mathrm{p}}
\newcommand{\q}{\mathrm{q}}
\newcommand{\s}{\mathrm{s}}
\renewcommand{\t}{\mathrm{t}}
\renewcommand{\u}{\mathrm{u}}
\newcommand{\yD}{y_\mathrm{D}}

\renewcommand{\B}{\mathrm{B}}

\newcommand{\K}{\mathrm{K}}

\newcommand{\Z}{\mathrm{Z}}

\newcommand{\dbar}{\overline{\mathrm{d}}}

\newcommand{\pbar}{\overline{\mathrm{p}}}
\newcommand{\qbar}{\overline{\mathrm{q}}}
\newcommand{\sbar}{\overline{\mathrm{s}}}

\newcommand{\ubar}{\overline{\mathrm{u}}}

\newcommand{\sg}{\tilde{\mathrm{g}}}
\newcommand{\sq}{\tilde{\mathrm{q}}}
\newcommand{\sqd}{\tilde{\mathrm{d}}}
\newcommand{\squ}{\tilde{\mathrm{u}}}
\newcommand{\sqs}{\tilde{\mathrm{s}}}
\newcommand{\st}{\tilde{\mathrm{t}}}
\newcommand{\schi}{\tilde{\chi}}

{\end{list}}
\newcounter{enumct}
\newenvironment{Enumerate}{\begin{list}{\arabic{enumct}.}%
{\usecounter{enumct}\setlength{\topsep}{0.2mm}%
\setlength{\partopsep}{0.2mm}\setlength{\itemsep}{0.2mm}%
\setlength{\parsep}{0.2mm}}}{\end{list}}

\newlength{\abstwidth}
\newlength{\captivewidth}
\newcommand{\captive}[1]{\rule{5mm}{0mm}%
\begin{minipage}{\captivewidth}%
\caption[small]{#1}\end{minipage}}

\begin{document}

\sloppy

\pagestyle{empty}

\begin{flushright}
LU TP 02--46\\
hep-ph/0212264\\
December 2002
\end{flushright}

\vspace{\fill}

\begin{center}
{\LARGE\bf Baryon Number Violation}\\[3mm]
{\LARGE\bf and String Topologies}\\[10mm]
{\Large T. Sj\"ostrand\footnote{torbjorn@thep.lu.se} and %
P. Z. Skands\footnote{zeiler@thep.lu.se}} \\[3mm]
{\it Department of Theoretical Physics,}\\[1mm]
{\it Lund University,}\\[1mm]
{\it S\"olvegatan 14A,}\\[1mm]
{\it S-223 62 Lund, Sweden}
\end{center}

\vspace{\fill}

\begin{center}
{\bf Abstract}\\[2ex]
\begin{minipage}{\abstwidth}
In supersymmetric scenarios with broken $R$-parity, baryon number
violating sparticle decays become possible. In order to search for
such decays, a good understanding of expected event properties is
essential. We here develop a complete framework that allows detailed
studies. Special attention is given to the hadronization phase,
wherein the baryon number violating vertex is associated with the
appearance of a junction in the colour confinement field. This
allows us to tell where to look for the extra (anti)baryon directly
associated with the baryon number violating decay.
\end{minipage}
\end{center}

\vspace{\fill}

\clearpage
\pagestyle{plain}
\setcounter{page}{1}

\section{Introduction}

There are good reasons to consider supersymmetry (SUSY) as the next
logical step in formulating a consistent theory of particle physics.
It complements the Standard Model by the possibility to secure the
Higgs potential against excessive radiative corrections, and provides
a framework for understanding electroweak symmetry breaking. SUSY is
the largest symmetry, within the requirements of conventional field
theory, where internal symmetry groups can be combined with the
space--time Poincar\'e group. It thereby points the way towards
a unified theory also including gravity, e.g. based on superstrings.
For this and other reasons SUSY has been a prime scenario in the
search for manifestations of physics beyond the Standard Model.
Furthermore, in planning future experiments, SUSY offers a broad
range of distinctive signatures that form convenient targets for
detector performance and analysis strategies.

In the Minimal Supersymmetric extension of the Standard Model (MSSM),
the standard particle content, extended to two Higgs doublets, is
doubled up by the presence of superpartners to all normal particles.
A multiplicative quantum number called $R$-parity may be defined
by $R = (-1)^{2S + 3B + L}$, where $S$ is the particle spin, $B$
its baryon number and $L$ its lepton number, such that normal
particles have $R = 1$ while the superpartners have $R = -1$. If
$R$-parity is conserved, supersymmetric particles are pair produced,
and the Lightest Supersymmetric Particle (LSP) is stable and hence
a Dark Matter candidate. However, there is at present no deep theoretical
motivation why $R$-parity should not be broken. This paves the way
both for baryon number violating (BNV = baryon number
violation/violating in the following) and lepton number violating
(LNV) processes. If both of these are allowed, proton decay would be
extremely rapid, unless the relevant couplings are abnormally tiny
\cite{hinchliffe95}.
As a rule, phenomenology is therefore based on assuming either BNV
or LNV, but not both.

The supersymmetric particles that could be produced in high-energy
colliders normally are not directly observable: either they decay
rapidly or they are weakly interacting and escape detection. With
$R$-parity conserved, the signals would consist of jets, leptons and
missing $E_{\perp}$ from escaping neutrinos and LSP's. In scenarios
with BNV the main decay product is jets, with only few leptons or
neutrinos, and so observability above QCD backgrounds becomes far from
trivial at hadron colliders such as the Tevatron or the LHC. In order
to carry out realistic studies it is therefore necessary to have a
detailed understanding of the properties of both signal and background
events. The prime tool for achieving such an understanding is to
implement the relevant processes in event generators, where simulated
events can be studied with all the analysis methods that could be used
on real events. Main generators used for
SUSY studies include \textsc{Isajet} \cite{Isajet}, \textsc{Herwig}
\cite{Herwig}, \textsc{Susygen} \cite{Susygen}, and \textsc{Pythia}
\cite{Pythia}. Traditionally the \textsc{Pythia} framework assumes
$R$-parity conservation, but recently LNV processes have been
implemented and some first studies carried out \cite{Peterthesis,Peterlviol}.

In the past, BNV has been modelled \cite{Herwigmodel,Herwigsusy} and studied
\cite{Herwigstud} in detail in the \textsc{Herwig} framework, with
emphasis on the perturbative aspects  of the production process. In
this article we present a corresponding implementation in
\textsc{Pythia}
where a special effort is dedicated to the non-perturbative aspects,
i.e.\ what happens with the (anti)baryon number generated by the BNV.
This allows us to address the possibility to obtain the ``smoking-gun''
evidence that a BNV decay has occurred, with questions such as
\textit{Could the presence of a violated baryon number be directly
observed?} and \textit{If so, what strategy should be used?}. A brief
presentation
of this work can be found in \cite{skandssusy02}.

In addition, many differences exist between the \textsc{Pythia}
and \textsc{Herwig} physics scenarios, for parton showers and
underlying events, allowing useful cross-checks to be
carried out and uncertainties to be estimated.

The outline of the article is as follows. In section 2, we briefly summarize
the short distance physics associated with BNV SUSY and
give an account of its implementation into \textsc{Pythia}. Next, in section
3, we turn our attention to a parton-shower description of
gluon emission in BNV decays. In section 4,
the special aspects related to the presence of
a net baryon number in the hadronization process are described
and an approach based on the Lund model of QCD strings is developed.
In section 5 we concentrate on tests of
the model, with special attention to the fate of the
(anti)baryon produced by BNV. Some (semi)realistic studies are collected
in section 6. Finally, section 7 summarizes and gives an outlook.

\section{The Baryon Number Violation Scenario \label{s:bnv}}

In Supersymmetric theories, it is usually convenient to work at the
superpotential level rather than at the Lagrangian level, since
the former has a simpler structure and is in any case related to the latter
by straightforward manipulations, see e.g.~\cite{tata97}. The most
general superpotential which can be written down for the MSSM includes 4
$R$-parity odd terms:
\begin{equation}
W^{\mathrm{MSSM}}_{\mathrm{RPV}} =
\frac12\lambda_{ijk}\epsilon^{ab}L_a^iL_b^j\bar{E}^k +
\lambda'_{ijk}\epsilon^{ab}L_a^{i}Q_{b}^{j\alpha}\bar{D}_{\alpha}^{k} +
\frac12\lambda''_{ijk}\epsilon^{\alpha_1\alpha_2\alpha_3}
\bar{U}^i_{\alpha_1}\bar{D}^j_{\alpha_2}\bar{D}^k_{\alpha_3}
+ \kappa_{i}L^i_a H_2^a \label{eq:wmssm}
\end{equation}
where $i,j,k$ run over generations, $a,b$ are SU(2)$_L$ isospin indices,
and $\alpha_{(i)}$ runs over colours. $L$ and $Q$ are
the SU(2) doublet (left-handed) chiral superfields for (s)leptons and
(s)quarks, respectively, and $E$, $U$, and $D$ are the SU(2) singlet
(right-handed) superfields for charged (s)leptons, (s)up type, and (s)down
type quarks, respectively. The last term represents a mixing between the
left-handed (s)leptons and one of the Higgs superfields.

There are two noteworthy aspects about the third term in
eq.~(\ref{eq:wmssm}), which we shall henceforth refer to as the
``UDD'' term: 1) it is the only one which has a non-zero baryon
number (all the others violate lepton number),
and 2) the colour indices have a Levi-Civita tensor structure. Of course,
these two aspects are really one and the same. They express the simple fact
that, since SU(3) of colour is unbroken, the only way to make a colour
singlet is to combine the three colours antisymmetrically, e.g.\ as used in
constructing ordinary baryon wavefunctions.

In a $B$-conserving theory like the SM or the
$R$-conserving MSSM, there is no colour antisymmetric perturbative
interaction term, i.e.\ no term with a colour structure like that of the UDD
term. Apart from extreme occurrences, like
knocking two valence quarks out of the same proton in different directions,
by two simultaneous but separate interactions, normal high-energy events
would therefore not fully display the antisymmetric colour structure of
the proton. Instead, it is normally enough to consider a baryon as consisting
of a colour triplet quark and a colour antitriplet
diquark, where the internal structure
of the latter need not be specified. So what is different about the UDD term
is that it allows the production of three colour carriers at large momentum
separation, without the creation of corresponding anticolour carriers.
It is the necessary SU(3) gauge connection between these three partons
 that will lead us in the development of the nonperturbative framework.

A further point about the UDD term is that the contraction of the
$\epsilon$ tensor with $\bar{D}^j\bar{D}^k$ implies that $\lambda''_{ijk}$
should be chosen antisymmetric in its last two indices, since a
$(j,k)$-symmetric part would cancel out.

The part of the Lagrangian
coming from the UDD superpotential term in which we are interested is:
\begin{equation}
\mathcal{L}_{\mathrm{BNV}} = {\textstyle\frac12}\lambda''_{ijk}
\epsilon^{\alpha_1\alpha_2\alpha_3}
\left(\bar{u}^i_{R\alpha_1}(\tilde{\d}^*)^{j}_{R\alpha_2}(\d^c)_{R\alpha_3}^k
+\bar{\d}^j_{R\alpha_1}(\tilde{\u}^*)^i_{R\alpha_2}(\d^c)^k_{R\alpha_3}
- (j\leftrightarrow k)
\right)+ h.c. \label{eq:bnvlag}
\end{equation}
where we have made the choice of not yet using any of the antisymmetry
requirements, so that the ordinary Einstein summation convention applies.
Superscript $c$ implies charge conjugation and $\sq^*$ denotes a
charge (=complex) conjugate squark.
From this, the possible lowest-order
BNV 3-point functions can immediately be read
off. In this paper, we only consider sparticle decays; BNV production
mechanisms are ignored. This is first and foremost a conservative approach,
since any additional sparticle production could only increase the observable
signal. Secondly, the underestimation of the sparticle production cross
sections is small as long as 1) the BNV couplings are small compared to the
gauge couplings and 2) the squarks are not so heavy that single-squark
production via BNV is significantly enhanced over the ordinary pair
production processes. For discussions of single squark production, see
\cite{allanach99}.

Combining the vertices in eq.~(\ref{eq:bnvlag})
with the full MSSM Lagrangian, also decays involving one or
more gauge couplings are clearly possible, e.g.\ neutralino decay via
$\schi^0\to\tilde{\q}_i(\to \bar{\q}_j\bar{\q}_k)\bar{\q}_i$. The BNV
SUSY decay processes currently implemented in \textsc{Pythia}, with Born
level matrix elements as calculated by
\cite{Herwigmodel}, are:\vspace*{1mm}\\
\begin{tabular}{llclr}
1)\hspace*{4mm}&$\sqd_{jn}$&$\to$&$\bar{\u}_i\bar{\d}_k$ & (36)
\vspace*{1mm}\\
2)&$\squ_{in}$&$\to$&$\bar{\d}_j\bar{\d}_k$ & \hspace*{5mm}(18)
\vspace*{1mm}\\
3)&$\schi^0_n$&$\to$&$\u_i\d_j\d_k$ & (144) \vspace*{1mm}\\
4)&$\schi^+_n$&$\to$&$\u_i\u_j\d_k$ & (30) \vspace*{1mm}\\
5)&$\schi^+_n$&$\to$&$\bar{\d}_i\bar{\d}_j\bar{\d}_k$ & (14) \vspace*{1mm}\\
6)&$\sg$&$\to$&$\u_i\d_j\d_k$ & (36) \vspace*{1mm}\\
\end{tabular}\\
where $n$ runs over the relevant mass eigenstates: $n\in\{L,R\}$ for the
 first two generations of squarks, $n\in\{1,2\}$ for the third generation
 squarks and the charginos, and
 $n\in\{1,...,4\}$ for the neutralinos. The numbers in
brackets are the number
of modes when summed over $n$, $i$, $j$, and $k$, and over charge conjugate
modes for the Majorana particles.

The matrix elements for these processes, as implemented in \textsc{Pythia},
are not quite identical to those used in \textsc{Herwig}. Most
importantly, \textsc{Pythia} uses running masses
and couplings at some points
where \textsc{Herwig} does not. See \cite{Peterlviol} for a list of these
differences.

When calculating the partial widths (and hence also the rates)
into these channels, we integrate the matrix elements over the
full phase space with massive $\b$ and $\t$ quarks, 
and massive sparticles. All other particles are only treated as
massive when checking whether the decay is kinematically allowed or not,
i.e.\ they are massless in the phase space integration.

A feature common to both programs is how double-counting in the BNV three-body
modes is avoided. The diagrams for these modes contain intermediate squarks
which may be either on or off the mass shell, depending on the other masses
involved in the process. If a resonance can be on shell, we risk doing double
counting since \textsc{Pythia} is then already allowing the process, in the
guise of two sequential $1\to 2$ splittings. Technically, this means that
the list of $1\to3$ BNV widths obtained by a call to
\texttt{PYSTAT(2)} only represent the non-resonant contributions, the resonant
ones being accounted for by sequences of $1\to 2$ splittings in
other parts of the code.

In the description of the momentum distribution in a
three-body resonance decay, the default \textsc{Pythia} procedure is to assume
an isotropic phase space, i.e.\ the matrix-element information used above to
obtain partial widths is neglected here. This should not be a bad approximation
when the intermediate squark propagators are far from mass shell over the full
phase space and therefore do not vary much, but could be a problem when a
squark mass (plus the mass of the corresponding quark) is only slightly above
the gaugino one.
In section \ref{s:modtest}
we return to this issue, comparing with \textsc{Herwig}, where the full,
unintegrated matrix elements are used to give a more correct phase space
population.

\section{Parton Showers}

The production and decay of unstable particles is normally associated
with brems\-strahlung emission of gluons and/or photons as applicable.
This radiation is conveniently described in the parton-shower language,
where the effects of multiple emissions are explicitly included by
iterative applications of the relevant splitting kernels, such as
$\q \to \q \g$, $\g \to \g \g$, $\g \to \q \qbar$ and $\q \to \q \gamma$.
Even though existing showering algorithms typically only fully resum the
leading logarithms, most of the effects of next-to-leading logs are also
included in the form of energy--momentum conservation, recoil effects,
scale choice in $\alphas$, coherence effects, and matching of energetic
emissions to higher-order matrix elements. At LEP the \textsc{Pythia},
\textsc{Herwig} and \textsc{Ariadne} \cite{Ariadne} radiation routines
have been well tested. All three are found to describe the 
data well, although
some problems exist \cite{LEPstudy}. Since \textsc{Ariadne} has not
been used for SUSY studies, we restrict the continued discussion to the
former two.

The \textsc{Herwig} algorithm is based on an evolution variable that
ensures angular ordering;
subsequent emissions occur at
smaller and smaller angles. Thereby coherence effects are respected,
i.e.\ double-counting of emission is avoided. The colour flow of the
hard process defines cones around the partons, within which all
emissions occur. In the limit where each emitted gluon is much
softer than the hard-scattering partons, and much softer than all
preceding gluons emitted, this approach can be shown to give the correct
emission rate, and also when going away from this limit
it should provide a good
overall description. However, a consequence of the way the kinematics of the
algorithm is constructed is that ``dead zones'' occur, within
which no emission at all is possible \cite{Sey}. Thus, in
$\e^+\e^- \to \q\qbar\g$ events, the region of energetic gluon emission
at large angles to both the $\q$ and $\qbar$ is completely empty.
The solution \cite{Sey} is to combine two classes of events, both
$\q\qbar$ and $\q\qbar\g$ ones, with the latter picked in
the ``dead zone'' region of the three-body phase space according to the
relevant matrix elements. All events, i.e.\ from either class, are then
allowed to shower further. However, such corrections have
only been worked out for a few cases, e.g.\ $\e^+\e^- \to \q\qbar\g$
and top decay \cite{Cor}.

The \textsc{Pythia} algorithm is based on evolution in the virtuality
of the emitting parton, with the requirement of angular ordering
enforced by vetoing non-angular-ordered emissions. This gives a less
exact description of coherence in the soft-gluon limit. Yet it does allow
the full three-body phase space to be populated, simply by not
imposing any angular ordering constraint for the first emission of each
of the two original partons. As it turns out, the incoherent addition of
radiation from these two sources tends to give an overestimation of the
$\q\qbar\g$ rate in the hard, non-collinear gluon region, while the
soft and collinear regions give what they should. It is therefore
straightforward to introduce a correction factor, whereby a fraction
of the emissions are vetoed, so that the remaining fraction agrees
with the desired three-jet rate \cite{Mats}. Some years ago, the
relevant gluon-emission corrections were calculated and implemented
for most two-body decays within the SM and the MSSM, with $R$-parity
conserved \cite{Emanuel}.

Normal processes are characterized by unbroken colour lines: the colours
present in the initial state will also be there in the final state,
with the proviso that opposite colours and anticolours may be
pair-produced or annihilated. This information is used in parton showers,
e.g. to set up emission cones. BNV processes are different in this
respect: in a $\sq \to \qbar\qbar$ branching a blue colour line may end
and antired and antigreen begin. Therefore the standard rules need to be
supplemented. In \textsc{Herwig} an effort has been made to study the
different new topologies and use the soft-gluon eikonal, i.e.\
spin-independent, expressions
to set up the relevant maximum emission cones in the different
processes \cite{Herwigmodel}. Still,
the ``dead zone'' issue is not
addressed, meaning that the rate of energetic, wide-angle gluon emission
is almost certainly underestimated.

By contrast, we shall here allow emission over the full phase space, like
before. However, not having calculated the corresponding matrix-element
correction factors, one should not expect the correct rate in the
hard-gluon region. Technically, in the lack of further knowledge, the
$\sq \to \qbar\qbar$ process is corrected to the eikonal
answer for a colour singlet decaying to a triplet
plus an antitriplet \cite{Emanuel}. For three-body decays such as
$\schi \to \q\q\q$ or $\sg \to \q\q\q$, no matrix element corrections
are available.

With three (or more) primary partons to shower, one is left with the
issue how the kinematics from the on-shell matrix elements should be
reinterpreted for an off-shell multi-parton configuration. We have made
the arbitrary choice of preserving the direction of motion of each parton
in the rest frame of the system, which means that all three-momenta are
scaled down by the same amount, and that some particles gain energy at
the expense of others. Mass multiplets outside the allowed phase space
are rejected and the evolution continued.

In principle, radiation in the initial and final states may
interfere, thereby requiring the inclusion of further coherence
conditions. The extent of such interference critically depends on the
width of the intermediate resonance: only gluons with energies below
this width are emitted coherently from initial and final partons
\cite{reswidth}. We here assume that the resonances are sufficiently
narrow that such interference effects can be neglected; see further the
discussion in section \ref{subsec:morejunc}.

The bottom line is that \textsc{Pythia} likely overestimates the
hard-gluon emission rate, whereas \textsc{Herwig} underestimates it.
This makes it interesting to compare the two descriptions, and
possibly to use differences as a first estimation of systematic
uncertainties in our current description of BNV processes.

\section{String Topologies}

Up till now we have considered short-distance processes, where
perturbation theory provides a valid description in terms of quarks,
gluons and other fundamental particles. At longer distances, the
running of the strong coupling $\alphas$ leads to confinement and a
breakdown of the perturbative description of QCD processes. Instead
there is a transition to a nonperturbative r\'egime, characterized by
hadrons rather than by partons. In the lack of an exact approach, this
hadronization process must be modelled. The perhaps most successful
and frequently used approach is the Lund string fragmentation model
\cite{Lundstring}.

This approach has not before been applied to the colour topologies
encountered in BNV. We therefore here extend the string model by the
introduction of a junction, where three string pieces come together.
Effectively, it is this junction that carries the (anti)baryon number
that is generated by a BNV process. The hadronization in the region
around the junction will therefore be of special interest.

\subsection{The normal string}

In this subsection we summarize some properties of the ordinary
Lund string model, highlighting the concepts that will be needed for
the developments in the remainder of the article. Readers who
are already familiar with the fundamentals of the string model may
wish to proceed directly to the next subsection.

To illustrate the string model, it is useful to start with the
simplest possible system, a colour-singlet $\q \qbar$ 2-jet event,
as produced in $\e^+\e^-$ annihilation. Here lattice QCD studies
lend support to a linear confinement picture in the absence
of dynamical quarks, i.e.\ in the quenched approximation \cite{ESW}.
Thus the energy stored in the colour dipole field between
a charge and an anticharge increases linearly with the separation
between the charges, if the short-distance Coulomb term is
neglected. This is quite different from the behaviour in QED,
and is related to the non-Abelian character of QCD. The dynamical
mechanisms involved are not fully understood, however.

The assumption of linear confinement provides the starting point for
the string model. As the $\q$ and $\qbar$ partons move apart from
their common production vertex, the physical picture is that of a
colour vortex line, or maybe a colour flux tube, being stretched
between the $\q$ and the $\qbar$. (The difference between these two
terminologies is related to whether the QCD vacuum more resembles a
type II or a type I superconductor. This is an interesting question
by itself, with the vortex line analogy somewhat favoured by theoretical
prejudice, but it will not be crucial for the continued discussion here.)
The transverse dimensions of the tube are of typical hadronic sizes,
roughly 1 fm. If the tube is assumed to be uniform along its length,
this automatically leads to a confinement picture with a linearly
rising potential. In order to obtain a Lorentz covariant and causal
description of the energy flow due to this linear confinement,
the most straightforward approach is to use the dynamics of the
massless relativistic string with no transverse degrees of freedom.
The mathematical, one-dimensional string can be thought of as
parameterizing the position of the axis of a cylindrically symmetric
flux tube or vortex line. From hadron spectroscopy, the string constant,
i.e.\ the amount of energy per unit length, is deduced to be
$ \kappa \approx 1$~GeV/fm.

For a massless $\q\qbar$ pair moving out along the $\pm z$ axis,
the original energy and momentum of the quark is then reduced with
time according to
\begin{equation}
E_{\q}(t) = p_{z\q}(t) = E_{\q}(0) - \kappa t ~,
\end{equation}
with a corresponding equation for $E_{\qbar}(t) = -p_{z\qbar}(t)$.
This classical equation of motion obviously does not take into
account the quantum mechanical uncertainty relation; in this sense
one may conceive of the (unmeasured) space--time picture as a subordinated
intermediate step in the derivation of the (measured) energy--momentum
picture.

If the quark instead has a nonvanishing rest mass $m_{\q}$ the
equation of motion is
\begin{equation}
\frac{\d p_{z\q}(t)}{\d t}
= \frac{\d p_{z\q}(t)}{\d E_{\q}(t)} \, \frac{\d E_{\q}(t)}{\d t}
= \frac{E_{\q}(t)}{p_{z\q}(t)} \, \left( - \kappa \,
   \frac{\d z}{\d t} \right)
= \frac{E_{\q}(t)}{p_{z\q}(t)} \, \left( - \kappa \,
   \frac{p_{z\q}(t)}{E_{\q}(t)} \right)
= - \kappa  ~,
\label{eq:dpdt}
\end{equation}
which has the solution
\begin{eqnarray}
p_{z\q}(t) & = & p_{z\q}(0) - \kappa t ~;\\
E_{\q}(t) & = & \sqrt{E_{\q}^2(0) - 2 \kappa t \, p_{z\q}(0) +
\kappa^2 t^2 } ~.
\end{eqnarray}
A key aspect, which we will make use of for the junction phenomenology
below, is that the momentum loss per unit of time is independent of the
quark mass. 

In the back-to-back configuration the string does not carry
any net momentum, but acquires energy proportional to its total length
as the $\q$ and $\qbar$ move apart. Once the endpoint quarks have lost
their momenta, they will be pulled back by the string tension and
re-acquire momenta in the opposite direction, at the same time
as the string length shrinks back. When the quarks meet again the
force flips sign and a second  half-period of the oscillation begins.

Such stable ``yo-yo'' modes are used as simple representations of mesons.
For a high-invariant-mass $\q \qbar$ system, the fragmentation into
lower-mass mesons proceeds through the production of new $\q' \qbar'$ pairs,
corresponding to dynamical quarks in lattice QCD. Thereby the original
system first splits into two colour-singlet systems, $\q \qbar'$ and
$\q' \qbar$. If the invariant mass of either of these string pieces is
large enough, further breaks occur. In the Lund model, the
string break-up process is assumed to proceed until only on-mass-shell
hadrons remain, each hadron corresponding to a small piece of string
with a quark in one end and an antiquark in the other. This process is
illustrated in Fig.~\ref{fig:stringfrag}.

\begin{figure}
\begin{center}
\begin{picture}(250,160)(-110,-10)
\SetScale{0.7}
\SetWidth{2}
\LongArrow(160,0)(220,0)\Text(168,0)[]{$z$}
\LongArrow(160,0)(160,60)\Text(112,56)[]{$t$}
\Vertex(95,105){3}
\Vertex(55,85){3}
\Vertex(25,75){3}
\Vertex(-15,65){3}
\Vertex(-50,80){3}
\Vertex(-95,105){3}
\ArrowLine(0,0)(150,150)\Text(95,75)[]{$\q$}
\Line(150,150)(140,160)\Line(140,160)(180,200)
\Line(95,105)(75,125)\Line(75,125)(135,185)\Line(135,185)(120,200)
\Line(55,85)(35,105)\Line(35,105)(75,145)\Line(75,145)(55,165)
\Line(55,165)(90,200)
\Line(25,75)(-5,105)\Line(-5,105)(45,155)\Line(45,155)(15,185)
\Line(15,185)(30,200)
\Line(-15,65)(-65,115)\Line(-65,115)(-45,135)\Line(-45,135)(-95,185)
\Line(-95,185)(-80,200)
\Line(-50,80)(-120,150)\Line(-120,150)(-100,170)\Line(-100,170)(-130,200)
\Line(-95,105)(-190,200)
\ArrowLine(0,0)(-150,150)\Text(-95,75)[]{$\qbar$}
\Line(95,105)(190,200)
\Line(55,85)(115,145)\Line(115,145)(95,165)\Line(95,165)(130,200)
\Line(25,75)(65,115)\Line(65,115)(45,135)\Line(45,135)(85,175)
\Line(85,175)(65,195)\Line(65,195)(70,200)
\Line(-15,65)(35,115)\Line(35,115)(5,145)\Line(5,145)(55,195)
\Line(55,195)(50,200)
\Line(-50,80)(-30,100)\Line(-30,100)(-80,150)\Line(-80,150)(-60,170)
\Line(-60,170)(-90,200)
\Line(-95,105)(-75,125)\Line(-75,125)(-145,195)\Line(-145,195)(-140,200)
\Line(-150,150)(-140,160)\Line(-140,160)(-180,200)
\SetWidth{3}
\Line(-25,25)(25,25)
\Line(-50,50)(50,50)
\Line(-75,75)(-25,75)\Line(-5,75)(75,75)
\Line(-100,100)(-70,100)\Line(-50,100)(-30,100)\Line(0,100)(20,100)
\Line(40,100)(50,100)\Line(70,100)(100,100)
\Line(-125,125)(-115,125)\Line(-95,125)(-75,125)\Line(15,125)(25,125)
\Line(75,125)(95,125)\Line(115,125)(125,125)
\Line(-150,150)(-140,150)\Line(-120,150)(-100,150)\Line(-80,150)(-60,150)
\Line(10,150)(40,150)\Line(60,150)(70,150)\Line(100,150)(110,150)
\Line(140,150)(150,150)
\Line(-165,175)(-155,175)\Line(-125,175)(-105,175)\Line(-85,175)(-65,175)
\Line(25,175)(35,175)\Line(65,175)(85,175)\Line(105,175)(125,175)
\Line(155,175)(165,175)
\end{picture}
\end{center}
\captive{%
The breakup of an original $\q \qbar$ system into a set of mesons,
each represented by a yo-yo state. For simplicity quarks are here
assumed massless, and so move along lightcones. The broken horizontal
lines illustrate the string pieces at a few discrete times.
\label{fig:stringfrag}}
\end{figure}

In general, the different string breaks are causally disconnected.
This means that it is possible to describe the breaks in any convenient
order, e.g.\ from the quark end inwards. One therefore is led to
formulate an iterative scheme for the fragmentation, as follows.
Assume, as above, an initial quark $\q$ moving out along the $+z$ axis,
with the antiquark going out in the opposite direction.
By the production of a $\q_1 \qbar_1$ pair, a meson with flavour content
$\q \qbar_1$ is produced, leaving behind an unpaired quark $\q_1$.
A second pair $\q_2 \qbar_2$ may now be produced, to give a new meson
with flavours $\q_1 \qbar_2$, etc. At each step the produced
hadron takes some fraction of the available energy and momentum.
This process may be iterated until all energy is used up, with some
modifications close to the $\qbar$ end of the string in order to
make total energy and momentum come out right, since all hadrons are required
to be on mass shell.

The choice of starting the fragmentation from the quark end is
arbitrary, however. A fragmentation process described in terms of
starting at the $\qbar$ end of the system and fragmenting towards
the $\q$ end should be equivalent. This ``left--right'' symmetry
constrains the allowed shape of the fragmentation function $f(z)$,
where $z$ is the fraction of the remaining light-cone momentum
$E \pm p_z$ (+ for the $\q$ jet, $-$ for the $\qbar$ one) taken by
each new particle. The resulting ``Lund symmetric fragmentation
function'' has two main free parameters, which are determined from data.

Viewed in time, the fragmentation process actually starts near the
middle of the event and spreads outwards. The $\q'\qbar'$ production
vertices on the average occur along a hyperbola of constant invariant
time $\tau$, $\tau^2 = t^2 - z^2$, so even if the string is boosted
along the $z$ axis it is still the slow particles in this new frame
that are the ones produced first. In this sense, a Lorentz-covariant
``inside-out'' cascade can technically be described by an ``outside-in''
iteration scheme. As an order-of-magnitude,
$\langle\tau\rangle \approx 1.5$~fm or
$\langle\kappa\tau\rangle \approx 1.5$~GeV.

In order to generate the quark--antiquark pairs $\q' \qbar'$ which
lead to string break-ups, the Lund model invokes the idea of
quantum mechanical tunnelling. This gives a flavour-independent
Gaussian spectrum for the $\pT$ of $\q' \qbar'$ pairs.
Since the string is assumed to have no transverse excitations,
this $\pT$ is locally compensated between the quark and the
antiquark of the pair. The total $\pT$ of a hadron is made
up out of the $\pT$ contributions from the quark and
antiquark that together form the hadron. Some contribution of
soft  unresolved perturbative gluon emission may also effectively
be included in this description.

The tunnelling picture implies a suppression of heavy-quark
production, $\u : \d : \s : \c \approx 1 : 1 : 0.3 : 10^{-11}$.
Charm and heavier quarks hence are not expected to be produced in
the soft fragmentation, but only in perturbative parton-shower
branchings $\g \to \q \qbar$.

When the quark and antiquark from two adjacent string breaks are
combined to form a meson, it is necessary to invoke an algorithm to
choose between the different allowed possibilities, notably
between pseudoscalar and vector mesons.
Here the string model is not particularly predictive. Qualitatively one
expects a $1 : 3$ ratio, from counting the number of spin states,
multiplied by some wave-function normalization factor, which should
disfavour heavier states.

A tunnelling mechanism can also be used to explain the production of
baryons. In the simplest possible approach, a diquark in a colour
antitriplet state is just treated like an ordinary antiquark, such
that a string can break either by quark--antiquark or
antidiquark--diquark pair production \cite{diquark}.
A more complex scenario is the ``popcorn'' one \cite{popcorn}, where
diquarks as such do not exist, but rather quark--antiquark pairs
are produced one after the other. This latter picture gives a less
strong correlation in flavour and momentum space between the
baryon and the antibaryon of a pair.

In all fairness, it should be said that the description of baryon
production is one of the less predictive aspects of the Lund
model, with many free parameters for the flavour composition that
need to be determined from the data. Most single-baryon spectra are
well described after such a tuning, except possibly at very
large momentum fractions \cite{fragrev}. A reasonable description is
also obtained for baryon--antibaryon correlations, although with some
disagreements. For the aspects that we will study here, and within
the precision allowed by other considerations, the baryon production
model should be fully adequate, however.

\begin{figure}
\begin{center}
\begin{picture}(450,225)(-210,-80)
\SetWidth{1.5}
\LongArrow(160,-70)(200,-70)\Text(210,-70)[]{$z$}
\LongArrow(160,-70)(160,-30)\Text(160,-20)[]{$x$}
\LongArrow(0,0)(200,0)\Text(210,0)[l]{$\q$ ($r$)}
\LongArrow(0,0)(-160,120)\Text(-160,132)[]{$\g$ ($\overline{r}b$)}
\LongArrow(0,0)(-190,-62)\Text(-190,-75)[]{$\qbar$ ($\overline{b}$)}
\SetWidth{1}
\DashLine(80,0)(-64,48){5}
\DashLine(-64,48)(-76,-25){5}
\SetWidth{2}
\DashLine(160,0)(-128,96){5}
\DashLine(-128,96)(-152,-50){5}
\end{picture}
\end{center}
\captive{%
The string motion in a $\q \qbar \g$ system, neglecting hadronization.
The $\q$, $\qbar$ and $\g$ move out from the common origin, all with
the speed of light and along straight lines, in the limit that quark
masses are neglected. The connecting dashed lines illustrate the
string pieces at two times. A possible colour assignment is indicated
within brackets.
\label{fig:qqg}}
\end{figure}

If several partons are moving apart from a common origin, the details
of the string drawing become more complicated. For a $\q \qbar \g$
event, a string is stretched from the $\q$ end via the $\g$ to the
$\qbar$ end, Fig.~\ref{fig:qqg}, i.e.\ the gluon is a kink on the
string, carrying energy and momentum.
As a consequence, the gluon has two string pieces attached, and
the ratio of gluon to quark string force is 2, a number which
can be compared with the ratio of colour-charge Casimir operators,
$N_C/C_F = 2/(1-1/N_C^2) = 9/4$. In this, as in other
respects, the string model can be viewed as a variant of QCD
where the number of colours $N_C$ is not 3 but infinite.
Note that the factor 2 above is independent of the actual 
kinematical configuration: 
a smaller opening angle between two partons corresponds to a smaller
string length being drawn out per unit time, but also to a larger
transverse velocity of the string piece, 
thereby increasing its energy density. In
fact, these two factors exactly cancel so 
that the energy loss per unit time indeed becomes the same.

The $\q \qbar \g$ string will fragment along its length. To first
approximation this means that there is
one fragmenting string piece between
$\q$ and $\g$ and a second one between $\g$ and $\qbar$. One hadron
is straddling both string pieces, i.e.\ sitting around the gluon
corner. The rest of the particles are produced as in two simple
$\q \qbar$ strings, but strings boosted with respect to the overall
CM frame. When considered in detail, the string motion and
fragmentation is more complicated, with the appearance of
additional string regions during the time evolution of the system
\cite{stringfrag}. These corrections are especially important for soft
and collinear gluons, since they provide a smooth transition between
events where such radiation took place and events where it did not.
Therefore the string fragmentation scheme is ``infrared safe'' with
respect to soft or collinear gluon emission.

\subsection{The junction}

As we saw above, meson states can be represented by little yo-yo's,
with a string piece stretched between the $\q$ and $\qbar$. What
would be the corresponding representation for a baryon? In normal
string fragmentation the issue is not so interesting, since the
hadron size is consistent with the string width, meaning the
internal structure is not really resolved. Thus the mesonic yo-yo
is more of a convenient concept for book-keeping, and it is only
for higher-mass systems that the elongated string structure really
becomes of relevance. The equivalent situation for a baryon state
would then be when several valence quarks are kicked out of an
incoming proton or, as here, when baryon number is violated in
Supersymmetry. In its simplest form, it can be illustrated by
the decay $\schi_1^0 \to \u_i\d_j\d_k$, or equivalently a proton in which
all valence quarks are kicked out in different directions at the
same time.

\begin{figure}
\begin{center}
\begin{picture}(450,225)(-210,-80)
\SetWidth{1.5}
\LongArrow(160,-70)(200,-70)\Text(210,-70)[]{$z$}
\LongArrow(160,-70)(160,-30)\Text(160,-20)[]{$x$}
\LongArrow(0,0)(200,0)\Text(210,0)[l]{$\u_i$ ($r$)}
\LongArrow(0,0)(-160,120)\Text(-160,132)[]{$\d_j$ ($g$)}
\LongArrow(0,0)(-190,-62)\Text(-190,-75)[]{$\d_k$ ($b$)}
\SetWidth{1}
\LongArrow(0,0)(-116,28)\Text(-125,30)[]{J}
\DashLine(-46,11)(80,0){5}
\DashLine(-46,11)(-64,48){5}
\DashLine(-46,11)(-76,-25){5}
\SetWidth{2}
\DashLine(-92,22)(160,0){5}
\DashLine(-92,22)(-128,96){5}
\DashLine(-92,22)(-152,-50){5}
\end{picture}
\end{center}
\captive{%
The string motion in a junction system, neglecting hadronization.
The $\u_i$, $\d_j$ and $\d_k$ move out from the common origin, all with
the speed of light and along straight lines, in the limit that quark
masses are neglected. The thin arrow indicates the resulting motion
of the junction J. The connecting dashed lines illustrate the Y-shaped
string topology at two discrete times. A possible colour assignment is
indicated within brackets.
\label{fig:Ytopology}}
\end{figure}

Our solution here is to introduce the concept of a junction, in
which a string piece from each of the three quarks meet, i.e.\ a
Y-shaped topology, Fig.~\ref{fig:Ytopology}. Each of the three
strings are of the normal kind already encountered. For instance,
the string from the blue $\d_k$ quark acts as if there were an
antiblue antiquark in the junction position. Recall that the
colours of the two other quarks add like $ g + r = \overline{b}$,
given that the colour representation of the quarks is totally
antisymmetric so as to make the state a colour singlet.
Correspondingly for the other two string pieces. Only in the
immediate neighbourhood of the junction would the field topology
be more complicated.

\begin{figure}
\begin{center}
\begin{picture}(300,175)(-115,-75)
\Line(0,0)(10,0)\Line(0,0)(-8,6)\Line(0,0)(-8,-6)
\Vertex(10,0){1}\Vertex(-8,6){1}\Vertex(-8,-6){1}
\Text(10,-10)[]{$\q_4$}\Text(-8,16)[]{$\q_7$}\Text(-6,-16)[]{$\q_9$}
\Line(30,0)(50,0)\Vertex(30,0){1}\Vertex(50,0){1}
\Text(30,-10)[]{$\qbar_4$}\Text(50,-10)[]{$\q_3$}
\Line(70,0)(90,0)\Vertex(70,0){1}\Vertex(90,0){1.5}
\Text(70,-10)[]{$\qbar_3$}\Text(90,-10)[]{$\qbar\qbar_2$}
\Line(110,0)(130,0)\Vertex(110,0){1.5}\Vertex(130,0){1}
\Text(110,-10)[]{$\q\q_2$}\Text(130,-10)[]{$\q_1$}
\Line(150,0)(170,0)\Vertex(150,0){1}\Vertex(170,0){1}
\Text(150,-10)[]{$\qbar_1$}\Text(170,-10)[]{$\u_i$}
\Line(-24,18)(-40,30)\Vertex(-24,18){1}\Vertex(-40,30){1}
\Text(-24,28)[]{$\qbar_7$}\Text(-40,40)[]{$\q_6$}
\Line(-56,42)(-72,54)\Vertex(-56,42){1}\Vertex(-72,54){1}
\Text(-56,52)[]{$\qbar_6$}\Text(-72,64)[]{$\q_5$}
\Line(-88,66)(-104,78)\Vertex(-88,66){1}\Vertex(-104,78){1}
\Text(-88,76)[]{$\qbar_5$}\Text(-104,88)[]{$\d_j$}
\Line(-24,-18)(-40,-30)\Vertex(-24,-18){1}\Vertex(-40,-30){1}
\Text(-22,-28)[]{$\qbar_9$}\Text(-38,-40)[]{$\q_8$}
\Line(-56,-42)(-72,-54)\Vertex(-56,-42){1}\Vertex(-72,-54){1}
\Text(-54,-52)[]{$\qbar_8$}\Text(-70,-64)[]{$\d_k$}
\end{picture}
\end{center}
\captive{%
Hadronization by $\q'\qbar'$ production in a junction string topology.
The figure is in an abstract ``flavour space'', to be related to a
space--time topology like Fig.~\ref{fig:Ytopology}, with flavours
belonging to the same hadron connected by string pieces. The
labelling of new quarks is arbitrary. The $\q_4\q_7\q_9$ hadron
carries the original baryon number, while $\q_1\q\q_2$ and
$\qbar\qbar_2\qbar_3$ correspond to the possibility of
baryon--antibaryon pair production in the hadronization process,
and the rest represents normal meson production.
\label{fig:Yfrag}}
\end{figure}

In the hadronization, ordinary $\q'\qbar'$ pair production will
break up each of the three string pieces. This leads to meson
production, and in diquark/popcorn scenarios also to
baryon--antibaryon pair production, but not to the production of
any net baryon number. Instead it is the three $\q'$'s nearest to
the junction, one from each string piece, that team up to form
the baryon that is associated with the original baryon number,
Fig.~\ref{fig:Yfrag}.
In this sense the junction, by its colour topology, becomes the
carrier of the baryon number of the system, rather than the three
original quarks. This baryon typically comes to have a small
momentum in the rest frame of the junction, as we shall see.
This holds also in the rest frame of a typical
$\schi_1^0 \to \u_i\d_j\d_k$ decay, where the three quarks move out in
widely separated directions, since the junction then has a small
velocity.

The concept of a junction is not new. It was introduced in the
early string model of hadrons \cite{artrujunc,earlyjunc,mrv80},
and has been used in confinement studies to set up the colour
structure of baryons in the same way as we do here \cite{confjunction}.
Indeed, the simplest locally gauge invariant operator of baryon number
unity that can be constructed in SU(3) is \cite{mrv80}:
\begin{equation}
B_{i_1i_2i_3} = \epsilon^{\alpha_1\alpha_2\alpha_3}
\prod_{n=1}^3 P\!\left[e^{ig\int_{\mathcal{P}(x,x_n)} A_\mu
 \d x^\mu}q_{i_n}(x_n)\right]_{\alpha_n}
\end{equation}
where $i_n$ and $\alpha_i$ are indices for flavour and colour, respectively,
$A_\mu$ is the gluon field, and $P$ expresses ordering along the path
$\mathcal{P}$ from $x$ to $x_n$. Physically, this operator represents
three quarks sitting at $x_1$, $x_2$, and $x_3$, respectively, each
connected via the gluon field to the point $x$, which we may identify
as corresponding to the locus of the string junction. Thus, from this
point of view, QCD does appear to select the junction topology.

The junction concept has also been applied to
the understanding of the fragmentation function of baryons in
low-$\pT$ hadronic collisions \cite{lowptjunction},
providing a more sophisticated variant of the popcorn mechanism
whereby the two quarks of a diquark need not end up in the same
hadron. Multiple perturbative parton--parton interactions have
been suggested as a mechanism to activate the junction in proton
beams \cite{Zijl}. And junctions have been proposed for $\q\qbar\g$
configurations, in scenarios where a separate kind of colour octet
gluon string exists (which requires a string tension less than twice
that of the normal colour triplet string, or else it would be
energetically unfavourable) \cite{montjunction}.

A simplified and unofficial
implementation of junctions has existed in \textsc{Pythia} since
many years, and been used e.g. from \textsc{Susygen}. To the best
of our knowledge, however, what we present here is the first
complete scenario for the hadronization of generic junction
topologies in high-energy interactions.

It could have been interesting to contrast the junction concept
with some alternatives, but we have been unable to conceive of any
realistic such, at least within a stringlike scenario of confinement.
The closest we come is a V-shape topology, with two string pieces,
similar to the $\q\qbar\g$ topology in Fig.~\ref{fig:qqg}. This
would be obtained if
one e.g. imagined splitting the colour of the upper left quark
in Fig.~\ref{fig:Ytopology} as $g = \overline{r}\overline{b}$. In such
a scenario the baryon would be produced around this quark, and could
be quite high-momentum. Of course, such a procedure is arbitrary,
since one could equally well pick either of the three quarks to be in
the privileged position of producing the key baryon. Further,
with two string pieces now being pulled out from one of the quarks,
the net energy stored in the string at a given (early) time is
larger than in the junction case, meaning the Y junction is
energetically favoured over the V topology. For these reasons, the
V scenario has not been pursued.

It should be noted that, in our junction scenario, it can happen that
one quark has a low initial momentum in the junction rest frame, and
therefore never strays very far away. The string between this quark
and the junction need then not break, so that the quark ends up in the
central baryon together with the junction, i.e.\ the quark and the
junction effectively act as a unit. Seemingly this is akin to the
scenario sketched above but, in the junction topology framework, it
could only happen when the quark has low momentum in the junction
rest frame, so again the key baryon would be low-momentum.

Other alternatives appear even more far-fetched. For instance, we
see no possibility for a $\Delta$-shaped topology, i.e.\ connecting
each quark directly with the other two. Not only would it be even
more energetically unfavourable relative to the junction topology,
but it is also difficult to conceive of a colour assignment that would
at all allow it. One would then need to introduce a novel kind of
strings, that do not obey the normal QCD colour charge and flux
quantization conditions.

\subsection{Junction motion}
\label{ss:jmot}

In the rest frame of the junction the opening angle between any pair of
quarks is 120$^{\circ}$, i.e.\ we have a perfect Mercedes topology.
This can be derived from the action of the classical string
\cite{artrujunc}, but follows more directly from symmetry arguments.
What is maybe not so obvious is that the 120$^{\circ}$ condition also
is fulfilled if the quarks are massive. However, in the junction rest
frame each string piece is pulled straight out from the origin, without
any transverse motion, so the string pieces do not carry any momentum.
Then overall momentum conservation gives that
\begin{equation}
\frac{\d \mathbf{p}_{\mathrm{tot}}}{\d t} =
 \sum_{\q} \frac{\d \mathbf{p}_{\q}}{\d t} =
-  \kappa \, \sum_{\q} \mathbf{e}_{\q} = \mathbf{0} ~.
\label{eq:dpsum}
\end{equation}
Here we have used eq.~(\ref{eq:dpdt}) to give the momentum loss of
quarks of arbitrary mass, where $\mathbf{e}_{\q}$ is a unit vector
along the direction of motion of each of the three quarks.

The flow of energy and momentum is akin to that in the normal $\q\qbar$
system, where the intermediate string absorbs and cancels the momentum
of the receding $\q$ and $\qbar$. 
With respect to the
$\q\q\q$ case, 
it is implicit in eq.~(\ref{eq:dpsum}) that no net momentum is accumulated by
the junction. Instead, it acts as a switchyard for the momentum flowing 
in the system, thereby
cancelling the momentum given up by the three endpoint quarks.
In this way, the junction itself remains with a vanishing
four-momentum, which obviously holds in an arbitrary reference
frame. In a general frame the string pieces would have a transverse
motion as well, however, and thereby carry a nonvanishing momentum.
Then eq.~(\ref{eq:dpsum}) would need to be generalized to include
these additional terms and would become less transparent.

The rest frame of the junction can easily be found for the case of
three massless quarks (and no further gluons), but the general massive
case admits no analytical solution. A convenient numerical solution
can be obtained as follows. Imagine three quarks with four-momenta
$p_1$, $p_2$ and $p_3$. The four-products $a_{ij} = p_i p_j$ are
Lorentz invariants, and thus the boosted momenta $p'_i$ obey
\begin{equation}
a_{ij} = p'_i p'_j = E'_i E'_j - \mathbf{p}'_i \mathbf{p}'_j  =
E'_i E'_j - |\mathbf{p}'_i| \, |\mathbf{p}'_j| \, \cos\frac{2\pi}{3}
= E'_i E'_j + \frac{1}{2} \, |\mathbf{p}'_i| \, |\mathbf{p}'_j|
\end{equation}
in the junction rest frame. This leads us to introduce
\begin{equation}
f_{ij} \equiv f(|\mathbf{p}'_i| , |\mathbf{p}'_j|; m_i, m_j, a_{ij}) =
\sqrt{|\mathbf{p}'_i|^2 + m_i^2} \sqrt{|\mathbf{p}'_j|^2 + m_j^2}
 + \frac{1}{2} \, |\mathbf{p}'_i| \, |\mathbf{p}'_j| -  a_{ij} ~.
\end{equation}
Note that $f_{ij}$ is a monotonically increasing function of each of its
first two arguments. If we choose e.g.\ to let $|\mathbf{p}'_1|$ vary
freely within the kinematically allowed region, the requirements
$f_{12}=0$ and $f_{13}=0$ can then be uniquely solved to give the
other two momenta,
\begin{equation}
|\mathbf{p}'_j| = \frac{2 E'_1 \sqrt{4 a_{1j}^2 - m_j^2
(4{E'_1}^2 - |\mathbf{p}'_1|^2)} - 2 |\mathbf{p}'_1| \, a_{1j}}%
{4{E'_1}^2 - |\mathbf{p}'_1|^2}~,
\end{equation}
which both decrease with increasing
$|\mathbf{p}'_1|$. Therefore also $f_{23}$ is monotonically
decreasing if viewed as a function of $|\mathbf{p}'_1|$,
$f_{23} = f(|\mathbf{p}'_2|(|\mathbf{p}'_1|),%
|\mathbf{p}'_3|(|\mathbf{p}'_1|))$
 The final requirement $f_{23} = 0$ now gives a unique solution
$|\mathbf{p}'_1|$. This solution can be obtained by an iterative
interpolating procedure. In the massless case, the solution simplifies
to $E'_i = |\mathbf{p}'_i| = \sqrt{2 a_{ij} a_{ik} / 3 a_{jk}}$.

Once the energies in the junction rest frame are known, the construction
of a boost is straightforward. A first boost with
$\boldsymbol{\beta}^{\mathrm{CM}} = - \sum \mathbf{p}_i / \sum E_i$
brings the three quarks to their rest frame, with known momenta
$p_i^{\mathrm{CM}}$. A second boost $\boldsymbol{\beta}'$, that brings
the system to the junction rest frame, now obeys the three equations
$\gamma' E_i^{\mathrm{CM}} + \gamma'\boldsymbol{\beta}'%
\mathbf{p}_i^{\mathrm{CM}} = E'_i$. After division by $E_i^{\mathrm{CM}}$,
and subtraction of two of the equations from the third, one obtains e.g.\
\begin{equation}
\gamma'\boldsymbol{\beta}' \left(
\frac{\mathbf{p}_1^{\mathrm{CM}}}{E_1^{\mathrm{CM}}} -
\frac{\mathbf{p}_j^{\mathrm{CM}}}{E_j^{\mathrm{CM}}} \right) =
\frac{E'_1}{E_1^{\mathrm{CM}}} -
\frac{E'_j}{E_j^{\mathrm{CM}}} ~.
\end{equation}
Since the three vectors $\mathbf{p}_i^{\mathrm{CM}}$ lie in a plane,
the boost vector can be parameterized as a linear sum of the two
difference vectors defined by the above equation for $j=2$ and 3, which
gives a simple linear equation system for $\gamma'\boldsymbol{\beta}'$.
Finally, the overall boost is obtained by combining
$\boldsymbol{\beta}^{\mathrm{CM}}$ with $\boldsymbol{\beta}'$.

In the general case, there does not have to exist a solution with
perfect symmetry. Such troublesome events, characterized e.g.\ by
$f_{23} < 0$ for all $|\mathbf{p}'_1|$, are very rare, however.
We have only encountered them when a massive quark is almost at rest
in the frame that comes closest to giving a symmetrical topology.
We here accept an imperfect solution somewhere in the neighbourhood
of this singular point. Since it is the size of the boost that will
matter in the hadronization description, not the 120$^{\circ}$ opening
angles \textit{per se}, this should be acceptable.

\begin{figure}
\begin{center}
\begin{picture}(260,220)(-120,-110)
\SetWidth{1} \LongArrow(60,-90)(100,-90)\Text(110,-90)[]{$z$}
\LongArrow(60,-90)(60,-60)\Text(60,-40)[]{$x$}
\LongArrow(0,0)(100,0)\Text(110,0)[l]{$\u_i$ ($r$)}
\Gluon(0,0)(-50,87){5}{7}\Text(-50,100)[]{$\g$ ($g\overline{r}$)}
\LongArrow(0,0)(-80,60)\Text(-90,60)[r]{$\d_j$ ($r$)}
\Gluon(0,0)(-50,-87){5}{7}\Text(-50,-100)[]{$\g$
($b\overline{g}$)}
\LongArrow(0,0)(-80,-60)\Text(-90,-60)[r]{$\d_k$ ($g$)}
\SetWidth{2} \DashLine(0,-1)(100,-1){5}
\DashLine(0,0)(-50,87){5}\DashLine(-50,87)(-80,60){5}
\DashLine(0,0)(-50,-87){5}\DashLine(-50,-87)(-80,-60){5}
\end{picture}
\end{center}
\captive{%
String topology of a $\schi_1^0 \to \u_i\d_j\d_k$ decay, where the two
$\d_{(j,k)}$ quarks each has radiated a gluon. The event is drawn in
the rest frame of the junction at early times, where the $\u$ and the
two gluons are separated by 120$^{\circ}$, and the string topology
is shown by dashed lines. A possible colour assignment is indicated
within brackets.
\label{fig:Ywithrad}}
\end{figure}

So far we have assumed that the junction remains in uniform motion.
When gluon emission is included, this need no longer be the case.
Consider e.g.\ an event like the one in Fig.~\ref{fig:Ywithrad}.
Here the two $\d_{(j,k)}$ quarks each radiated a gluon, and so the strings
from these quarks to the junction are drawn via the respective gluon,
cf.\ Fig.~\ref{fig:qqg}. It is the direction of these gluons, together
with the $\u$ quark, that determines the junction motion at early times,
and the directions of the $\d_{(j,k)}$ quarks themselves are irrelevant.
As a gluon moves out from the junction origin, it loses energy at a rate
$\d E / \d t = -2\kappa$, where the 2 comes from it being attached to
two string pieces. So after a time $E_{\g}(0)/2\kappa$ it has lost
all its original energy, and after a further equal time this information
has propagated back to the junction. From then on it would be the
direction of the respective $\d_{(j,k)}$ quark, and not of the gluon, that
defines the pull on the junction. In the example of
Fig.~\ref{fig:Ywithrad}, the junction would originally be at rest but
later on start to move leftwards. In the generic configuration, each
of the outgoing quarks could radiate several gluons, that would be
colour-connected in a chain stretching from the endpoint quark inwards
to the junction. The innermost gluon on each of the three chains would
together determine the original motion, with the next-innermost taking
over after the innermost lost its energy, and so on, spreading outwards.
As a consequence the junction would ``jitter around''.

Now, all of this is moot if the string starts to fragment before the
innermost gluon lost its energy, because a string break will sever the
inward flow of momentum. As we noted above, the typical scale for this
to happen is of order $\langle \kappa\tau \rangle \approx 1.5$~GeV.
Therefore an innermost gluon with energy much above this scale by itself
defines the junction motion, whereas a gluon with energy much below
would act on the junction too short a time to matter.

Rather than trying to trace the junction jitter in detail --- which
anyway will be at or below the limit of what it is quantum mechanically
meaningful to speak about --- we define an effective pull of each string
on the junction as if from a single particle with a four-momentum
\begin{equation}
p_{\mathrm{pull}} = \sum_{i=1}^n \, p_i \, \exp \left( -
{\textstyle \sum}_{j=1}^{i-1}
E_j / E_{\mathrm{norm}} \right) ~.
\label{eq:ppull}
\end{equation}
Here $i=1$ is the innermost gluon, $i=2$ is the next-innermost one, and
so on till the endpoint quark $i=n$. The energy sum in the exponent runs
over all gluons inside the one considered (meaning it vanishes for $i=1$),
and is normalized to a free parameter $E_{\mathrm{norm}}$, which by default
we associate with the $\langle \kappa\tau \rangle$ above. Note that the
energies $E_j$ depend on the choice of frame. A priori, it is the energies
in the rest frame
of the junction which should be used in this sum, yet since these are
not known to begin with, we use an iterative procedure, starting with
the (known) energies in the CM of the string system, calculating the
corresponding pull vectors in this frame
and from them the candidate junction rest
frame, calculating the pull vectors in this new frame and so forth
until the Lorentz factor of the last boost is
$\gamma_{\mathrm{last}}<1.01\gamma_{\mathrm{tot}}$.

\subsection{Junction hadronization \label{ss:jhad}}

As we have noted above, hadronization begins in the middle of an event
and spreads outwards. In the junction rest frame the junction baryon
would thus be the one to form first, on the average, and have a rather
small momentum. Thereafter, each of the three strings would fragment
pretty much as ordinary strings, e.g. as in a back-to-back $\q\qbar$ pair
of jets. Also extensions to systems with multiple gluon emissions should
closely follow the corresponding pattern for normal events. With or
without gluon emission, we shall speak of three strings connected at the
junction, where a string may consist of several string pieces between
adjacent (i.e.\ colour-connected) partons.

In particular, if we consider events where each of the three outgoing
quark jets have large energies in the junction rest frame, the production
of high-momentum particles inside a jet should agree with the one of a
corresponding jet in an ordinary two-jet event. This can be ensured by
performing the fragmentation from the outer end of the strings inwards,
just like for the $\q\qbar$ string. Thus an iterative procedure can be
used, whereby the leading $\q$ is combined with a newly produced $\qbar_1$,
to form a meson and leave behind a remainder-jet $\q_1$, which is fragmented
in its turn. Flavour rules, fragmentation functions and handling of
gluon-emission-induced kinks on the string are identical with the ones
of the ordinary string.

While these hadronization principles as such are clear, and give the bulk
of the physics, unfortunately there is a catch: if all three strings are
fragmented until only little energy and momentum remain in each, and then
these remainders are combined to a central baryon, what guarantees that
this baryon obtains the correct invariant mass it should have?

The same problem exists in simple $\q\qbar$ events, that also there energy
and momentum is not guaranteed to come out right, if the event is fragmented
independently from the two ends and then joined by a single hadron in the
middle. In this case, the overall energy--momentum conservation is obtained
by the ``area law'' \cite{Lundstring}, which couples
the production of \textit{all} the hadrons in the event. It is therefore
difficult to simulate exactly, but for large remaining invariant masses
it simplifies to the iterative framework already introduced. For
small masses a pragmatic approximation is used. Firstly, not one but two
hadrons are used to join the jets. A two-hadron system has a continuous
mass spectrum, so one may construct consistent kinematics if the
normal fragmentation is stopped when the remaining invariant mass has
dropped below some value. Since the final two hadron masses are not known
beforehand, sometimes the remaining mass drops below the two-body threshold,
in which case the fragmentation is re-started from the beginning.
Secondly, by a random choice of producing the next hadron either off the
$\q$ end or off the $\qbar$ one, the final joining does not always occur
exactly in the middle of events, thereby smearing remaining imperfections
of the joining procedure proper.

For the fragmentation of a junction topology, we attempt to retain as much
as possible the known good features of the existing approach to $\q\qbar$
events, although this involves conflicting interests, as follows. In order
to describe
the production of high-momentum particles, fragmentation should be allowed
to proceed from all three string ends inward. But, in order not to bias the
junction baryon overly, the joining for energy-momentum conservation should
not always have to influence this hadron. Briefly put, the solution is to
fragment two of the three strings inwards, thereafter combine their leftovers
to an effective diquark, and finally fragment the string between this diquark
and the third end in much the same spirit as described for $\q\qbar$ events
above, i.e.\ by fragmentation at random off both ends of the system.
Put in the context of Fig.~\ref{fig:Yfrag}, imagine first tracing the chains
$\d_j - \q_5 - \q_6 - \q_7$ and $\d_k - \q_8 - \q_9$, then forming a
diquark $\q_7\q_9$, and thereafter fragmenting the string between this
diquark and $\u$ from both ends.

There are a number of technical details, as follows. 
The hadronization process itself
is normally carried out in the rest frame of the colour singlet system
under consideration, but is Lorentz covariant, so another choice would be
no problem. Information on the junction motion is encoded in its velocity
four-vector $v_{\mathrm{jun}} = (\gamma', - \gamma'\boldsymbol{\beta}')$,
with the $\boldsymbol{\beta}'$ defined in subsection \ref{ss:jmot}.

For normal string pieces, spanned between two partons, the invariant mass
of the piece is defined by the four-momenta of the two endpoint partons.
Since the junction does not carry a momentum, there is no corresponding
definition for the string pieces spanned between the junction and each of
its three nearest colour-connected partons. Instead the junction end is
here represented by a fictitious parton, specific to each string, opposite
to the $p_{\mathrm{pull}}$ vector of eq.~(\ref{eq:ppull}), as viewed in the
junction rest frame, which gives $p_{\mathrm{opp}} = - p_{\mathrm{pull}} %
+ 2 (v_{\mathrm{jun}} p_{\mathrm{pull}}) v_{\mathrm{jun}} $.

Two of the three strings are fragmented from the respective end inwards,
towards a fictitious other end as defined above. In order to have a
large-mass system left for the system in which the joining occurs, we
prefer to pick these two to be the ones with lowest energy, as defined
in the junction rest frame, i.e.\ with lowest
$E'_{\mathrm{str}} = v_{\mathrm{jun}} \sum_i p_i^{\mathrm{CM}}$. Here
$p_i$ are the four-momenta of the partons belonging to the given string,
excluding the fictitious junction one. As the hadrons are successively
produced, their summed energy $E'_{\mathrm{had}}$ (in the same frame) is
also updated. Once the hadronic energy exceeds the string one,
$E'_{\mathrm{had}} > E'_{\mathrm{str}}$, the process has gone too far,
i.e.\ passed the junction point of the string system, so it is stopped
and the latest hadron is rejected.

The random choices made in the fragmentation function allows the
energy in this latest hadron to fluctuate significantly. It can therefore
be that, after its removal, the energy
$\delta E' = E'_{\mathrm{str}} - E'_{\mathrm{had}}$ remaining of a string can
be quite significant. This is particularly dangerous if it happens in both
of the strings considered, since then the leftovers would be combined to
give a momentum intermediate in direction to the two strings, and thereby
maybe give a jet where none originally existed. Therefore the two hadronic
chains are rejected and new ones are generated in their place if both
$\delta E'$ are larger than a parameter $\delta E'_{\mathrm{min}}$, by
default 1~GeV. One of the two can have a larger energy than this, since
then the combined leftovers would still essentially agree with the direction
of this original string, but the chains are also rejected if this one has an
energy above $\delta E'_{\mathrm{min}} + R \, \delta E'_{\mathrm{max}}$, where
$\delta E'_{\mathrm{max}}$ by default is 10~GeV and $R$ is a random
number uniformly selected between 0 and 1. The two parameters $\delta
E'_{\mathrm{min}}$ and $\delta E'_{\mathrm{max}}$ are ``tuned''
by the requirements of a consistent description, see below. In order to
avoid infinite loops, at most 10 attempts of the above kind are rejected.

When two acceptable hadronic chains have been found, the remaining
four-momenta from the respective two strings are combined into a single
parton, which then replaces the junction as endpoint for the third string.
(Actually, technically, the whole of the two string four-momenta is
assigned to a parton, but then the fragmentation process is assumed
already to have produced the given set of hadrons. This is almost
equivalent, apart from some minor details of transverse momentum handling.)
The new effective parton may have a larger momentum than energy, and thereby
nominally be spacelike. If only by a little, it normally would not matter,
but in extreme cases the whole final string may even come to have a negative
squared mass. Therefore additional checks are made to ensure that the final
string mass is above the threshold for string fragmentation. If so, the
fragmentation procedure is identical with that of an
ordinary string from here on, else repeated
attempts are made, starting over with the first two strings.

A further aspect is the flavour properties. Except at the junction, the
normal rules of string fragmentation are used. At the junction, the two
strings fragmented first each define a leftover quark, which are combined
into a diquark. This diquark is assigned a spin 0 or 1 according to the
same relative probabilities as normally. Since also ordinary fragmentation
can produce diquark--antidiquark pairs, it could be that either string
ends with a leftover antidiquark rather than quark. Such configurations are
not acceptable, and have to be rejected in favour of new tries. Once a
diquark is defined, this can fragment further as usual. In the simplest
scenario, it will next fragment to produce a baryon and a leftover
antiquark. If popcorn baryon production is allowed, however, a meson may
alternatively be split off to produce a new diquark. That is, the baryon
number may then migrate to higher energies than otherwise, but will still
be rather centrally produced.

\begin{figure}
\begin{center}
\begin{picture}(220,135)(-110,-70)
\SetWidth{1}
\LongArrow(0,0)(70,50)\Text(80,50)[l]{$\d_m$($r$)}
\LongArrow(0,0)(70,-50)\Text(80,-50)[l]{$\d_n$($g$)}
\LongArrow(0,0)(-70,50)\Text(-80,50)[r]{$\dbar_j$($\overline{r}$)}
\LongArrow(0,0)(-70,-50)\Text(-80,-50)[r]{$\dbar_k$($\overline{g}$)}
\SetWidth{2}
\DashLine(-20,0)(20,0){5}
\DashLine(20,0)(70,50){5}\DashLine(20,0)(70,-50){5}
\DashLine(-20,0)(-70,50){5}\DashLine(-20,0)(-70,-50){5}
\Text(0,-60)[]{(a)}
\end{picture}
\begin{picture}(220,130)(-110,-70)
\SetWidth{1}
\LongArrow(0,0)(70,50)\Text(80,50)[l]{$\d_m$ ($r$)}
\LongArrow(0,0)(70,-50)\Text(80,-50)[l]{$\d_n$ ($g$)}
\LongArrow(0,0)(-70,50)\Text(-80,50)[r]{$\dbar_j$ ($\overline{r}$)}
\LongArrow(0,0)(-70,-50)\Text(-80,-50)[r]{$\dbar_k$ ($\overline{g}$)}
\SetWidth{2}
\DashLine(-70,50)(70,50){5}\DashLine(-70,-50)(70,-50){5}
\Text(0,-60)[]{(b)}
\end{picture}
\end{center}
\captive{%
(a) Schematic string topology of a
$\e^+\e^- \to \st \st^* \to \dbar_j \dbar_k \d_m \d_n$ decay.
The strings are shown as dashed lines. A possible colour
assignment is indicated within brackets. The string between
the two junctions would then be blue--antiblue.
(b) Alternative string drawing for the same process, without any
junctions.
\label{fig:twojun}}
\end{figure}

\subsection{Other string topologies \label{subsec:morejunc}}

So far we have considered the production of string systems containing a
single junction, well illustrated by the case of neutralino decays, but
also possible in other processes. There is a second possibility, however,
namely that of a system consisting of two junctions, or, more precisely,
of a junction and an antijunction, where the former is associated with
a baryon number $+1$ and the latter $-1$. An example would be
$\e^+\e^- \to \st \st^* \to \dbar_j \dbar_k \d_m \d_n$, containing
two BNV decays, $\st \to \dbar_j \dbar_k$ and $\st^* \to \d_m \d_n$.
This kind of topology is illustrated in Fig.~\ref{fig:twojun}a.
There are now two quark ends, two antiquark ones, and five strings,
including the one between the two junctions. Each of these strings
could contain an arbitrary number of intermediate gluons, where gluons
in between the two junctions are related to emission in the production
process $\e^+\e^- \to \st \st^*$ and the others to the decay
processes (with the standard ambiguities, especially relevant for the
soft gluons).

The junction motion and string fragmentation follows by simple
generalization of what has already been discussed for configurations
with a single junction. Like before, two of the three strings from a
junction --- the two not connected to the other junction --- are first
fragmented, and the leftovers combined to an effective (anti)diquark.
When this has been performed for both junctions, what remains is a
diquark--antidiquark string spanned between the two junctions, which
then is fragmented. Thus the hadronization will produce one baryon and
one antibaryon, associated with the junctions, plus potentially
additional baryon--antibaryon pairs as usual.

The same checks as above have to be applied at
various stages, but no new ones. The one ambiguity is how to calculate
the $p_{\mathrm{pull}}$ of eq.~(\ref{eq:ppull}) for the third string
of a junction, hooking up to the other junction. Here gluons on the string
between the two junctions are considered as normally, while partons on
the far side of the other junction are tracked down each of the two
strings without including any exponential suppression from energies
in the other of the strings.

Since the net baryon number vanishes in the above topology, an alternative
string scenario is illustrated by Fig.~\ref{fig:twojun}b. Here two strings
are pulled directly between a $\d_{(m,n)}$ and a $\dbar_{(j,k)}$ quark.
The junctions are gone, and so such an event would not have to contain
any baryons or antibaryons at all, and definitely not any associable
with the BNV processes as such. This is the approach adopted in
\textsc{Herwig}.

A priori, one does not know which of the two above scenarios would
be the correct one. It would not even have to be a unique answer,
valid for all events. A possibility would be that the topology with
minimal string length is selected dynamically. Then events of the
general kind shown in Fig.~\ref{fig:twojun} would obtain the 2-junction
topology of Fig.~\ref{fig:twojun}a when the $\d_m\d_n$ and
$\dbar_j\dbar_k$ opening angles are small, while the 0-junction
one of Fig.~\ref{fig:twojun}b would result if instead the
$\d_m\dbar_j$ and $\d_n\dbar_k$ opening angles were the small ones.

Such a simple picture would become more complicated by the addition of
gluons in the production process, i.e.\ the ones we above put on the
string between the two junctions. It now becomes necessary to subdivide
such gluons into two separate colour lines, one between each of the
outgoing $\d_{(m,n)}\dbar_{(j,k)}$ pairs. In general, we would expect
the 0-junction topology to become more disfavoured when there are
several gluons being emitted.

Furthermore, in the limit of large $\st$ lifetime, the string between the
$\st$ and $\st^*$ would have time to start to fragment before
the $\st$ or $\st^*$ has decayed. In such a case, the event would
consist of two separate colour singlets, that fragment independently of
each other. This does not have to mean timescales long enough to have
multiple string breaks and the formation of stop-hadrons: one single string
break is enough to make the production of a baryon and an antibaryon
unavoidable. At first glance a sufficiently large lifetime for this to
happen, $c\tau \gtrsim 1$~fm, would seem unlikely.
However, when only BNV decay channels are kinematically
allowed, the $\st$ lifetime is almost always non-negligible,
since, for massless d, s, and b quarks the decay length is roughly:
\begin{equation}
c\tau^{\mathrm{BNV}}_{{\st}} \approx  (100\ \mathrm{fm}) \times
\left(\frac{0.01}{|\lambda''_{3}|}\right)^{2} \times
\frac{100\ \mathrm{GeV}}{m_{\st}}\,
\label{eq:stoplife}
\end{equation}
where $|{\lambda}''_3|$ represents the average of the $|\lambda''_{3jk}|$
couplings. Strictly speaking, the above formula is valid for $\st_R$.
For $\st_1$ ($\st_2$), it is increased by
1/$|\sin\theta_{\st}|^2$ (1/$|\cos\theta_{\st}|^2$).
In passing, we note that event properties are likely to be similar
whether stop-hadrons have time to be produced or not: corresponding studies
for ordinary top quarks have shown that differences are restricted to the
region of low-momentum particles, and there mainly show up in angular
distributions, not averaged event properties \cite{valery}.

If gauge-interaction decay channels such as $\st \to \b \schi_1^+$ are
open, these will almost always dominate. Then, ignoring what happens to
the colour disconnected $\schi_1^+$, it would be doubly rare to
have both $\st$ and $\st^*$ decay with BNV. Experimental studies
would presumably have to concentrate on finding events where one decay is
BNV and the other not, and so we are back with a 1-junction topology.

Finally, we note that $\st\st^*$ production in hadron colliders would
predominantly come from $\g\g \to \st\st^*$ in the colour octet channel,
where thereby the $\st$ and $\st^*$ would belong to separate colour
singlet subsystems. Again the production of a baryon and an antibaryon
would then be inevitable.

Nevertheless, in order to test the consequences, we have developed an
alternative model. For a given event with two colour-connected BNV decays,
the total string length is evaluated under the assumption that the
event either is arranged in a 2-junction or in a 0-junction topology,
and the topology is chosen that corresponds to the smallest length.
Given this choice the subsequent hadronization is well-defined.

In more detail, the string length is defined by the so-called
$\lambda$ measure \cite{lambda}. For a simple two-parton system
\begin{equation}
\lambda = \ln \left( \frac{s}{m_0^2} \right) ~.
\end{equation}
Here $s$ is the squared invariant mass of the system and $m_0$ is
some hadronization reference scale of order
$\langle\kappa\tau\rangle$, just like the $E_{\mathrm{norm}}$
defined in section \ref{ss:jmot}. Particles are produced with a
flat rapidity distribution in the central regions of the string,
while the distribution falls off near the ends. Then
$y_{\mathrm{max}} \simeq \ln(\sqrt{s}/m_0)$ defines the effective
rapidity range, in the middle of the fall-off, such that the total
multiplicity on the average is proportional to $y_{\mathrm{max}}$.
For a generic string configuration, with many gluons between the
quark and antiquark, the complete $\lambda$ expression is rather
messy, but approximately it can be represented by the linear sum
of the $\lambda$ measure for each string piece,
\begin{equation}
\lambda = \sum_{i=1}^{n-1} \lambda_{i,i+1} =
\sum_{i=1}^{n-1} \ln \left( \frac{s_{i,i+1}}{m_0^2} \right) ~.
\end{equation}
Here the squared invariant mass $s_{i,i+1}$ is calculated with the
full momentum for the endpoint quarks but only half for the
intermediate gluons, which are shared between two adjacent string
pieces.

We now need to generalize the $\lambda$ measure to the case with
junctions, which has not been considered in the literature so far.
To begin, revert to a simple back-to-back $\q\qbar$ system, with
quark masses negligible. Then
\begin{equation}
\lambda = \ln \left( \frac{s}{m_0^2} \right) =
\ln \left( \frac{4E_{\q}E_{\qbar}}{m_0^2} \right) =
\ln \left( \frac{2E_{\q}}{m_0} \right) +
\ln \left( \frac{2E_{\qbar}}{m_0} \right) =
\ln \left( \frac{2 v p_{\q}}{m_0} \right) +
\ln \left( \frac{2 v p_{\qbar}}{m_0} \right) ~.
\end{equation}
The splitting into two terms can be seen as a separation of the
full rapidity range into one on the quark side of the event and
another on the antiquark one, smoothly matching at the origin.
At this stage the origin is arbitrary, i.e.\ $E_{\q} \neq E_{\qbar}$
represents an event boosted along the event axis; the individual
terms may thus be changed but the sum remains invariant. In the final
step the origin is represented by the four-vector $v=(1; 0, 0, 0)$,
to allow a Lorentz invariant form also for the split expression.
Alternatively to considering $E_{\q} \neq E_{\qbar}$ one may then
use a $v$ boosted along the event axis.

In a $\q_1\q_2\q_3$ event, if viewed in the rest frame of the junction,
there will be three rapidity plateaus extending from the origin, so
one can calculate a total rapidity range
\begin{equation}
\lambda =
\ln \left( \frac{2 v p_1}{m_0} \right) +
\ln \left( \frac{2 v p_2}{m_0} \right) +
\ln \left( \frac{2 v p_3}{m_0} \right) ~.
\label{lambdathree}
\end{equation}
The $v$ four-vector now is easily identified with the motion of the
junction.

We also need to consider the rapidity length of the string
between two junctions. Since the rapidity plateau extends all the
way to the junction, this is actually given by the rapidity difference
between the junctions themselves. Evaluating this in a frame where
the junctions are back-to-back,
$v_{1,2} = (\gamma; 0, 0, \pm \gamma\beta)$, one obtains
\begin{equation}
\lambda = 2 y_{\mathrm{max}} = 2 \frac{1}{2}
\ln \left( \frac{\gamma + \gamma\beta}{\gamma - \gamma\beta} \right) =
\ln \left( \frac{1 + \beta}{1 - \beta} \right) =
\ln \left( \frac{1 + \beta^2 + 2\beta}{1 - \beta^2} \right) ~,
\end{equation}
with $v_1 v_2 = \gamma^2 (1 + \beta^2) = (1 + \beta^2) / (1 - \beta^2)$
or $\beta^2 = (v_1 v_2 - 1)/(v_1 v_2 + 1)$.

We can now give the $\lambda$ of the two configurations in
Fig.~\ref{fig:twojun}a and \ref{fig:twojun}b. Since the number of
$m_0$ factors agrees between the two, it is more convenient to
introduce $\Lambda = m_0^4 \exp(\lambda)$, with expressions
\begin{eqnarray}
\Lambda_{\mathrm{2-junction}} & = & (2 v_1 p_m) (2 v_1 p_n) (2 v_2 p_j)
(2 v_2 p_k) \left( v_1 v_2 + \sqrt{(v_1 v_2)^2 - 1} \right)
~, \label{eq:lamjj_ng}\\
\Lambda_{\mathrm{0-junction}} & = &  (2 p_m p_j) (2 p_n p_k)
~,\label{eq:lamnj_ng}
\end{eqnarray}
where we have reused the generation indices of the quarks to
distinguish them, and $v_1$ ($v_2$) corresponds to the right (left)
vertex. The smaller of the two $\Lambda$ values now determines which
configuration would be the preferred one. Note that both expressions
are linear in each of the $p_i$, so the choice only depends on the
directions of motion of the outgoing quarks.

Now consider the complications caused by showering. Radiation
after the BNV decays is no problem: one only needs consider the
connection out to the colour produced in the BNV decays, i.e.\ to
the gluon nearest to the junction, since the subsequent colour
chain out to the endpoint quark would be unaffected by the colour
arrangement between the two BNV vertices. The $p_{j,k,m,n}$ in
eqs.~(\ref{eq:lamjj_ng})--(\ref{eq:lamnj_ng}) should be redefined
accordingly. Radiation patterns before the BNV decays can be divided
into two classes.\\
\textit{(i)} If the shower contains a $\g \to \q\qbar$ branching, the
colour flow in the system is automatically broken, and then the
production of a baryon--antibaryon pair is unavoidable. This is
the perturbative equivalent of the string between $\st$ and $\st^*$
starting to fragment before the $\st$ or $\st^*$ has decayed.
As one moves up from the kinematical threshold this probability
increases. Since the collinear and soft gluon singularities are
tamed by the large $\st$ mass and width, the total $\g \to \q\qbar$
probability is reasonably reliably predictable, although a cut-off
dependence e.g. from the assumed effective quark masses remains.\\
\textit{(ii)} If there are no $\g \to \q\qbar$ branchings, all the gluons
in the shower can be arranged in a single colour chain between the
$\st$ and $\st^*$. When the BNV decays occur, the anticolour of the gluon
that matches the colour of the $\st$ could no longer match the
anticolour of either $\st$ decay product in a junction-free topology,
but would rather have to be matched to the colour of either $\st^*$
decay product. Thus the 0-junction configuration would be with
two strings, one between one of the $\st^*$ decay product traversing
all the gluons from the original $\st$ edge to the $\st^*$ one and
then zig-zagging back to one of the $\st$ decay products, while the
other string would be between the two remaining decay
products. See Fig.~\ref{fig:jjlen} for an illustration of (a) the 2-junction
configuration and (b) the 0-junction one just discussed.
(Actually, since the stop mass reduces collinear emission, and
since the stop decay products are widely scattered, the zig-zag
pattern need not be as extreme as one might guess at first glance.)
Colour-suppressed topologies are neglected, such as where the gluons
split off into a closed loop and then two strings are stretched
directly between the $\st$ and $\st^*$ decay products.
When comparing string length with and without junctions, the string
internally between the intermediate gluons is common and can be
neglected. Labelling by $p_i$ the gluon closest to the $v_i$
vertex in colour space, in the chain between the two junctions, and
now putting $\Lambda = m_0^6 \exp(\lambda)$, one obtains
\begin{eqnarray}
\Lambda_{\mathrm{2-junction}} & = & (2 v_1 p_m) (2 v_1 p_n) (2 v_1 p_1)
(2 v_2 p_j) (2 v_2 p_k) (2 v_2 p_2) ~, \label{eq:lamjj}\\
\Lambda_{\mathrm{0-junction}} & = &  (2 p_m p_2) (2 p_j p_1)
(2 p_n p_k) ~,\label{eq:lamnj}
\end{eqnarray}
where which is which of $m$ and $n$ (and of $j$ and $k$) is chosen
at random.

\begin{figure}
\begin{picture}(220,130)(-110,-70)
\SetWidth{1}
\Line(60,0)(75,35)
\Line(75,35)(100,55)
\Line(60,0)(75,-35)
\Line(75,-35)(100,-55)
\Line(-60,0)(-75,35)
\Line(-75,35)(-100,55)
\Line(-60,0)(-75,-35)
\Line(-75,-35)(-100,-55)
\Line(-60,0)(-25,15)
\Line(-25,15)(25,15)
\Line(25,15)(60,0)
\Text(25,5)[]{$p_1$}
\Text(55,-10)[]{$v_1$}
\Text(83,28)[]{$p_m$}
\Text(83,-28)[]{$p_n$}
\Text(-25,5)[]{$p_2$}
\Text(-55,-10)[]{$v_2$}
\Text(-83,28)[]{$p_j$}
\Text(-83,-28)[]{$p_k$}
\Text(0,-60)[]{(a)}
\end{picture}
\begin{picture}(220,130)(-110,-70)
\SetWidth{1}
\Line(-75,35)(-100,55)
\Line(-75,35)(25,15)
\Line(25,15)(-25,15)
\Line(-25,15)(75,35)
\Line(75,35)(100,55)
\Line(-75,-35)(-100,-55)
\Line(-75,-35)(75,-35)
\Line(75,-35)(100,-55)
\Text(25,5)[]{$p_1$}
\Text(83,28)[]{$p_m$}
\Text(83,-28)[]{$p_n$}
\Text(-25,5)[]{$p_2$}
\Text(-83,28)[]{$p_j$}
\Text(-83,-28)[]{$p_k$}
\Text(0,-60)[]{(b)}
\end{picture}\\
\captive{%
Illustration of (a) the 2-junction topology and (b) a 0-junction topology,
for identical parton configurations.
Indices have been chosen so as to correspond directly
to the expressions in eqs.~(\ref{eq:lamjj}) and (\ref{eq:lamnj}). Naturally,
the topology in (b) only corresponds to one possible
choice of which is which of ($m$,$n$) and ($j$,$k$).
\label{fig:jjlen}}
\end{figure}

If anything, the approach above could overestimate the probability of
no baryon--antibaryon production, since it only considers the gain in
string length once one has arrived at a new colour arrangement,
whereas there could be additional dynamic suppressions on the way
between the `original' 2-junction topology and the `final' 0-junction
one. Nevertheless, it should offer some realistic estimates how big
a loss of baryon signal could at worst be. To quantify this to some extent,
we have taken a
closer look at isolated decays of $\st\st^*$ pairs (colour connected to each
other to form an overall colour singlet) in the CM of the pair.

For high-momentum stops, $\gamma_{\st}\equiv \ECM/{2m_{\st}}\gg 1$,
the $\d_m\d_n$ pair lies at a large rapidity
separation from the $\dbar_j\dbar_k$ pair. In the 2-junction topology, this
rapidity range is spanned by \emph{one} string piece, the junction--junction
one, whereas, in the 0-junction topology, \emph{each} of the $\d\dbar$
string pieces must cross it. In fact, due to the see-saw nature of the
0-junction topology in the presence of gluon radiation from the stops, the
central rapidity range will be crossed more than twice if the stops radiate.
Thus, we expect the 2-junction topology to dominate for cases
$\gamma_{\st}\gg 1$.

On the other hand, when the stops are slow, $\gamma_{\st}\sim 1$, the
$\st$ and $\st^*$ decay products are widely spread by the decay kinematics.
Typically the invariant mass between a $\st$ decay product and a $\st^*$
one will be smaller than that between the two products of the same
$\st$ or $\st^*$, such that the 0-junction topology is guaranteed to
give a smaller string length. Also for somewhat larger $\gamma_{\st}$,
where one can start to speak of an event axis, decay products are often
thrown into the opposite hemisphere, such that the central rapidity region
need not be crossed at all in the 0-junction alternative. Thus, we expect
the 0-junction topology to dominate for reasonably low $\gamma_{\st}$.

The rapidity arguments presented in the above two paragraphs closely follow
the here defined string length measures,
eqs.~(\ref{eq:lamjj_ng})--(\ref{eq:lamnj}). However, we may alternatively
use an argument based on the string motion itself. Recall that a junction
is at rest when the opening angles between its three attached strings is
$120^\circ$. Thus, if the opening angles between the motion of the
$\st$ and its two decay products is $60^\circ$ each, and correspondingly
for the $\st^*$, we are exactly at a balancing point. For stops faster
than this, the smaller decay angles will force the two junctions to move
away from each other, and so it should be less likely for them to meet and
annihilate. For slower stops, the junctions actually move \emph{towards}
each other (if assumed produced at a distance apart), making it more
believable that they can annihilate. Although these deliberations
are separate from the string length minimization procedure described
above, and do not give the same answer event by event, they do corroborate
it by leading to the same expected dominant topologies as $\gamma_{\st}$
is varied. The two approaches are about equally credible, but the string
length argument has the advantage of leading to tractable answers also for
quite complicated topologies.

\begin{figure}
\center
\begin{tabular}{cc}
\hspace*{-8mm}\includegraphics*[scale=0.66]{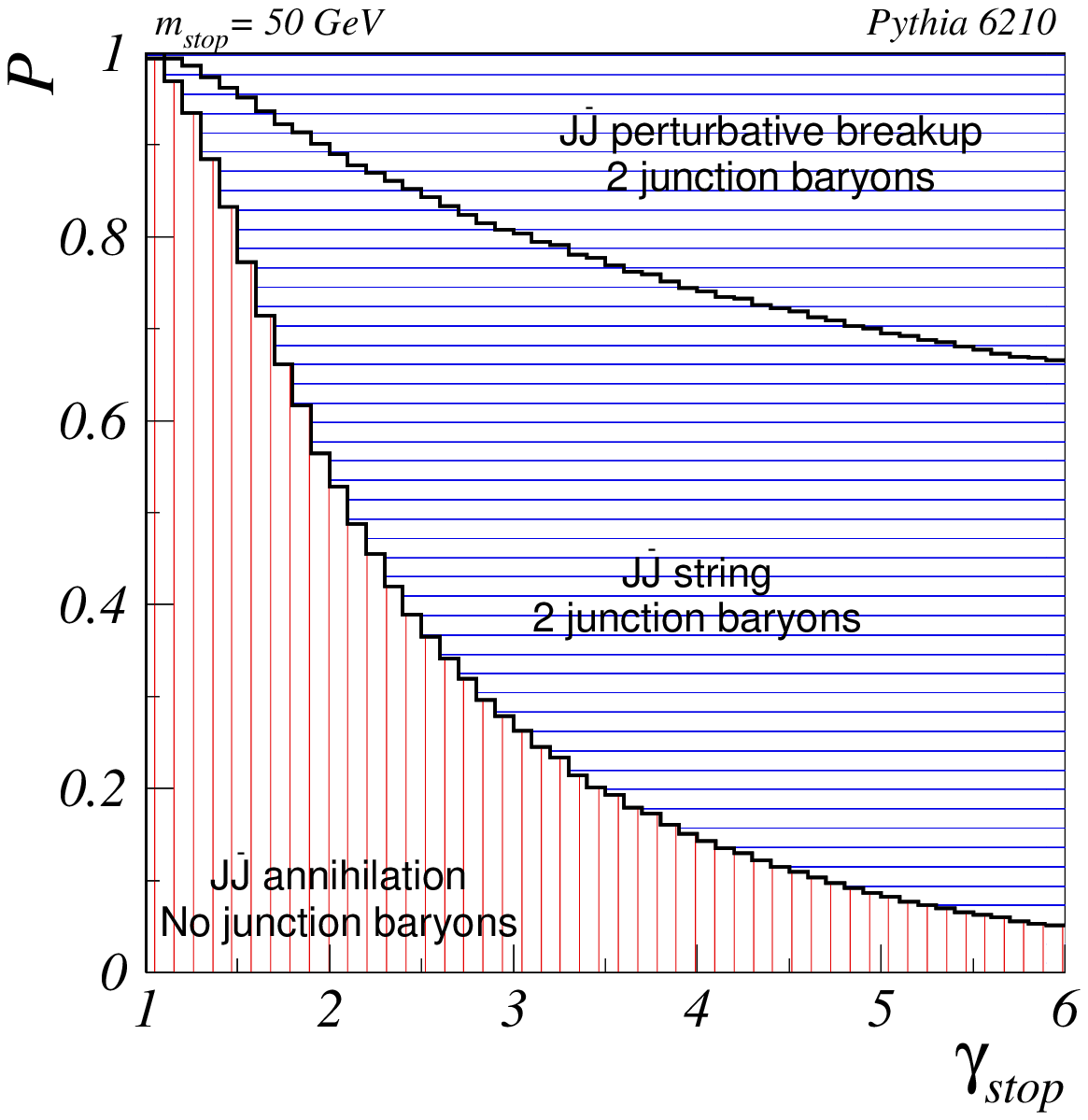}%
\hspace*{-10mm}&\hspace*{-10mm}
\includegraphics*[scale=0.66]{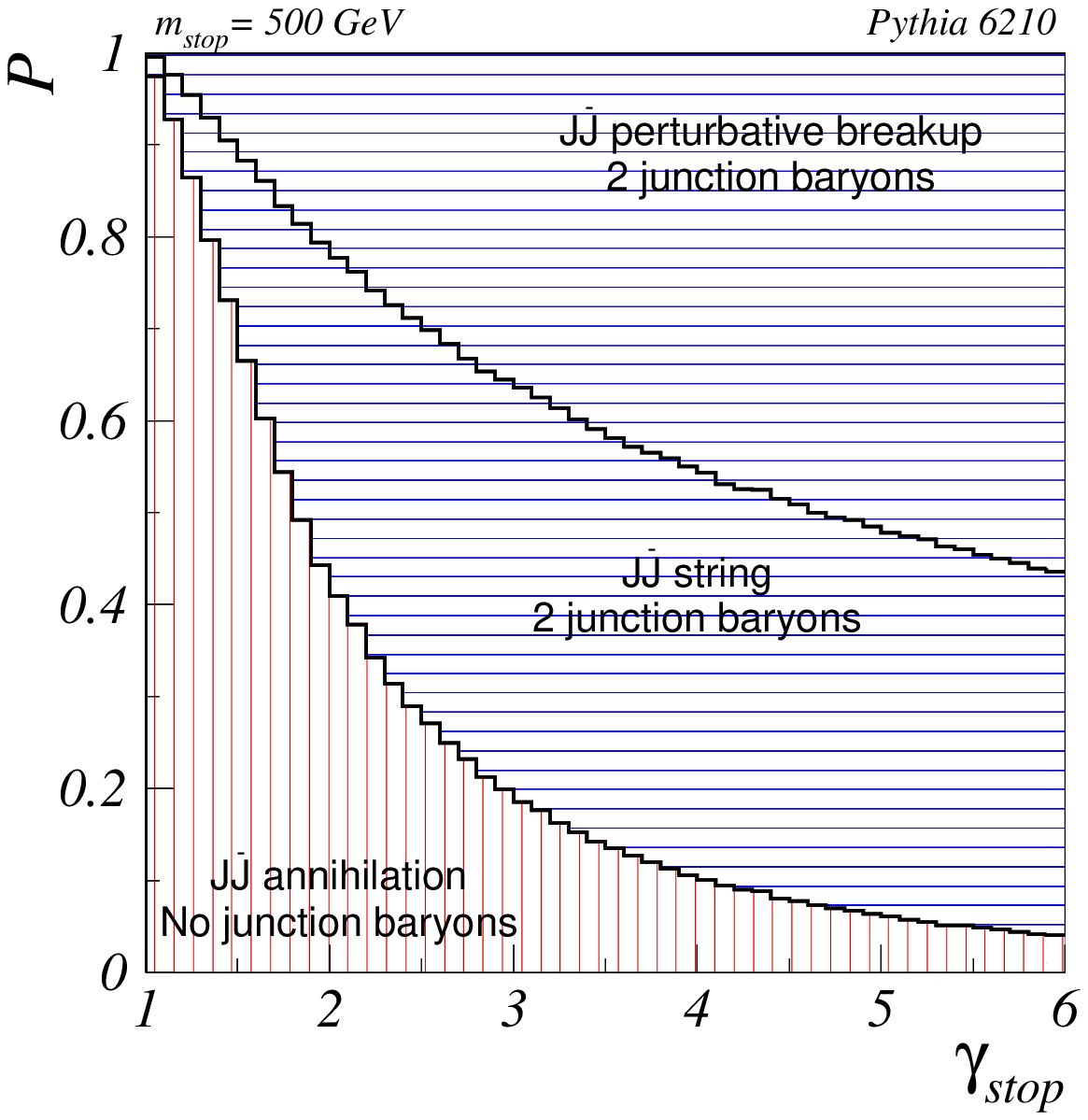}\hspace*{-8mm}\vspace*{-7mm}\\%
(a)&(b)
\end{tabular}
\captive{Probabilities for 2-junction and 0-junction configurations as
functions of $\gamma_{\st}=\frac{\ECM}{2m_{\st}}$ for
(a) $m_{\st}=50$~GeV and (b) $m_{\st}=500$~GeV. \label{fig:2j0j}}
\end{figure}

Using now eqs.~(\ref{eq:lamjj_ng})--(\ref{eq:lamnj}) to calculate
total string lengths for two different stop masses, and a range of
stop boosts $1 \le \gamma_{\st} \le 6$, gives the
plots shown in Fig.~\ref{fig:2j0j}. As is apparent from the discussion
above, we expect the ratio of 2-junction to 0-junction string
topologies to depend mainly on $\gamma_{\st}$ and only weakly
on the stop mass as such, and this agrees with what is shown
in the relation between the lower two regions of the plots. The feature
which \emph{does} change significantly is the rate of perturbative breakups
of the 2-junction systems, i.e.\ $\g\to\q\qbar$ splittings in the parton
shower initiated before the BNV decays, point \textit{(i)} above.
The $\g\to\q\qbar$ rate increases with larger total energy, also
for fixed $\gamma_{\st}$: while the primary emission rate of gluons off the
$\st\st^*$ system is rather constant, the phase space for further
cascading of these gluons is increased with energy.

\section{Model tests \label{s:modtest}}

In this section we will present some distributions that illustrate the
basic properties of our model, and show that it works as expected. In
several cases we will compare with results obtained with \textsc{Herwig},
the other main complete implementation of BNV phenomenology. The
distributions in this section are not necessarily directly observable;
we will return to experimental tests in the following section.

For calculating the sparticle mass and coupling spectrum,
we use the mSUGRA parameters
of Snowmass points 1a and 1b \cite{allanach02} input to \textsc{Isasusy}
\cite{Isajet} for RGE evolution. This is done
for both \textsc{Herwig} and \textsc{Pythia}, and so there can be no
artifacts created by non-identical EW scale superspectra when comparing
the output of the two programs. Apart from this aspect, most of the
topics we study here are not sensitive to the details of the SUSY
parameter set used.

For all studies, we use the ``factory presets'' for 
both \textsc{Pythia} and \textsc{Herwig}. No
tuning has been performed, and no demands e.g.\ of identical shower
cut-offs or fragmentation functions have been made. This does mean
that one must exercise a slight caution before drawing too strong
conclusions from the comparisons, since default
\textsc{Pythia} is not always directly comparable to default
\textsc{Herwig}. In the following, we will comment on these aspects
when necessary.

\subsection{Consistency checks}

\begin{figure}
\center\vspace*{-.75cm}
\includegraphics*[scale=0.8]{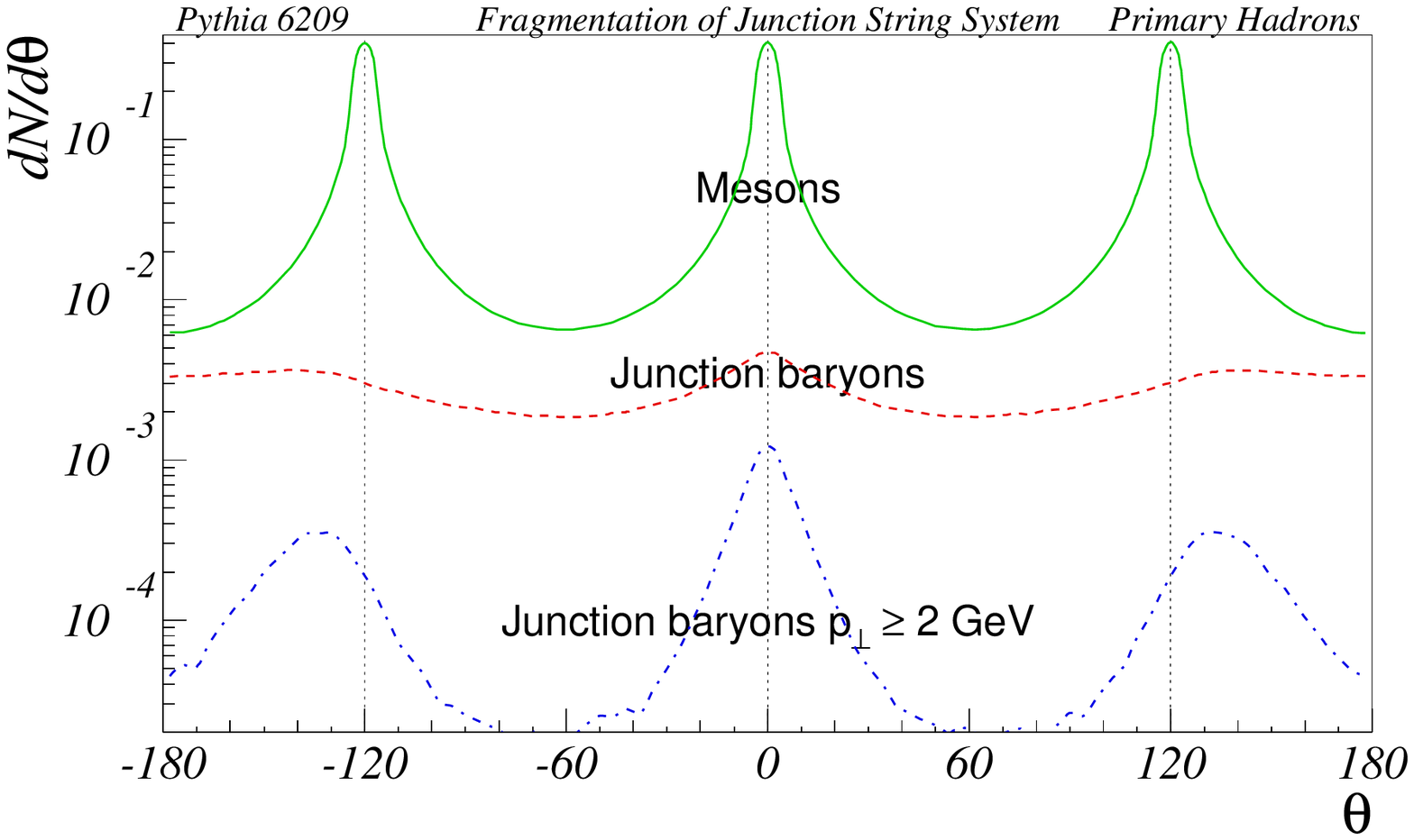}\\
\vspace*{-1.1cm}(a)\vspace*{-0.5cm}\\
\includegraphics*[scale=0.8]{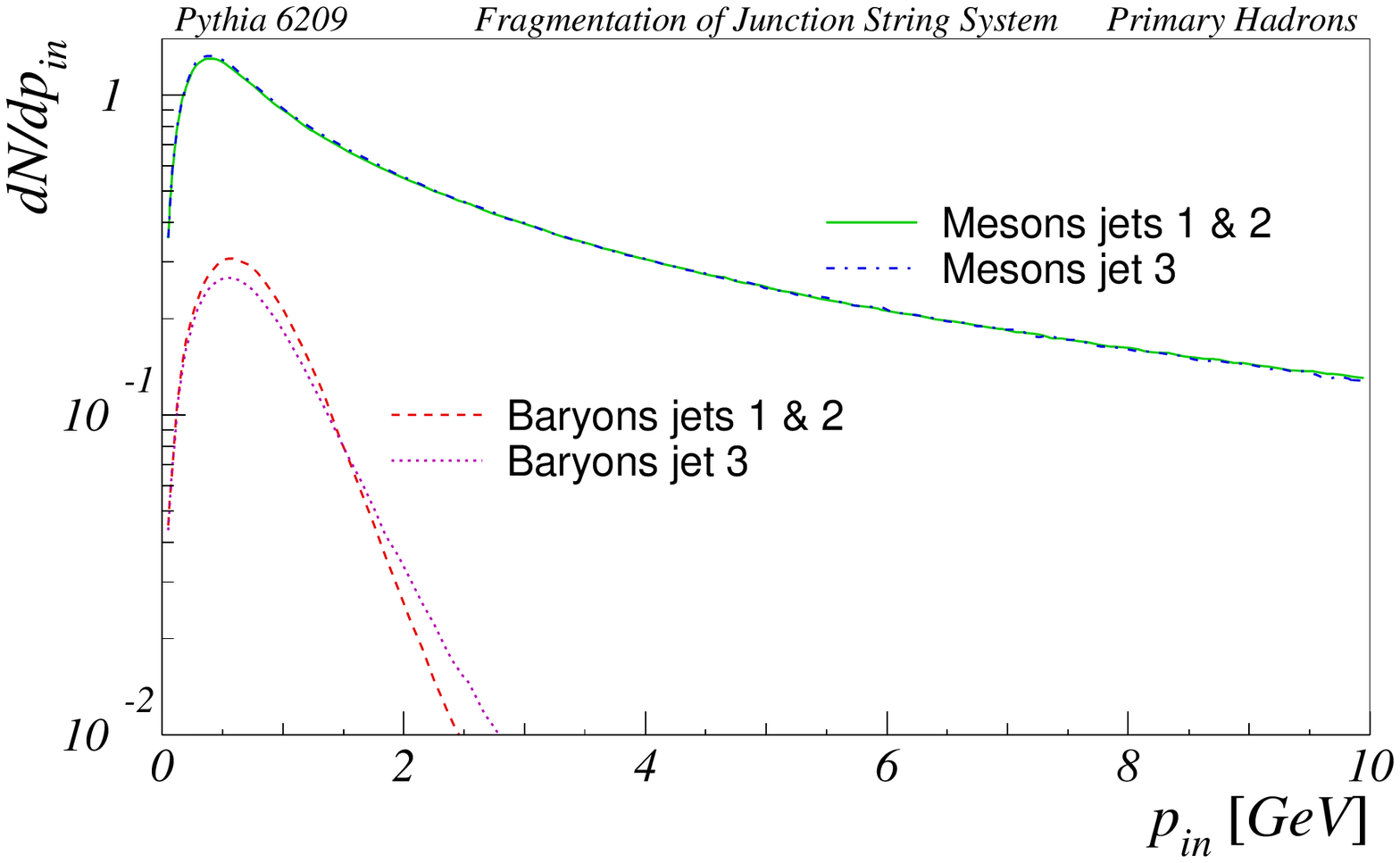}\\
\vspace*{-1.1cm}(b)\vspace*{2mm}\\
\captive{%
Fragmentation of a $\u\u\u$ parton configuration with 40 GeV in each
jet and opening angles 120$^{\circ}$ between partons. Only the primary
hadrons are shown, i.e.\ subsequent decays are disregarded. For
simplicity all normal baryon--antibaryon production is switched off,
as is the popcorn baryon possibility.
(a) The particle flow $\d N / \d \theta$ in the event plane, for
mesons, for the junction baryons and for the fraction of junction
baryons with momentum in the event plane above 2~GeV.  The jet at
around 0$^{\circ}$ is the last one to be considered in the
hadronization process.
(b) The meson and junction baryon momentum spectra per jet, shown
separately for the first two jets considered and for the last one.
Hadron assignment to jets is based on a simple division of the plane
into three equal 120$^{\circ}$ sectors.
\label{fig:symmflow}}
\end{figure}

The main technical simplification of the junction string fragmentation
scheme is the asymmetric fashion in which two of the three strings are
first fragmented inwards, with the remnants joined to an effective
diquark jet that is fragmented together with the third string. In
particular, the fear might be that the joined diquark could produce a
spurious new jet intermediate to the directions of the two strings from
which it inherits the leftover momenta. Therefore the fragmentation of
a symmetric three-parton configuration is studied in
Fig.~\ref{fig:symmflow}. For the production of hadrons other than the
junction baryon, an essentially perfect symmetry between the three jets
is obtained, both in angular and in momentum variables. As could be
expected, the situation is less perfect for the junction baryon itself.
Indeed, it has a tendency to be produced in a direction intermediate to
the first two jets considered, rather than along either of these two
separate directions, while baryons in the third jet line up the way one
would like. It should be noted that most of this effect
occurs for baryons of intermediate momenta $|\mathbf{p}| \simeq 1$~GeV;
at lower momenta everything is and should be close to isotropic, while
those few baryons that have large momenta tend to be better aligned with
the jet directions. The overall balance between the jets is good, with
32\% of the junction baryons found in the angular range around the third
jet, relative to 34\% in each of the other two. The momentum spectra in
the three jets also show reasonable agreement: although the third tends
to produce somewhat harder junction baryons than the first two, the
difference only corresponds to a mean momentum of 0.97 GeV rather than
of 0.87 GeV.

We may therefore conclude that the asymmetric algorithm does not seem
to induce any imperfections in the bulk of the hadron production, and
only rather modest ones for the junction baryon itself. These latter
imperfections should also largely average out in realistic simulations,
when the order in which jets have been hadronized is not known.
Specifically, it appears to be a stable and reliable prediction
of the model that the junction baryon should be found at low momentum,
$|\mathbf{p}| \lsim 2$~GeV, in the junction rest frame.

Further technical details, concerning the junction motion (subsection
\ref{ss:jmot}) and hadronization (subsection \ref{ss:jhad}), could have
been addressed in different approximations. Specifically, we have
investigated how the fragmentation spectra would be affected if we made the
following changes to our model:
\begin{Enumerate}
\item Switched the order of fragmentation of the three string pieces, so that
  the \emph{last} string to be fragmented would be the one containing
  \emph{least} energy, the reverse of the default behaviour.
\item Replaced eq.~(\ref{eq:ppull}), defining the junction pull vectors used
  in determining the junction rest frame, by a linear sum of 4-momenta instead
  of the default exponentially weighted sum, equivalent to setting
$E_{\mathrm{norm}}\to\infty$ in eq.~(\ref{eq:ppull}).
\item Enhanced the significance of soft gluon emission by setting
  $E_{\mathrm{norm}}=0.5$~GeV rather than the default $E_{\mathrm{norm}}
=1.5$~GeV.
\item Switched off the iterative procedure for finding the junction rest
  frame, so that the energies appearing in eq.~(\ref{eq:ppull}) are the
  string system CM energies rather than the energies in the junction rest
  frame.
\item Performed the fragmentation in the string system CM rather than in the
  junction rest frame.
\end{Enumerate}

\begin{figure}
\center\vspace*{-.75cm}
\includegraphics*[scale=0.89]{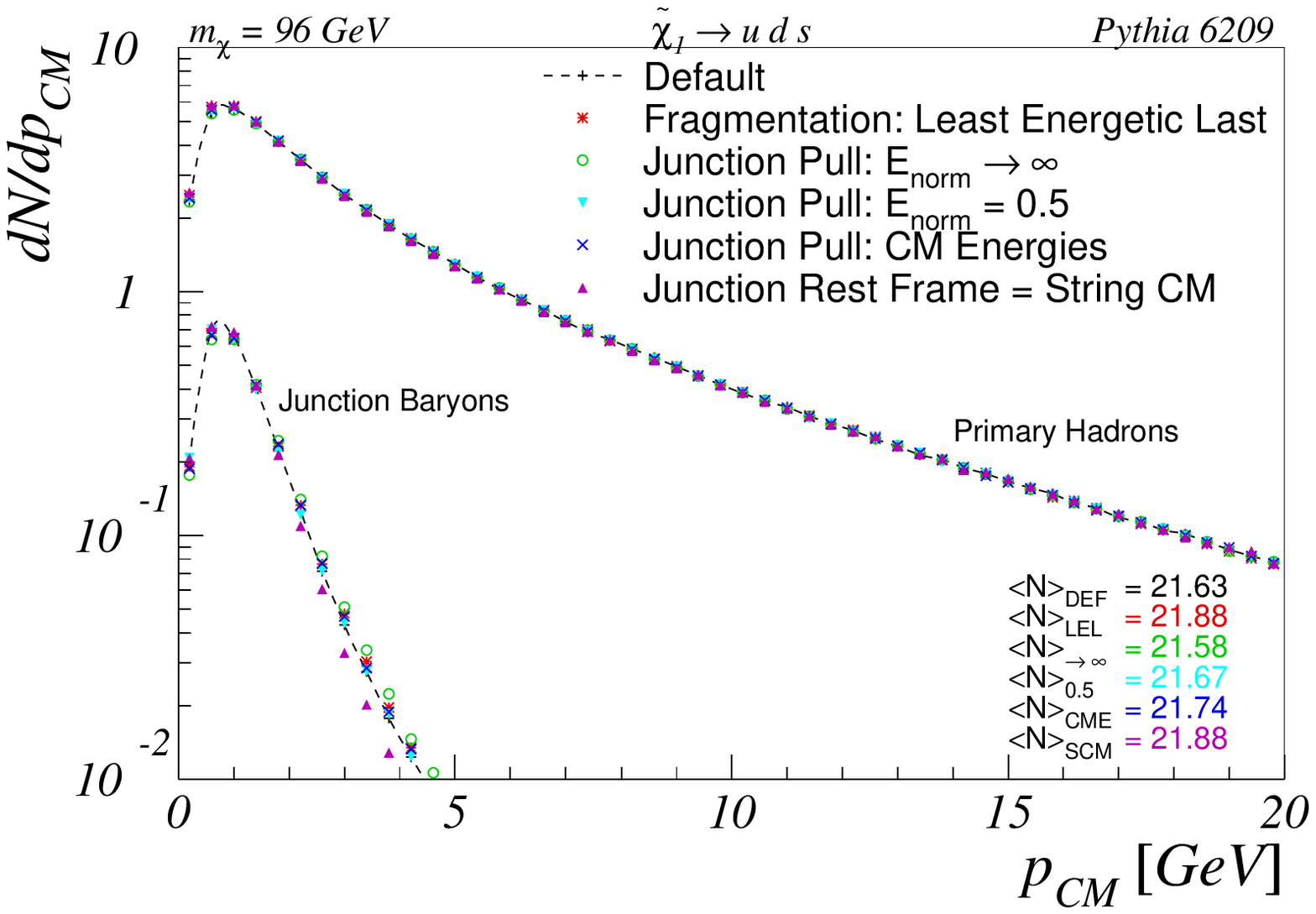}
\captive{%
Momentum spectra of primary hadrons in the decays of a 96~GeV neutralino.
Results with the default implementation are compared with five alternative
ones. Average multiplicities of primary hadrons are shown in the lower right
corner of the plot.
\label{fig:fragdiff}}
\end{figure}

The impact of these variations on the hadronization spectrum of a 96 GeV
neutralino decaying to light quarks
is shown in Fig.~\ref{fig:fragdiff}, where the rates of primary
hadrons and junction baryons produced in the fragmentation are plotted
as functions of their momenta in the CM of the decaying neutralino.
Average hadronic multiplicities are shown at the
lower right of the plot. As is readily observed, none of these variations
lead to a significant change in the spectra; hence we believe the systematic
uncertainties associated with these assumptions to be negligible to a good
precision.

As a final variation, in Fig.~\ref{fig:fragdiff} we also study
what happens if the junction is assumed to be at rest in the
string system CM frame (which agrees with the neutralino CM,
unless a $\g\to\q\qbar$ splitting occurred in the shower). This
is not intended as a realistic model variation, but indicates
that even such extreme scenarios would not  change the
qualitative picture of a low-momentum baryon. Here, however, we
are helped by the fact that the three jets in the neutralino
decay tend to be rather well separated in angle and that
therefore the boost between the junction rest frame and the
neutralino CM frame is not so large. Nevertheless, the more
correct description gives a somewhat harder junction baryon
spectrum, as should be expected from the boost.

\subsection{Shower and hadronization activity\label{ss:showhad}}

In section 3 we described the strategy for the generation of additional
gluon radiation in BNV decays. Thereby the number of reconstructed jets
can exceed the primary parton multiplicity. For our studies here we rely
on the Durham jet algorithm \cite{Durham}, with a distance measure
$y_{ij} = 2 \, \mathrm{min}(E_i^2,E_j^2) \,(1-\cos\theta_{ij}) /
E_{\mathrm{vis}}^2$ between two particles or clusters $i$ and $j$.
Results can depend on the algorithm used, but we expect most of the
phenomenology to come out similarly also for other algorithms
\cite{algcomp}.

\begin{figure}
\center\vspace*{-.75cm}
\includegraphics*[scale=0.8]{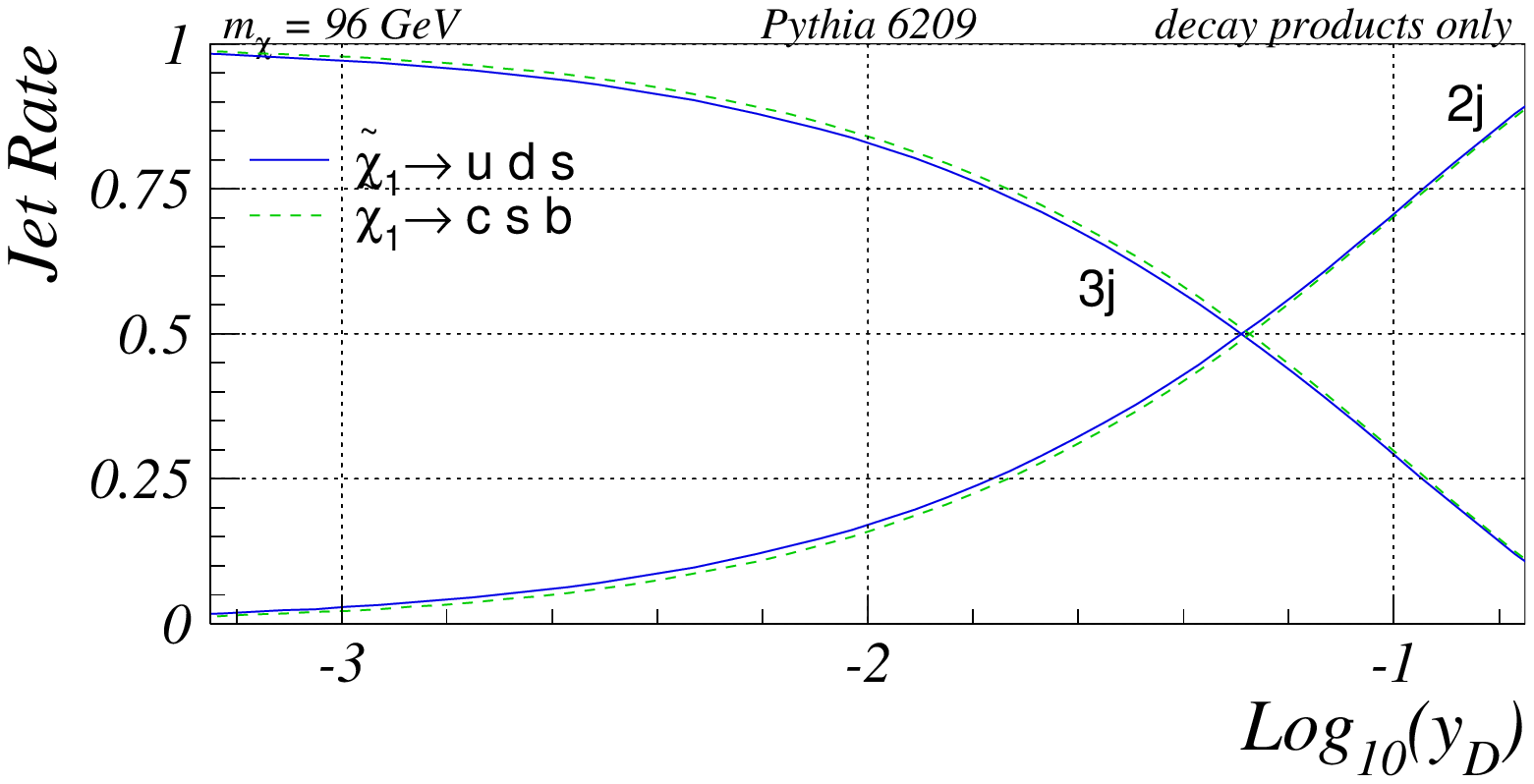}\\
\vspace*{-1.1cm}(a)\vspace*{-0.5cm}\\
\includegraphics*[scale=0.8]{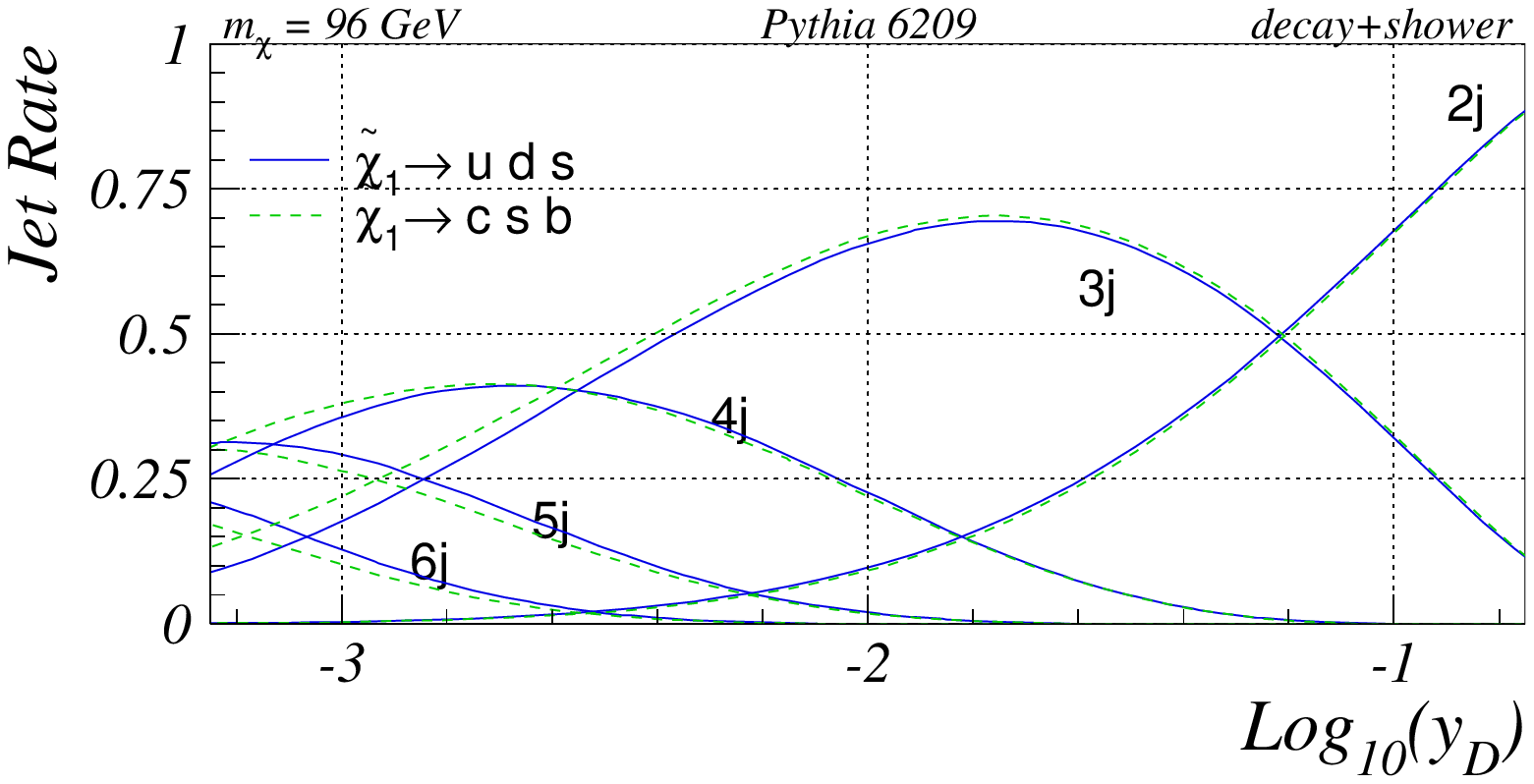}\\
\vspace*{-1.1cm}(b)\vspace*{-0.5cm}\\
\includegraphics*[scale=0.8]{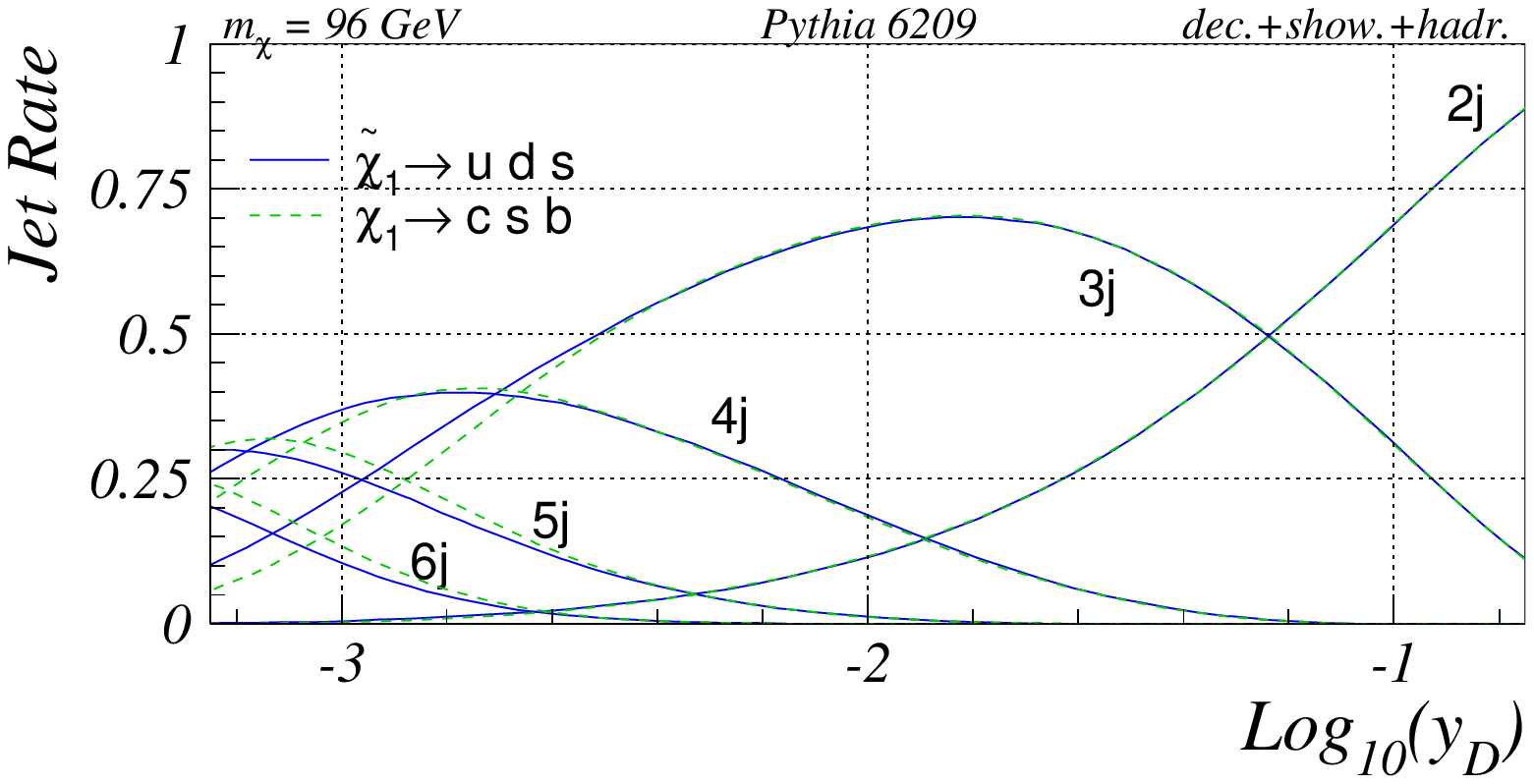}\\
\vspace*{-1.1cm}(c)\vspace*{0.5cm}\\
\captive{%
Jet rates in the decay of a 96~GeV neutralino resonance,
$\schi^0 \to \u_i \d_j \d_k$, as functions of $\yD$. 
$\schi^0 \to \u \d \s$ is shown with full lines and
$\schi^0 \to \c \s \b$ with dashed.
(a) Clustering only of the primary neutralino decay products,
(b) clustering after showering but before hadronization, and (c) clustering
after full event generation.
\label{fig:njets}}
\end{figure}

In Fig.~\ref{fig:njets}, jet rates are shown as functions of the
$\yD$ jet resolution parameter, with the jet clustering being performed
in the center of mass of the decaying neutralino.
The plots show a comparison between neutralino decay to light quarks
(full lines) and neutralino decay to heavy quarks (dashed lines) for
\textsc{Pythia}, at three stages of the event generation; (a) initial
decay, (b) after parton shower, and (c) after hadronization.

In the initial decay, phase space is isotropically populated, and so no
significant mass differences should be expected. However, since $y_{ij}$
contains energies rather than e.g.\ momenta in the numerator, and since
massive partons are assured nonvanishing energies, there are slightly more
3-jets for decays to heavy quarks. When the parton showers are included
(cf.\ Fig.~\ref{fig:njets}b), the
suppression of gluon radiation off massive quarks ensures that decays
into light quarks gives the larger number of further jets. Finally,
hadronization (cf.\ Fig.~\ref{fig:njets}c) once again flips the picture, at
small resolution scales $\yD$,
mainly by the larger $\pT$ kick impaired to hadrons in the $\B$ meson decays.
Such gluon radiation and hadronization effects are familiar from studies
e.g.\ at LEP \cite{LEPstudy}.

\begin{figure}
\center\vspace*{-.75cm}
\includegraphics*[scale=0.8]{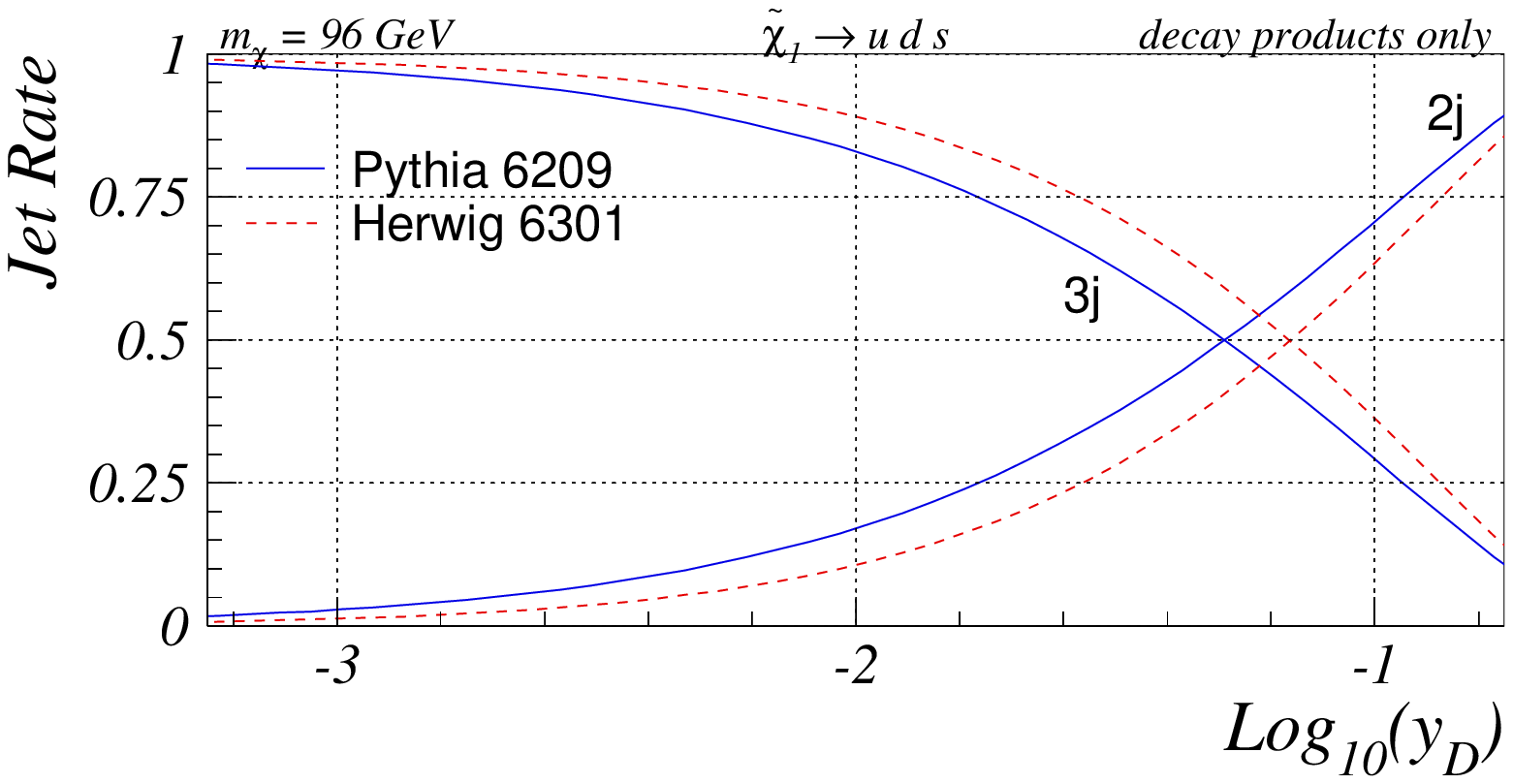}\\
\vspace*{-1.1cm}(a)\vspace*{-0.5cm}\\
\includegraphics*[scale=0.8]{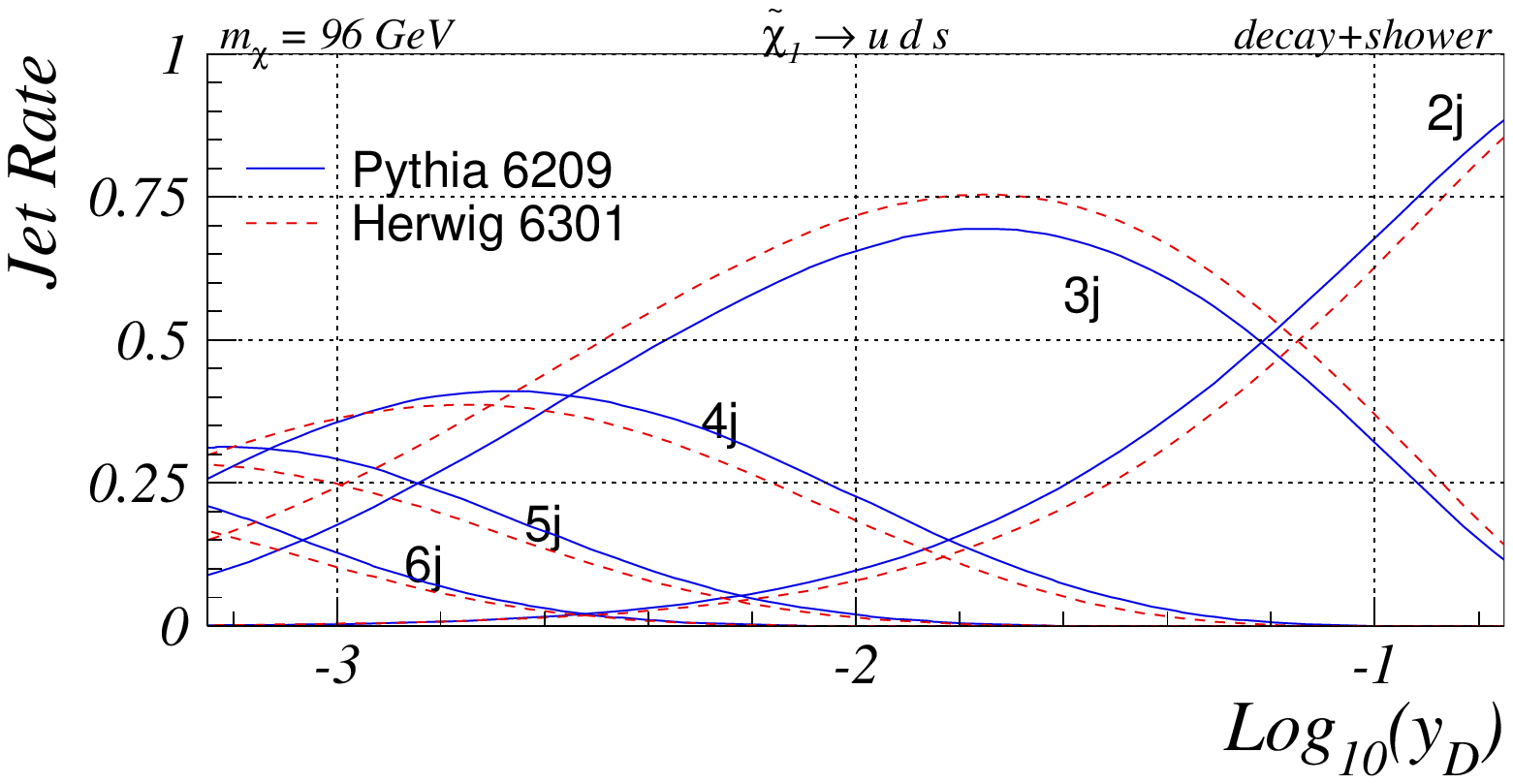}\\
\vspace*{-1.1cm}(b)\vspace*{-0.5cm}\\
\includegraphics*[scale=0.8]{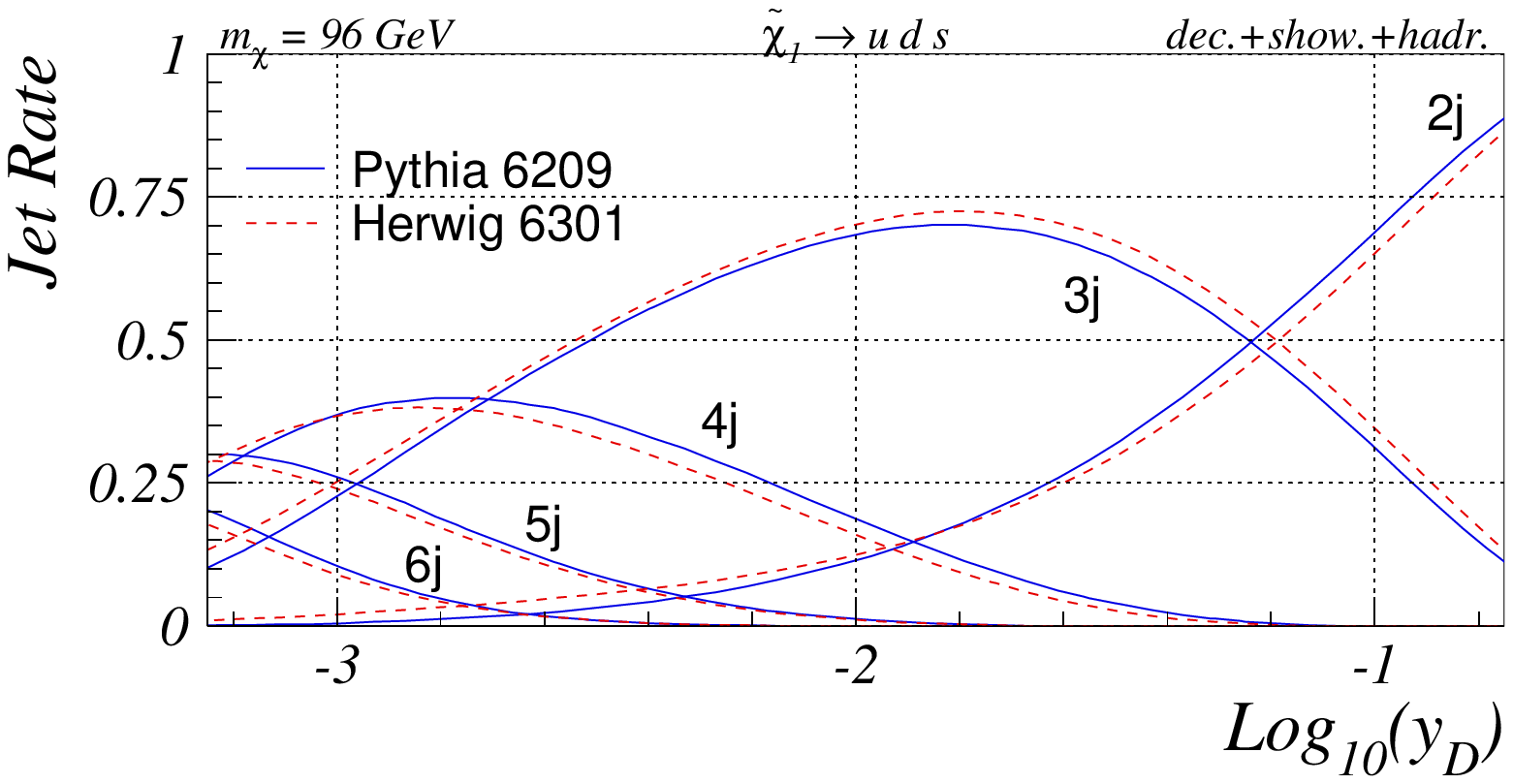}\\
\vspace*{-1.1cm}(c)\vspace*{0.5cm}\\
\captive{%
Jet rates in the decay of a 96~GeV neutralino resonance,
$\schi^0 \to \u \d \s$, as functions of $\yD$, 
as obtained with \textsc{Pythia} (full lines) and
\textsc{Herwig} (dashed lines).
(a) Clustering only of the primary decay products,
(b) clustering after showering but before hadronization, and
(c) clustering of the final event.
\label{fig:ndecpyhw}}
\end{figure}

In comparing with the\textsc{Herwig} implementation, we now concentrate
on decays to light quarks so as to obviate the further complication of extra
jets from heavy quark decays. This means that there are three main stages we
can check for differences: 1) the population of the initial 3-body phase
space, 2) the parton showers, and 3) the hadronization.
\begin{Enumerate}
\item Phase space population.
As mentioned above, \textsc{Pythia} produces an isotropic
phase space population whereas \textsc{Herwig} uses full matrix element
weighting. In Fig.~\ref{fig:ndecpyhw}a a comparison of the jet rates
between \textsc{Pythia} (full lines) and \textsc{Herwig} (dashed lines) is
shown for neutralino decay to light quarks. Only the three primary decay
products are used in the jet clustering here.
One notes that \textsc{Herwig} has more 3-jets than \textsc{Pythia} over
the whole $\yD$ range. This agrees with expectations that the matrix elements
should disfavour 2-jet configurations: in the 2-jet limit one
invariant mass vanishes, meaning that contributions from graphs with
that particular intermediate squark propagator also vanish. (A decay like
$\schi^0 \to \u \d \s$ receives contributions from the three intermediate
states $\u\squ^*$, $\d\sqd^*$, and $\s\sqs^*$.) There are no squark poles
in or near the phase space, however, so the variation of the matrix elements
is rather mild. Thus effects of the matrix-element weighting are significant,
but not dramatic.

\item Showering.
In Fig.~\ref{fig:ndecpyhw}b we show the jet rates after showering but
before hadronization. (For \textsc{Herwig} this also means before the
nonperturbative $\g \to \q\qbar$ branchings.) The difference produced
in the initial decay is reduced slightly, but otherwise the pattern at
high $\yD$ is the same as above, with \textsc{Herwig} producing more jets
than \textsc{Pythia}. However, when the four-or-more jet rates come into
play at lower values of $\yD$, 
\textsc{Pythia} tends to produce more jets than
\textsc{Herwig}, even if there is still qualitative agreement.

\begin{figure}
\center\vspace*{-.75cm}
\includegraphics*[scale=0.8]{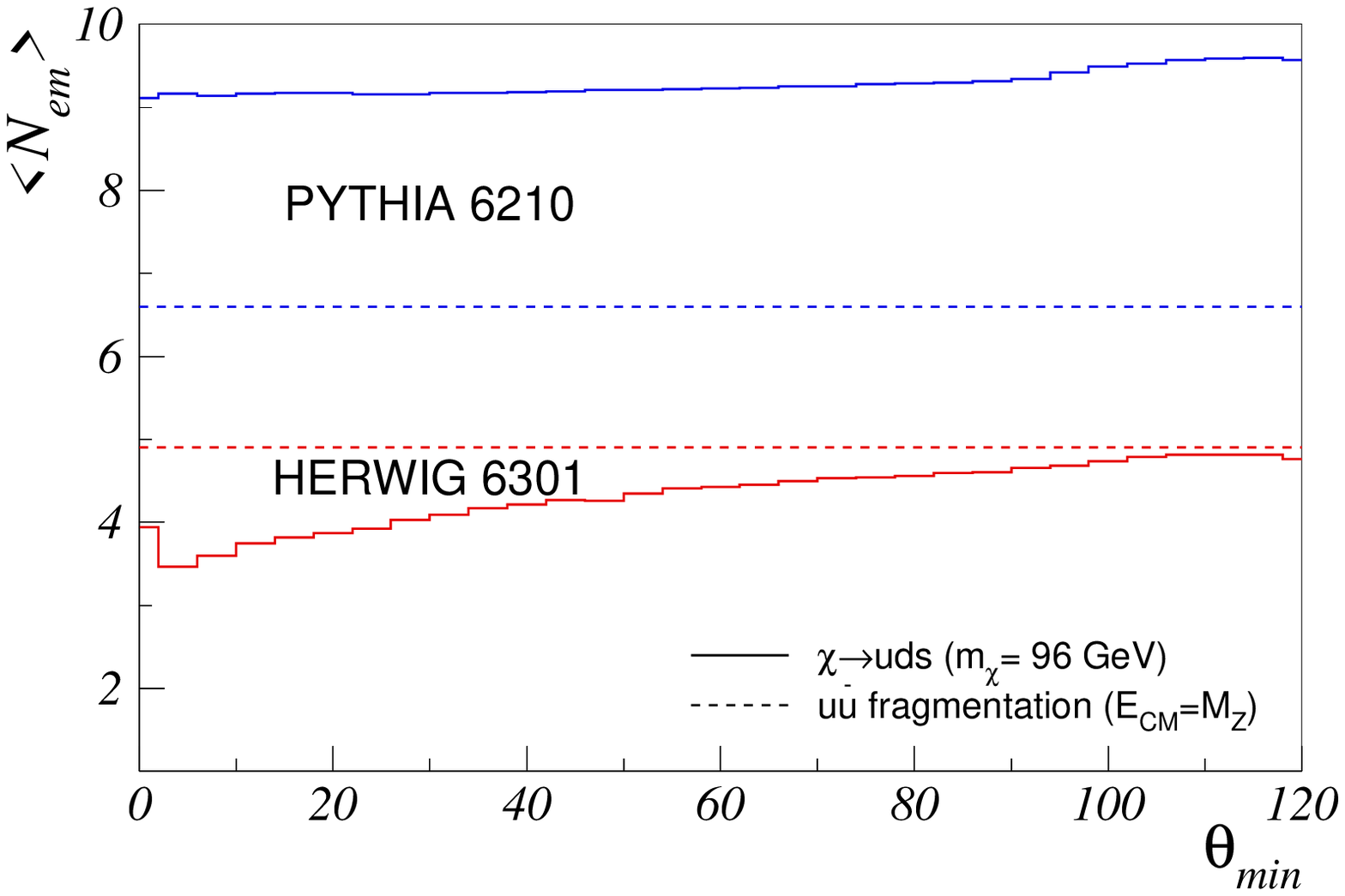}\vspace*{-.5cm}
\captive{%
The average number of partons emitted in the shower of $\schi\to \u\d\s$
decays, as a function of the smallest opening angle between the
neutralino decay daughters (solid lines; upper: \textsc{Pythia}, lower:
\textsc{Herwig}). The average number of partons emitted in $\u\ubar$
cascades at the $\Z^0$ resonance is shown for comparison (dashed lines).
\label{fig:nemis}}
\end{figure}

We explore this difference by considering the number of partons emitted
in the respective shower, Fig.~\ref{fig:nemis}. The variation from two-
to three-jetlike topologies is characterized by the minimum opening angle
between any two of the three original partons. Differences in the shower
algorithms, such as the choice of cutoff scale, imply that the number
of partons as such is not so informative. Therefore a convenient
reference is offered by the respective results for ordinary
$\Z^0 \to \q\qbar$ events, at a comparable energy.

In \textsc{Pythia},
the $\schi$ decays produce almost a factor $3/2$ more radiated partons
than does the $\Z^0$ ones. This is consistent with a rather constant
rate per radiating parton, somewhat reduced by the lower energy per parton
in the $\schi$ decays. Such a result is not unreasonable,
given that most gluons are emitted in the collinear regions around the
quark directions, according to a universal radiation pattern, i.e.\
independently of the colour flow to other quarks at wide angles. One
should note that the basic parton shower formalism of \textsc{Pythia}
tends somewhat to overestimate energetic wide-angle emission in
$\Z^0 \to \q\qbar$ events. This is then compensated by a rejection
factor that reduces this emission to what is expected from the
$\q\qbar\g$ matrix elements \cite{Mats,Emanuel}. Since no corresponding
matching is implemented for $\schi \to \q\q\q$ decays, it is quite
likely that the shower activity here is somewhat overestimated. Further,
in the limit that two partons become collinear, one should expect them
partly to screen each other and the overall activity thereby to drop down
towards the $\Z^0$ level. That this does not happen in \textsc{Pythia} is
an obvious shortcoming of the modelling, which is not set up to handle
these screening effects.

By contrast, in \textsc{Herwig} the variation with event topology
is more marked: the \textsc{Herwig} radiation cone is defined by
the opening angle to one of the other quarks picked at random
\cite{Herwigmodel}, meaning that two nearby partons may kill the
radiation from each other. The most notable feature, however, is that
the $\schi$ multiplicity is lower than the $\Z^0$ one, also for well
separated quark topologies, contrary to the arguments above.

\begin{figure}
\center\vspace*{-.75cm}
\includegraphics*[scale=0.8]{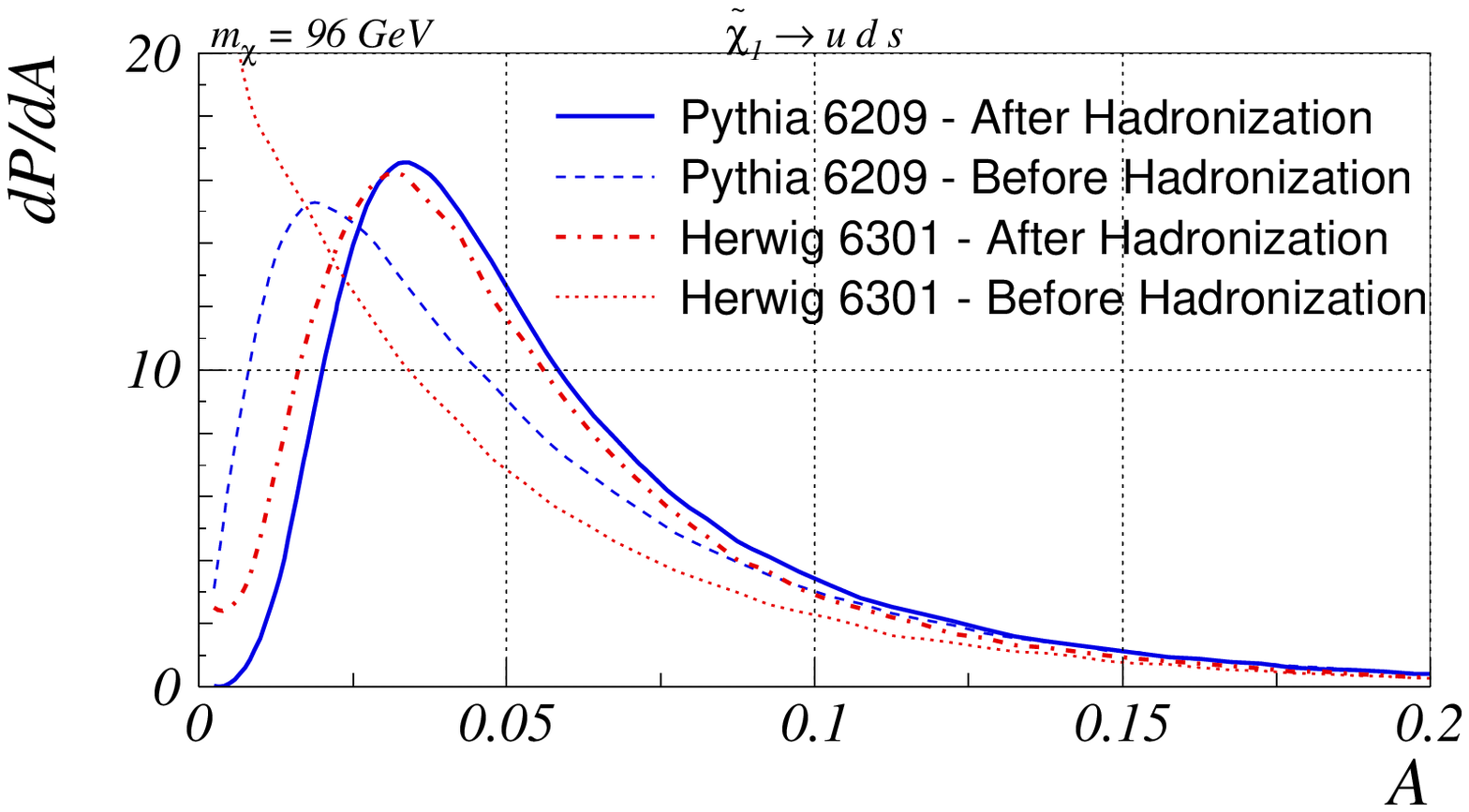}\vspace*{-.5cm}
\captive{%
Distribution of linear Aplanarity, comparing \textsc{Pythia} and
\textsc{Herwig}.
\label{fig:apan}}
\end{figure}

We do not claim to understand the low emission rate in \textsc{Herwig}.
One observation, that we will quantify below, is that $\schi$ decays
tend to have a large-mass central cluster, which could indicate a
shortfall of emissions at wider angles. Among several potential
differences between the two shower descriptions, it is then tempting
to suspect the procedures used to start up the showers
from some maximum virtuality. As we have already noted, the ``dead zone''
phenomenon is a well-known reason for \textsc{Herwig} to underestimate
multi-jet rates. If so, the differences should be larger for
out-of-the-event-plane distributions than for inclusive ones,
since the direction at $90^{\circ}$ to the event plane is as far
away as one can be from any of the three original quark directions.
In Fig.~\ref{fig:apan}, distributions of the linearized
Aplanarity event shape variable \cite{ellis81} are compared.
The curves relevant for showering are the dashed for \textsc{Pythia}
and dotted for \textsc{Herwig}. One observes that, also in the present
case, \textsc{Herwig} generates less activity out of the event
plane than does \textsc{Pythia}. The steep rise of the \textsc{Herwig}
curve towards small Aplanarities is caused by events where no
emissions harder than roughly 1~GeV or so occurred. The pattern is
consistent with the hypothesis of a shortfall of wide-angle emissions
in \textsc{Herwig}.

\item Hadronization.
The results of jet clustering after full event generation are shown in
Fig.~\ref{fig:ndecpyhw}c. One notes that \textsc{Herwig} still produces
more jets at high $\yD$, whereas \textsc{Pythia} has more jets at small $\yD$,
but overall the agreement between the two programs here is impressively
and unexpectedly good. In particular, the hadron-level
jet rate in \textsc{Herwig} goes up appreciably relative to the
 parton-level one, whereas there is less difference between the two in
\textsc{Pythia}.

Of special relevance here is the \textsc{Herwig}
treatment of the ``baryon cluster'' which carries
the net baryon number generated by the BNV decay. As we see from
Fig.~\ref{fig:hwclustermass}, this cluster has a rather different mass
spectrum than other clusters, with a much higher average
mass. (This is also noted in \cite{Herwigmodel}.) Such a large-mass
cluster is first fragmented into smaller clusters, which each then
decay to hadrons. The fragmentation is along an assumed ``string''
direction, as appropriate for simple $\q\qbar$ clusters. In the current
case, where three leftover quarks are to be considered, two of these,
picked at random, are combined to a diquark. The resulting system is
fragmented along a single quark--diquark axis. The extreme case is those
events, roughly 2\% of the total number, where the whole neutralino
becomes a single cluster. Then, as shown in Fig.~\ref{fig:hwjets}, no
well-separated third jet exists after hadronization, this information
having been erased by the cluster formation procedure! In a normal event,
effects tend in the opposite direction: some clusters appear along each
of the three quark directions as a consequence of the shower activity,
but then the baryon cluster also emits particles in a new ``jet'' direction
intermediate to the two quarks that get combined into a diquark. The net
result is an increase in the nonperturbative jet rate relative to the
perturbative one. To the extent that this effect is unintentional,
the large hadronization correction in \textsc{Herwig} is misleading.

\begin{figure}
\center\vspace*{-.75cm}
\includegraphics*[scale=0.8]{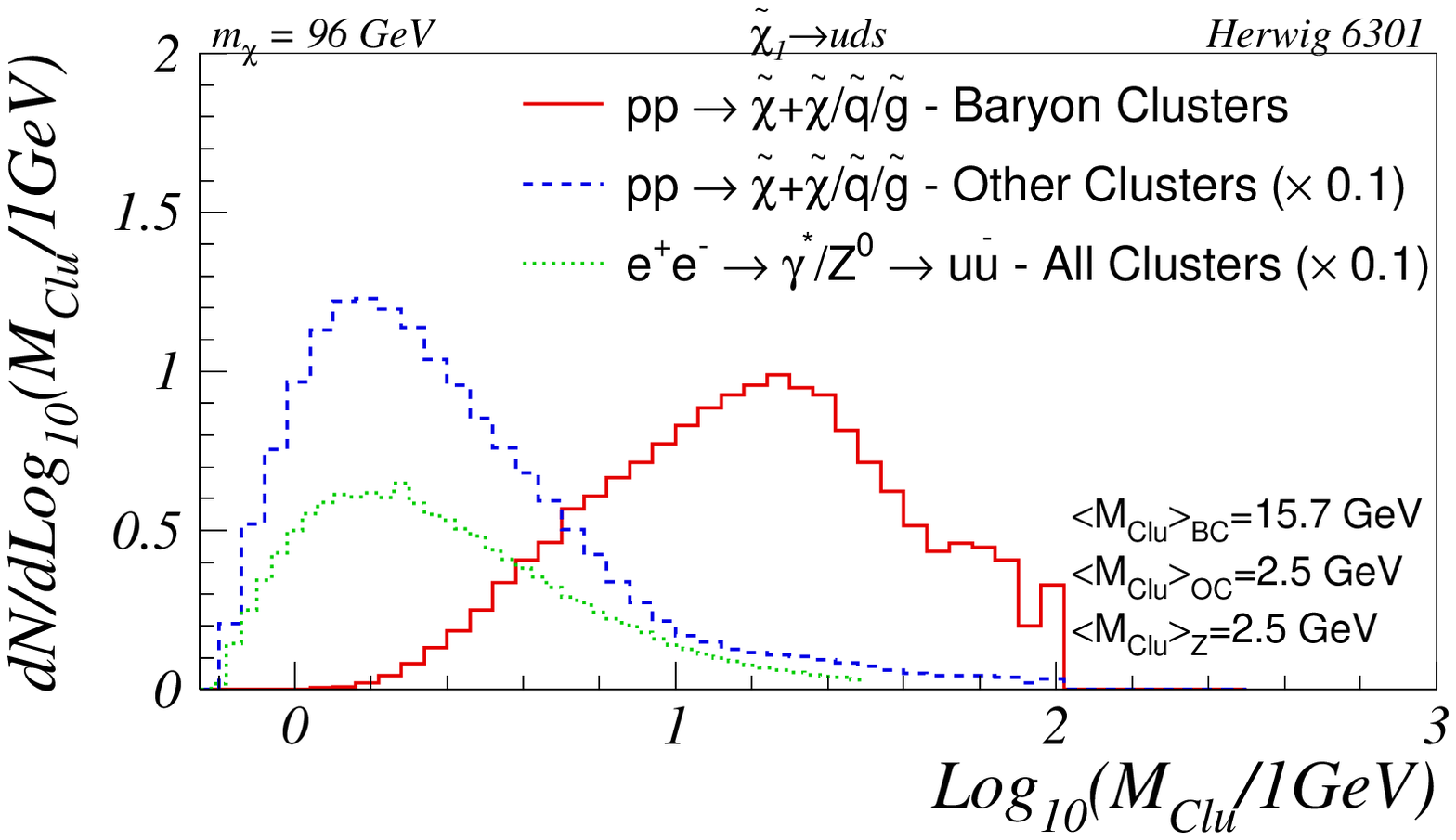}\vspace*{-.5cm}
\captive{%
\textsc{Herwig} cluster mass spectra for the clusters containing non-zero
baryon number (solid lines) as compared to the ordinary clusters from the
neutralinos and the additional perturbative activity in the event.
 For reference, the cluster spectrum for hadronic
$\Z^0$ decays at 90~GeV are also shown (dotted lines). Only the baryon
cluster spectrum is shown with its true normalization, the two other spectra
having been normalized by a factor 1/10 to fit them onto the same scale as the
baryon clusters.
\label{fig:hwclustermass}}
\end{figure}

\begin{figure}
\center\vspace*{-.75cm}
\includegraphics*[scale=0.8]{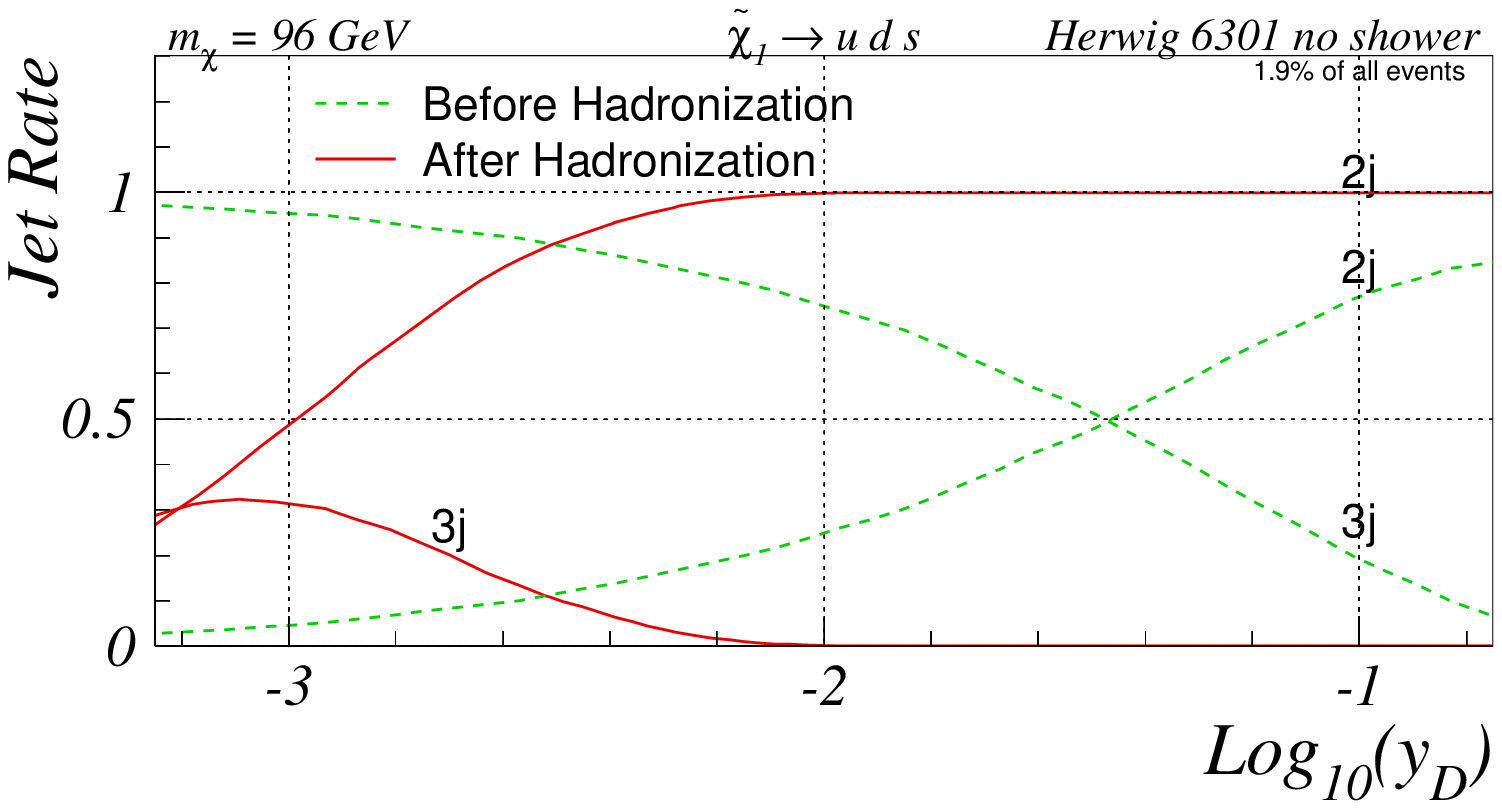}\vspace*{-.5cm}
\captive{%
2- and 3-jet rates before and after hadronization for \textsc{Herwig}
events with no gluon emission ($\sim$~2\% of the event rate for default
\textsc{Herwig}).
\label{fig:hwjets}}
\end{figure}
\end{Enumerate}
To summarize, we note that the final agreement in jet rates must largely be
viewed as
coincidental, since all three contributing physics components by themselves
differ between \textsc{Herwig} and \textsc{Pythia}.

\subsection{The junction baryon}

As discussed in subsection \ref{ss:jhad}, the non-perturbative collapse
of the colour wavefunction, which in our model results in a Y-shaped string
topology, essentially
``traps'' the baryon number around the locus of the string junction. Thus, in
the rest frame of the junction, we expect junction baryon momenta of the
order of the hadronization scale. For well-separated jets, we further expect
the junction rest frame to be only slightly different from the CM frame of
the decay. Thus a generic feature of our model is the prediction that the
baryon number generated by BNV decays predominantly
ends up in baryons which have small
momenta relative to the decaying particle.

\begin{figure}
\center\vspace*{-.5cm}
\includegraphics*[scale=0.8]{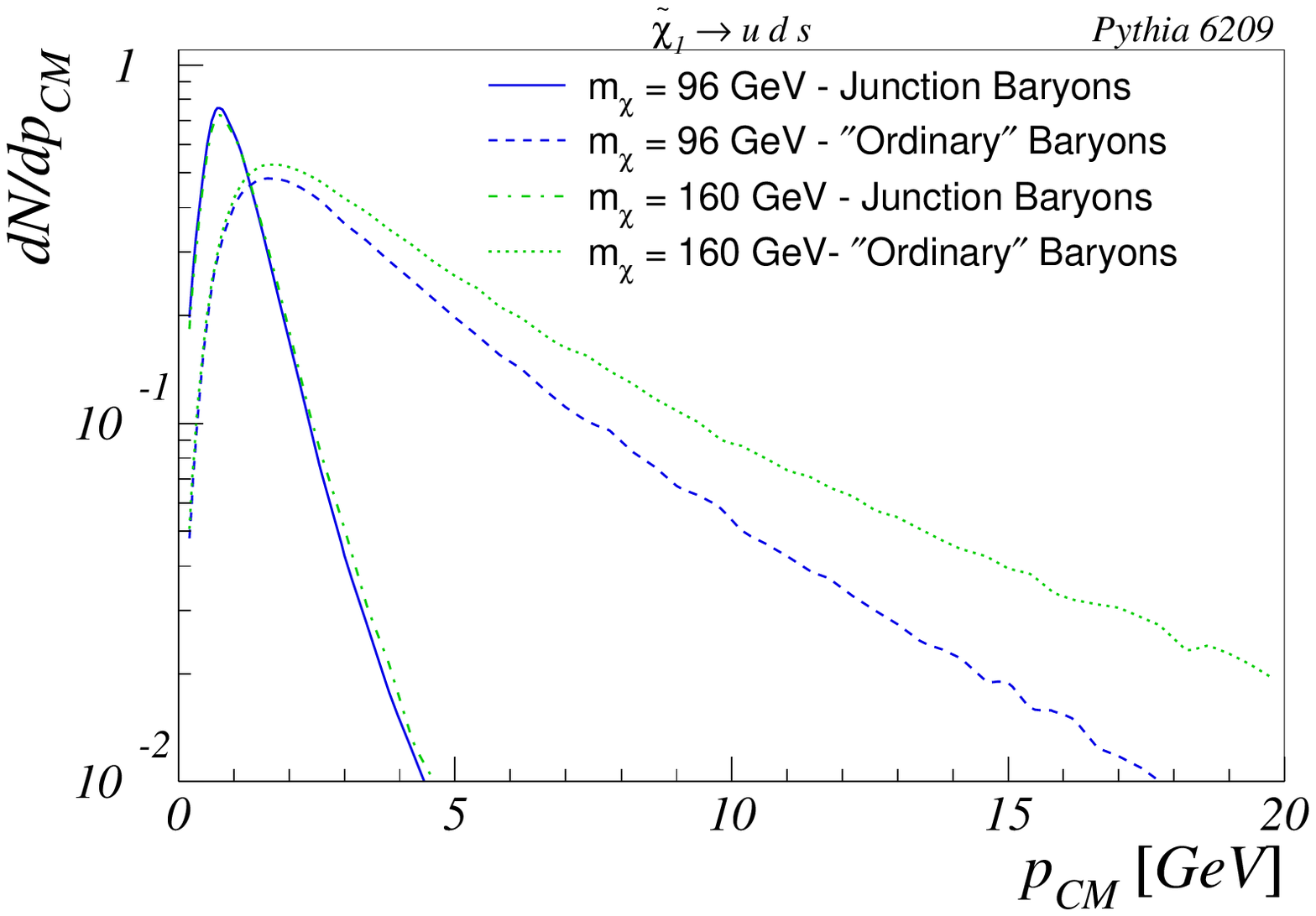}\\
\vspace*{-.75cm}(a)\vspace*{-0.2cm}\\
\includegraphics*[scale=0.8]{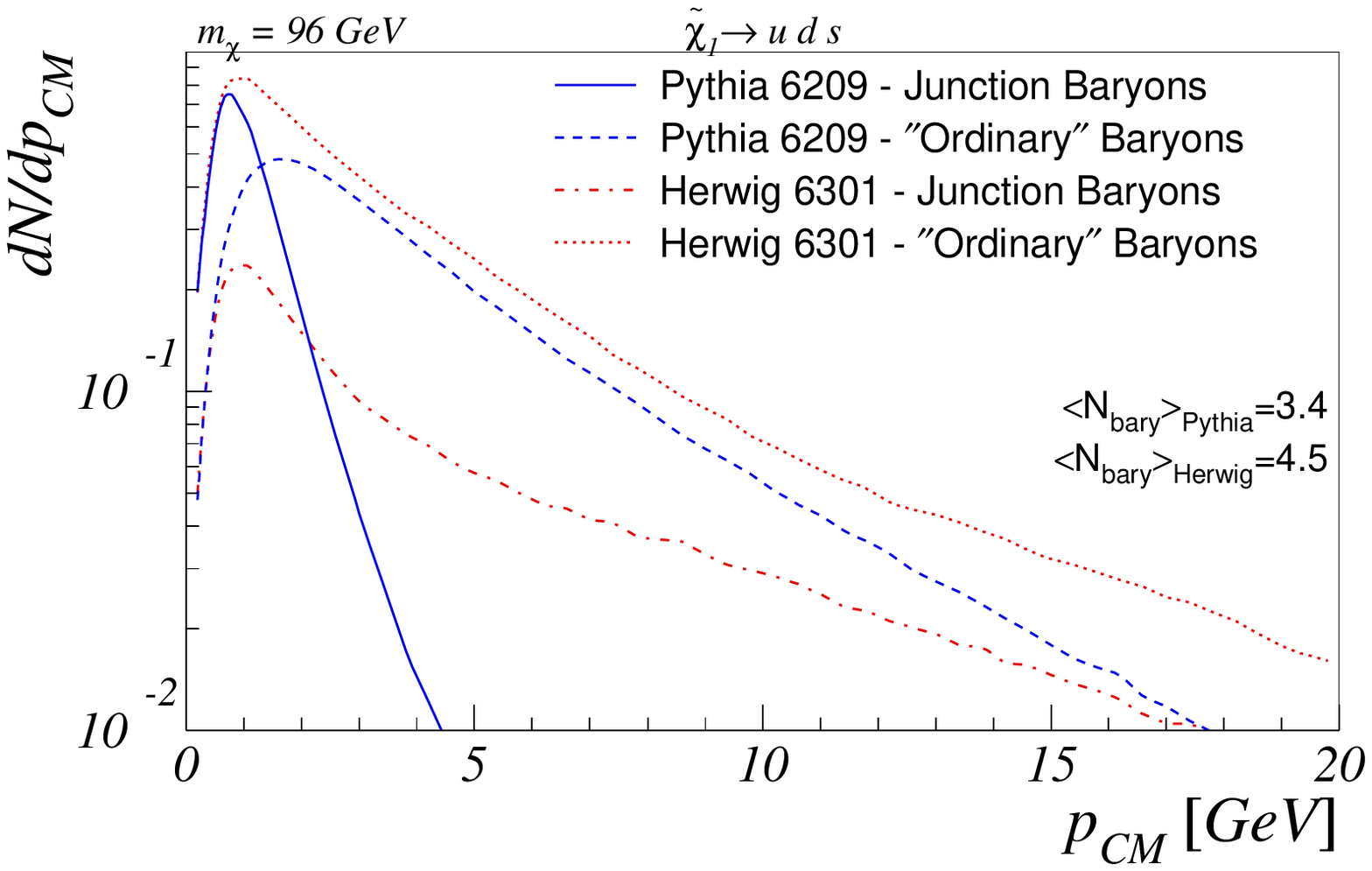}\\
\vspace*{-.75cm}(b)
\captive{%
CM momentum distributions in the decay of a 96 GeV neutralino for junction
(full line) and ordinary (dashed line) baryons (a) as compared
to the decay of a heavier neutralino (dot dashed and dotted line for
junction and ordinary baryons respectively), and (b) as compared to the same
process in \textsc{Herwig} (dot dashed and dotted line for
junction and ordinary baryons respectively).
\label{fig:jbmom}}
\end{figure}

In Fig.~\ref{fig:jbmom} the rates of both junction baryons and
``ordinary'' baryons are plotted as functions of their CM momenta for
$\schi\to\u\d\s$ for (a) two different neutralino masses (Snowmass points 1a
and 1b), and (b) comparison between \textsc{Pythia} and \textsc{Herwig} for
a 96 GeV neutralino.

As expected, one observes a much sharper peaking towards low
momenta for the junction baryon distributions as compared to the ordinary
baryon distributions in Fig.~\ref{fig:jbmom}a. Also, note that this
peaking does not depend on whether the neutralino is light or heavy.
A heavier neutralino gives rise to longer strings, i.e.\ larger momenta and
higher multiplicities, as can be seen by the shift in shape and normalization
of the ordinary baryon momentum distribution, but this does not significantly
affect how much energy will eventually be available to form the junction
baryon. The slight hardening of the junction baryon momentum for a higher
neutralino mass is attributable to the increased phase space for perturbative
gluon emission, which gives the junction a slightly higher average velocity
in the neutralino rest frame.

In Fig.~\ref{fig:jbmom}b we compare with the \textsc{Herwig}
implementation. Of course, \textsc{Herwig} does not have junctions and so no
junction baryons \emph{per se}, but it is nonetheless possible in most cases
to trace the BNV-associated baryon number, via a cluster with non-vanishing
baryon number, to a specific final-state baryon. And so, with a slight abuse
of nomenclature, we term this baryon the \textsc{Herwig} ``junction baryon''.
Naturally, both programs produce only one junction baryon per decay. However,
as shown in Fig.~\ref{fig:jbmom}b, the \textsc{Herwig} fragmentation produces
1.1 ordinary baryons more, on the average, than does
\textsc{Pythia}. Interpreted as a systematic uncertainty on the baryon
multiplicity in fragmentation, this difference is disturbingly
large. However, when comparing with the LEP experimental value at the $\Z^0$
peak, $\langle n_{\p} \rangle_{\mathrm{LEP}}=0.98\pm0.1$ \cite{LEPstudy}
protons per event (neutrons not being measured), default \textsc{Pythia} with
its 1.2 protons per event comes closer than default \textsc{Herwig} with its
1.5.  Presumably a better tuning to LEP than offered by the defaults would
also reduce the difference seen in neutralino decays.

The difference in the description of the junction baryon should persist,
however. In \textsc{Herwig} there is no particular reason why this baryon
should be slower than other baryons, and this results in a distribution for
junction baryons which looks more or less the same as that for ordinary
baryons, a situation quite at contrast to the shape of the junction baryon
spectrum predicted by the model presented here, cf.\ Fig.~\ref{fig:jbmom}b.
This means that the \textsc{Herwig} model offers less hope to find direct
evidence for the occurrence of BNV. We believe, however, that this model
is unrealistic, being flawed by an underestimated shower activity in
combination with an unphysical description of the fragmentation of the
baryon cluster, as explained above in section \ref{ss:showhad}.

\begin{figure}
\begin{center}
\includegraphics*[scale=0.8]{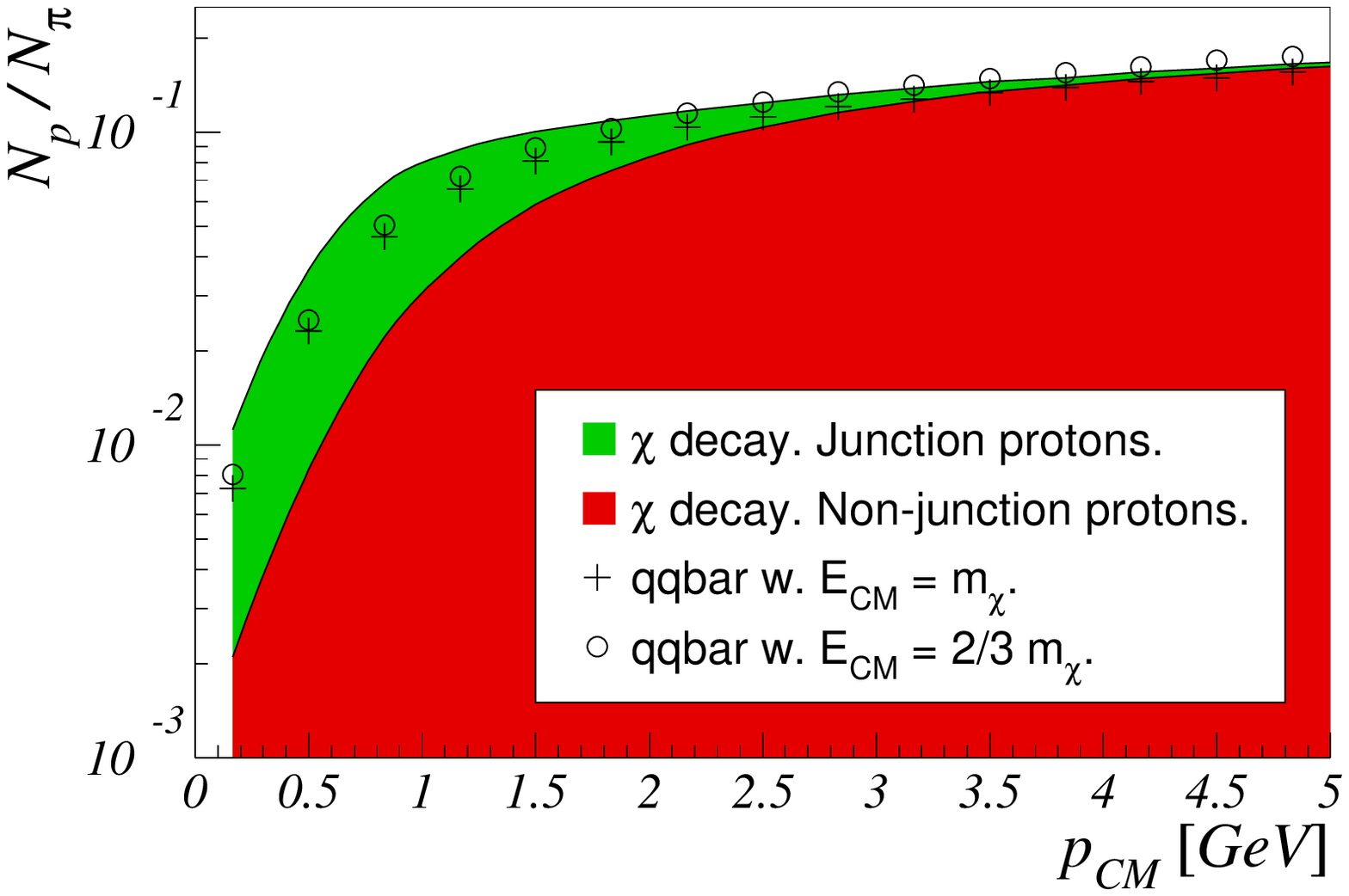}
\end{center}
\vspace*{-8mm}
\captive{%
Proton ($+$antiproton) CM-frame momentum spectrum normalized to the charged
pion one. Distributions are shown for 96 GeV neutralino decays to light
quarks (junction and
non-junction proton contributions shown separately) and for two cases of
ordinary $\q\qbar$ fragmentation, one (crosses) where the total CM energy is
the neutralino mass and one (circles) where it is only $\frac23 m_{\schi}$,
in order to make the energies of each piece of string equal.
\label{fig:baryonnumber}}
\end{figure}

When comparing baryon rates between physical processes, differences
in the overall multiplicity may confuse the issue. For instance, a
$\schi \to \q\q\q$ decay should have a larger multiplicity and softer
fragmentation spectra than a $\gamma^*/\Z^* \to \q\qbar$ decay of the
same mass, since the total energy is shared between more strings.
A decay to two gluons, such as $\hrm^0 \to \g\g$, would have an
even larger multiplicity, owing to the gluon energy being shared
between two strings and, by the same colour charge argument, also
radiating more in the perturbative phase. A more realistic measure is
the fraction of (anti)protons among charged particles, where the issue
of overall multiplicity  divides out, and quark and gluon jets of
different energies give almost identical results at low momenta.
In Fig.~\ref{fig:baryonnumber} we show the ratio of proton to
pion momentum spectra, which is basically equivalent with the above.
(Inclusion of kaons along with the pions would not affect the relative
difference between scenarios.) The mass differences and decay patterns
give a softer pion than proton spectrum, and therefore a rising $\p/\pi$
ratio. In the neutralino decay, the significant enhancement caused by the
junction baryon contribution is again visible. It is noteworthy that
the case of a $\q\qbar$ (or $\g\g$) decay gives a ratio intermediate to
the ones in the $\q\q\q$ decay without and with the junction baryon
included. The explanation is that the low-momentum region in the $\q\q\q$
topology is depleted from normal baryon production by the presence of the
junction baryon. Unfortunately, this means experimental signals are not
quite as dramatic as might have been wished for, as we shall see.

\subsection{Generic event properties}

We now turn our attention to a comparison of the expected
overall event features between \textsc{Pythia} and \textsc{Herwig},
specifically as characterized by the fragmentation spectra and
the multiplicity distributions for charged hadrons.

\begin{figure}
\center\vspace*{-.5cm}
\includegraphics*[scale=0.8]{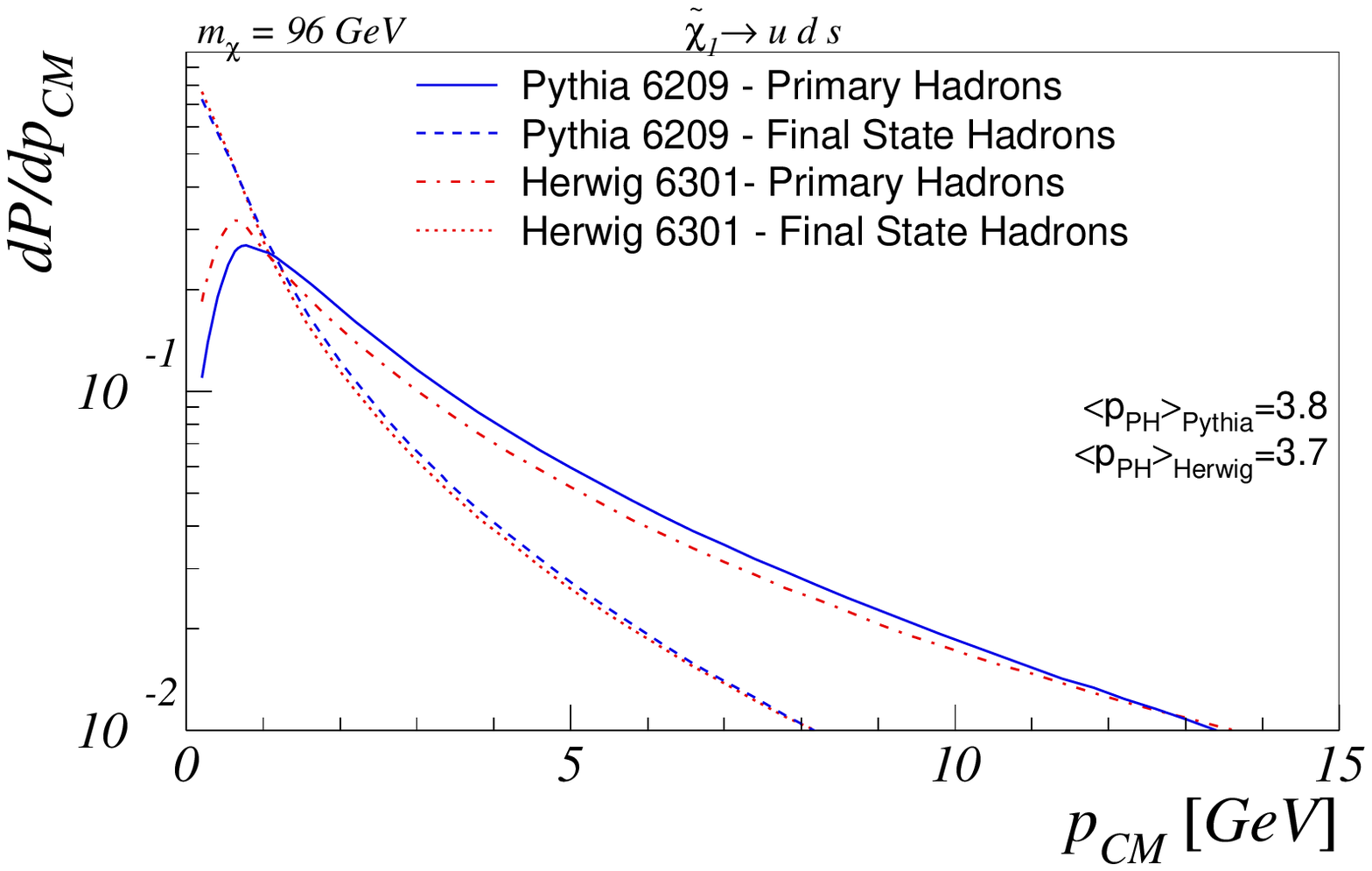}
\captive{%
CM momentum distributions in the decay of a 96 GeV neutralino for primary
hadrons (\textsc{Pythia} full lines, \textsc{Herwig} dot-dashed) and final
state hadrons (\textsc{Pythia} dashed lines, \textsc{Herwig} dotted). At the
bottom left, the mean primary hadron momenta are shown for each of the two
programs.
\label{fig:hadspec}}
\end{figure}

In Fig.~\ref{fig:hadspec} we show the hadronic
fragmentation spectra produced by the decay $\schi\to \u \d \s$. With
``primary hadrons'' and ``final state hadrons'' we understand hadrons
produced in the fragmentation of the strings, before and after decays take
place, respectively. There is good agreement, with slightly smaller primary
hadron momenta being produced by \textsc{Herwig} as compared to
\textsc{Pythia}.

\begin{figure}
\center\vspace*{-.5cm}
\includegraphics*[scale=0.8]{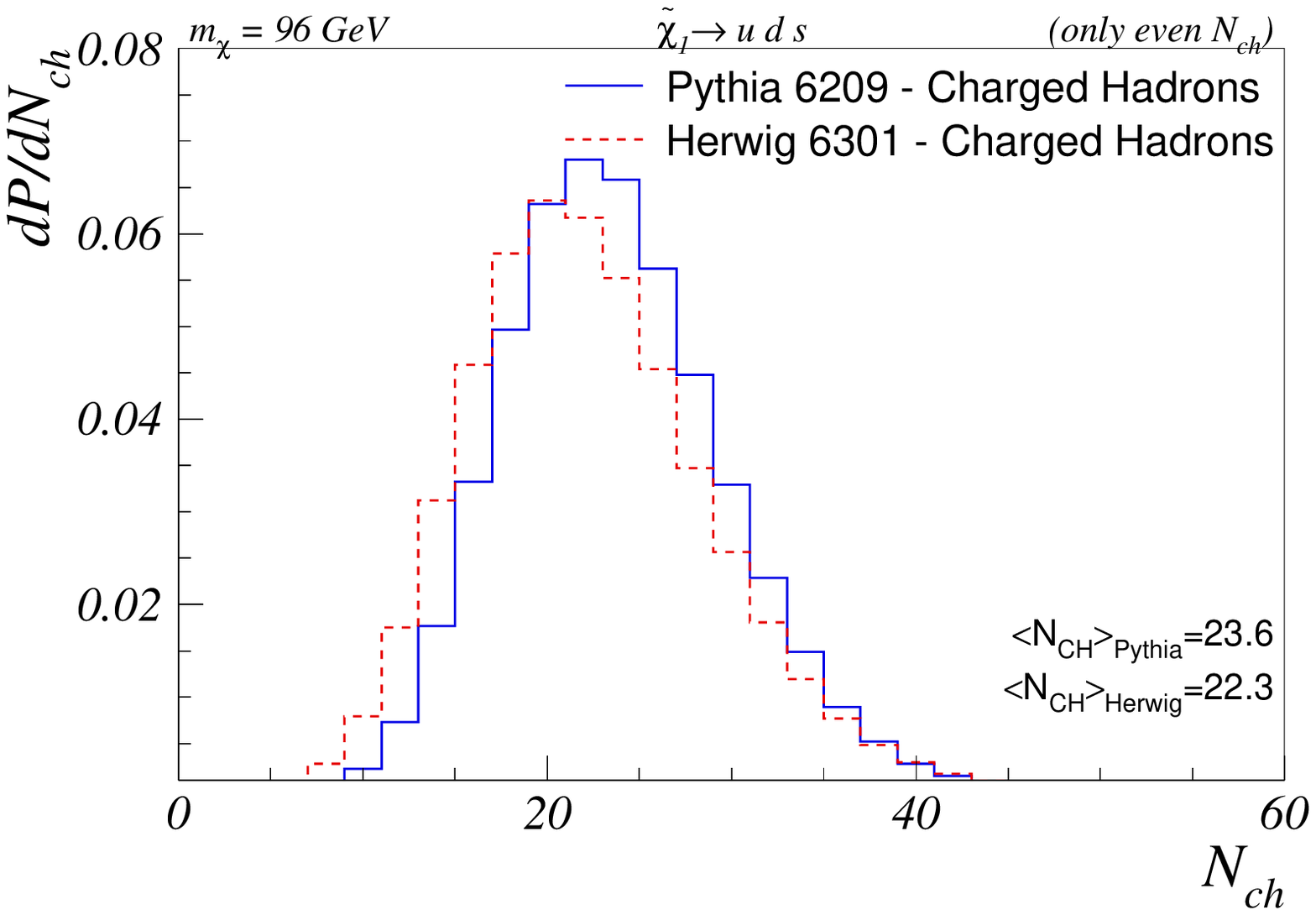}
\captive{%
Charged hadron multiplicities in the decay of a 96 GeV neutralino as
obtained with \textsc{Pythia} (full lines) and \textsc{Herwig} (dashed).
At the bottom left, the mean charged hadron multiplicities are shown for
each of the two programs. 
\label{fig:charhad}}
\end{figure}

In Fig.~\ref{fig:charhad} is compared the multiplicities of charged hadrons
as produced by \textsc{Pythia} and \textsc{Herwig}. An average difference
of almost 2 charged hadrons can be observed. When performing the
same comparison for the LEP1 process $\e^+\e^-\to\mathrm{hadrons}$, where
$\langle n_{ch}\rangle_{\mathrm{LEP}}=20.92\pm0.24$ \cite{LEPstudy}, the
\textsc{Pythia} 21.0 and \textsc{Herwig} 20.1 values also differ. This
mistuning partly explains the difference for neutralino decay between the two
programs. One should remember that the two hadronization models for this
kind of configurations are quite different from the respective description
of $\Z^0$ decay, however, so there is no strong reason for the two programs
to agree, neither with each other nor with the $\Z^0$ numbers.

Consider e.g.\ the simple string model, which predicts a logarithmic
increase of the multiplicity with jet energy,
$\langle n \rangle \propto \log (E / m_0)$, where $m_0$ is some typical
hadronic mass scale of order 1~GeV. Taking LEP events to consist of two jets
with half the energy each, and neutralino decays to be three jets with a
third each, the above LEP1 multiplicity would translate into something like
28 charged particles per neutralino decay. That the programs predict
multiplicities much below this indicates the importance of the details of the
event description.

\subsection{Alternative string topologies\label{subsec:altstring}}

Until now, we have considered only neutralino decay, since most of the salient
features of our model stand out more clearly in this case. Nevertheless, 
in regions of SUSY parameter space where the BNV couplings are
either larger than the gauge couplings, or where the gauge decays
of the lowest-lying squark mass eigenstate are kinematically
suppressed or forbidden, a special situation arises: the 
direct 2-body BNV decays listed in section \ref{s:bnv} dominate
for the lowest-lying squark mass eigenstate.
This gives us a unique chance to study the more exotic string
topologies discussed in section \ref{subsec:morejunc}:
junction--junction strings and junction--junction annihilation.

To quantify, we here take a generic ``light stop'' SUSY spectrum, Snowmass
point 5, with $m_{\st_1}=220$~GeV, and the decay 
$\st_1\to\schi^+_1 \b$ kinematically inaccessible.
Since we wish specifically to study 2-junction and 
0-junction topologies, we assume
third generation BNV couplings of order 0.1 or larger, 
so that the stop lifetime, cf.\ eq.~(\ref{eq:stoplife}), 
is sufficiently small that string breaks between
the stops will not normally occur before decay. Furthermore, we 
require that the stops be produced in an overall colour-singlet state, so that
their decays, leaving out the possibility of colour reconnections 
in the final state, are colour-connected to each other. This will generally 
not be the case at hadron colliders since, in processes like 
$\g\g\to\st_1\st_1^*$, each stop inherits its colour from a parton 
in the initial state rather than from the other stop. 

Consequently, we consider the process $\e^+\e^-\to\st_1\st_1^*$, with the
stops decaying to light quarks, $\st_1\to\dbar\sbar$. 
Again, differences with respect to the \textsc{Herwig}
implementation exist, but since we expect the main part of 
these to be covered already by the discussion in the preceding subsection, we
do not perform explicit comparisons here. Suffice it to note that the
hadronization model for BNV squark decays adopted in \textsc{Herwig} roughly 
parallels that of \textsc{Pythia} with only 0-junction configurations.

As was observed in subsection \ref{subsec:morejunc},
the probability for two connected 
junctions to annihilate is almost unity close to
threshold in our model. 
To ensure that the case where we let the choice between a
0-junction and a 2-junction topology be determined dynamically, i.e.\ by
eqs.~(\ref{eq:lamjj_ng})--(\ref{eq:lamnj}), indeed does differ from the pure
0-junction case, we choose a CM energy somewhat above threshold. 
For Snowmass point 5, the threshold is $2m_{\st_1} = 440$~GeV, thus 
a CM energy of 800~GeV is appropriate. 
This results in a junction--junction annihilation 
rate of about 0.6, cf.\ Fig.~\ref{fig:2j0j}. 

We again begin by considering the jet rates, for the specific 
process $\e^+\e^-\to\st_1(\to\dbar\sbar)\st_1^*(\to\d\s)$ at 800~GeV
CM energy. The jet clustering is performed using a Durham algorithm 
in the $\e^+\e^-$ CM frame, yet since 
we here make the simplification of neglecting initial-state radiation and
beamstrahlung, 
this frame is identical to the $\st_1\st_1^*$ CM. Note that this frame
need not generally coincide with the  $\st_1\st_1^*$ CM at decay time, due to
the possibility of the stop pair emitting final-state radiation.

\begin{figure}
\begin{center}
\center\vspace*{-.75cm}
\includegraphics*[scale=0.8]{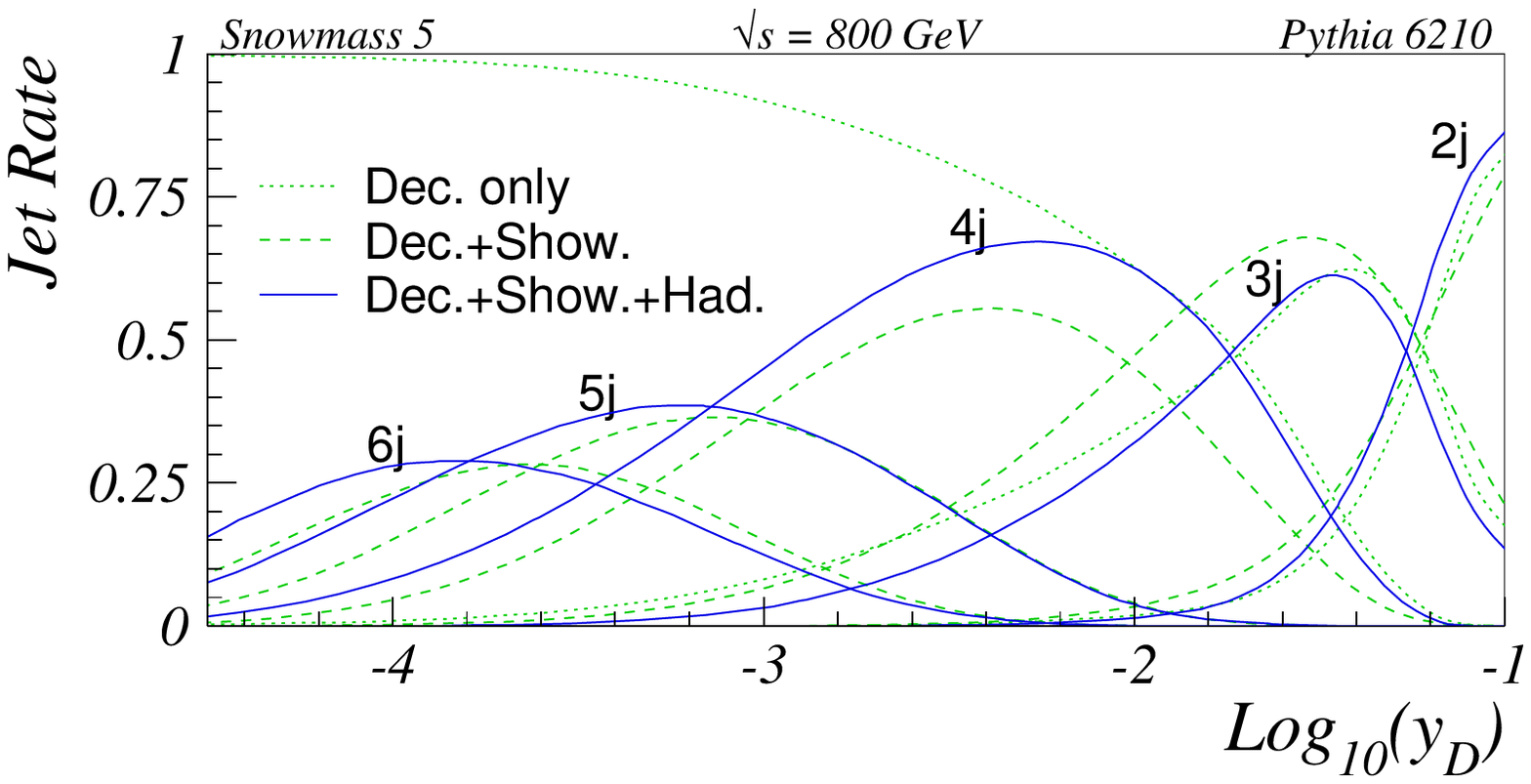}\vspace*{-0.3cm}\\
\end{center}
\captive{%
Jet rates for the process $\e^+\e^-\to\st_1(\to\dbar\sbar)\st_1^*(\to\d\s)$
at 800~GeV CM energy, as functions of $\yD$, 
for three stages of the event generation:
clustering only of the primary $\st_1\st_1^*$ decay products (dotted curves),
clustering after showering but before hadronization (dashed curves), 
and clustering after full event generation (solid curves). 
\label{fig:stoprates1}}
\end{figure}
In Fig.~\ref{fig:stoprates1} we show the jet rates for dynamical
selection between 2-junction and 0-junction topologies, 
as functions of $\yD$. 
To trace the evolution of the event, we plot the rates for
initial decay (dotted curves), after parton shower (dashed curves), and after
hadronization (solid curves). 

Since each stop decays to
a 2-body final state, we obtain four jets at leading order. 
As can be observed from the dotted curves in Fig.~\ref{fig:stoprates1}, the
initial decay daughters are actually clustered to fewer than 4 jets for a
non-negligible fraction of the events down to quite
low values of $\yD$, this 
due to asymmetric decays causing two daughters to end up close to each other
and so be clustered to one jet rather than two. 

\begin{figure}
\begin{center}
\center\vspace*{-.75cm}
\includegraphics*[scale=0.8]{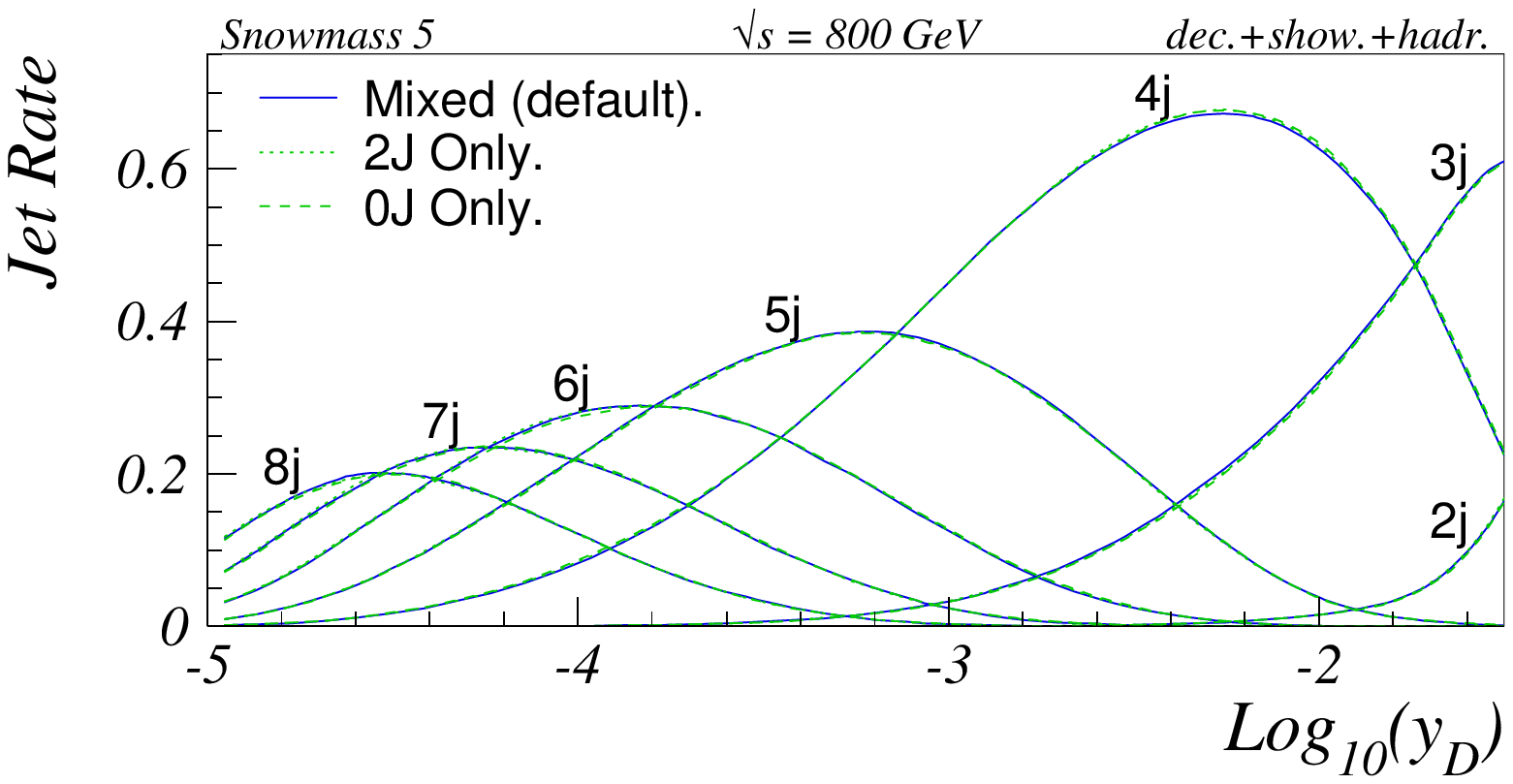}\vspace*{-0.3cm}\\
\end{center}
\captive{%
Jet rates for the process $\e^+\e^-\to\st_1(\to\dbar\sbar)\st_1^*(\to\d\s)$
at 800~GeV CM energy, as functions of $\yD$, 
for three choices of string topology:
mixed 0-junction and 2-junction (solid curves), 
2-junction only (dotted curves), and 0-junction only (dashed
curves). 
\label{fig:stoprates2}}
\end{figure}

The question now arises to what extent, if any, the choice of string topology
affects the jet rates. To study this, in 
Fig.~\ref{fig:stoprates2} we replot
the solid curves of Fig.~\ref{fig:stoprates1} together with the two extreme
cases where all events are forced to be either of the 2-junction (dotted
curves) or the 0-junction (dashed) type. Note that the latter sample has 
a small contamination of 2-junction events, from perturbative breakups in the
$\st_1\st_1^*$ shower. As can readily be observed,
the curves are identical to a high precision, bearing witness to the absence
of any resolvable structural difference between the topologies, at least for
the $\yD$ ranges and jet numbers we found it meaningful to include in the
analysis.

We thus do not believe that these semi-inclusive observables
could be used e.g.\ to measure the junction--junction annihilation rate. 
We return to the issue of whether exclusive quantities can do a better job in 
the more experimentally oriented study in subsection \ref{subsec:stop}.

\section{Experimental tests}

In this section we present some semi-realistic studies on how baryon
number violation could be pinned down experimentally. We do not propose
to recapitulate the extensive experience on SUSY search strategies,
with likelihood techniques or a succession of cuts and sideband
subtraction procedures, required to isolate signals relative to
backgrounds for different SUSY parameter sets. Rather, we assume that
a signal has already been found in multijet final states, with bumps
in the two- or three-jet invariant mass distributions consistent with
the presence of new particles. The key issue is then whether this is
evidence for SUSY with BNV or for some completely other physics signal.
The decision is likely to be based on a combination of different pieces
of information, but a crucial one would be to find the baryon
associated with the BNV itself. This is not possible on an event-by-event
basis, owing to the background of normal baryon production and the
non-identification of neutrons. What we can aim for is then to
demonstrate the existence of a statistically significant
excess (or not) of baryon production in the momentum region predicted by our
model.

This kind of demonstration will most likely be beyond the reach of the
Tevatron; therefore in the following we
concentrate on an LHC scenario with pp collisions at $\sqrt{s}=14$~TeV.
We take a crude
representation of the ATLAS
tracking and calorimetry capabilities (described
below) to be sufficiently detailed for the kind of analysis we will
consider, specifically
whether an excess of protons \`{a} la Fig.~\ref{fig:baryonnumber}
can still be seen for fully generated events in a semi-realistic
detector environment.

Of course, a similar exercise could have been performed
for CMS but, at our level of detail, we expect the differences between
the two experiments to be of negligible consequence. The exception
would be baryon identification strategies, see below.

With respect to the prospects at linear colliders, the comparatively
clean environment
and high luminosity imply that studies similar to the ones presented here
would be easier to carry out, 
although they would obviously have to wait for such a facility to be 
built. Therefore, we constrain our attention in this direction to one
special case, that of colour-connected BNV $\st_1\st_1^*$ decays, 
which we expect to be nigh impossible to study with hadron colliders.

\subsection{Detector layout and acceptance}
Based primarily on the ATLAS physics TDR \cite{atlastdr},
we assume a calorimeter that covers the region $|\eta|<5$ in
pseudorapidity and $0<\phi<2\pi$ in azimuth, with a granularity of
$\Delta\eta\times\Delta\phi = 0.1\times 0.1$. This reproduces the expected
ATLAS calorimeter granularity at mid-rapidity, but overestimates it at large
rapidities. Smearing due to finite resolution effects has not been simulated.

\paragraph{Proton and pion reconstruction.} For
$\p^\pm$ and $\pi^\pm$ reconstruction and identification we
use the region $|\eta|<2$, well inside the coverage of the ATLAS and
CMS tracker systems. Inside this region, all charged hadrons
with $p_\perp>1$~GeV are reconstructed. This is somewhat more
optimistic than what would be possible in the real world.
Particles with $p_\perp<1$~GeV predominantly come from the underlying
event, so the $p_\perp$ cut reduces the study to particles associated
with the hard physics of the event.

Ideally, we would like to identify all charged hadrons in the
above region, with complete separation between the $\p^\pm$
signal and the $\pi^\pm$ and $\K^{\pm}$ background. In reality
this will never be possible. Neither ATLAS nor CMS has yet
carried through detailed detector simulation of their proton
identification performance. Our physics scenario may represent
the first example where this capability is relevant in the study
of physics beyond the Standard Model, and so could offer a
convenient target for future detector studies. (QCD physics
obviously provides several reasons to have this capability, one
of which will be mentioned later.) In the meantime we exemplify
what could be possible.

The conservative approach is to restrict the search to the
$\Lambda^0\to \p^+\pi^-$ channel which, due to its two charged tracks
from a displaced vertex and narrow mass peak, offers a reasonably clean
signal. This channel also has the advantage that it doesn't cut down the
statistics much: the flavour composition of fragmentation in $\e^+\e^-$
annihilation is measured to give $N_{\Lambda^0}/N_{\p} \sim 0.4$
\cite{pdg} and the branching ratio of $\Lambda^0$ to $\p^+\pi^-$ is
about 60\% \cite{pdg}, so about one in every four protons comes from
a $\Lambda^0$ decay to $\p^+\pi^-$. Since the proton inherits most of
the momentum in the $\Lambda^0$ decay, the proton plots shown below
would also apply for $\Lambda^0$, with minor modifications.

Nevertheless, one may aggressively assume some proton identification
capability, to augment the statistics from the $\Lambda^0$ decay,
and maybe also to clean up this signal from the
$\K^0_{\mathrm{S}} \to \pi^+\pi^-$
background. For ATLAS, charged hadron identification is not a prime
objective for the Inner Detector, but the ionization loss, d$E$/d$x$,
in the Transition Radiation Tracker gives enough information to
separate protons from pions by more than 1$\sigma$ in the momentum range
$3$~GeV $\lesssim p\lesssim 20$~GeV \cite{atlastdr}. However, the
$\K^{\pm}/\p^{\pm}$ separation is very poor even in this range, so an
analysis purely based on this capability will contain a non-vanishing
misidentification background. Calorimetry provides auxiliary information,
especially for antiprotons, which release an additional GeV of energy by
their annihilation. This should be noticeable, at least for antiprotons
stopping in the electromagnetic calorimeter. For protons the mismatch
between energy and momentum goes in the other direction and is smaller.
We note that the hadrons discussed here are not ones found in the core
of jets, and anyway have so low momenta that they are significantly
deflected by the magnetic field. Therefore they are likely to be
isolated in the calorimeter.

The possibility to use d$E$/d$x$ in the CMS silicon tracker is still
under study, but in any case the tracker has only a few sampling points,
and so limited capability in this respect. On the other hand, there
is considerable empty space between the silicon layers. Therefore one
can imagine to install a time-of-flight system at some future date. This
would allow good proton identification in the lower part of the
interesting momentum range. As above, calorimetry could be used to
identify some of the antiprotons.

\paragraph{Jet reconstruction.} We use a cone algorithm over the full
fiducial volume of the calorimeter, $|\eta|<5$, with
a cone size of $\Delta R=\sqrt{\Delta\eta^2 + \Delta\phi^2}=0.4$ and a
minimum jet seed energy of 5 GeV. Reconstructed jets with a transverse energy
below 25 GeV are not used in the further analysis. No jet energy
recalibration procedure has been applied.

\subsection{Physics cuts}
As mentioned above, we assume that a relatively clean
sample of SUSY events has been isolated before the present analysis is
applied, hence we do not study SM backgrounds. The event sample thus consists
of a general SUSY simulation for our reference
point, Snowmass point 1a \cite{allanach02},
including all SUSY processes currently implemented in \textsc{Pythia}. Most of
these, however, are suppressed at hadron colliders so that the cross section
is dominated by gluino and squark production. We concentrate on the case
where the BNV couplings are smaller than the gauge couplings, thus sparticles
will only use the BNV channels as a ``last resort''. That is,
the ordinary MSSM pattern of cascade decays will persist, with the
modification that the LSP decays. Assuming that the lightest neutralino is
the LSP, we are thus looking at events of the type:
\begin{equation}
2(\q/\g) \to 2(\sq'/\sg)  \to 2(\q''/\g)(+X) + 2(\schi^0_1) \to 2(\q''/\g)(+X)
+2(\q_i\q_j\q_k),
\end{equation}
where the two initial partons come from the beam protons and $X$
signifies what, if anything, is split off in the cascades besides the
neutralinos. Since the neutralinos are not colour connected to the rest of
the event, we note that
this type of event is nothing but an ordinary $R$-conserving
MSSM cascade event
with the missing energy transformed into two separate $\schi^0_1$
decays.

\begin{figure}
\center
\includegraphics*[scale=0.7]{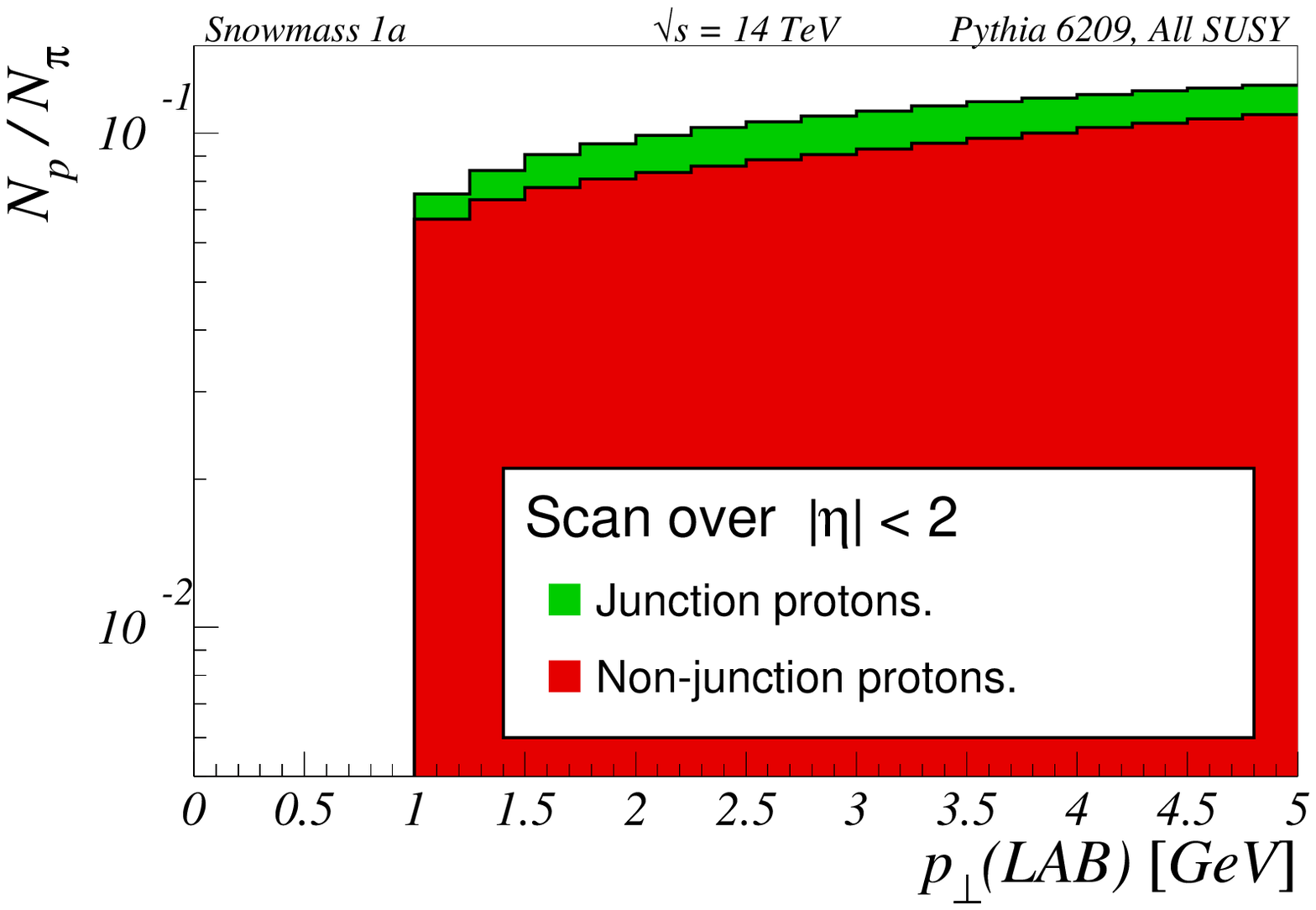}
\captive{Reconstructed $\pT$ spectrum for the
ratio of protons to charged pions in a simple scan over the detector
volume.
\label{fig:baryondet}}
\end{figure}

For such events, about 25\% of the junction protons lies 
within the detector acceptance. Without any further cuts, the
equivalent of Fig.~\ref{fig:baryonnumber} looks as depicted in
Fig.~\ref{fig:baryondet}. A rather depressing result, of course
due to the many protons which originate exterior to the neutralino decay
itself, and to the fact that we are looking at momenta
in the CM of the original collision rather than in the CM of the neutralino.

However, it is possible to do much better by placing cuts
designed to identify
the 3 jets from each of the $\schi^0_1$ decays,
reconstructing the presumed $\schi^0_1$ CM and looking
for low-momentum protons in that frame. In principle, it would be even better
to attempt to reconstruct the junction rest frame itself, based on the
directions of the three jets. Here, we note that for
well separated jets, as seen in the neutralino CM,
the junction rest frame is at any rate
closely approximated by the CM frame, and
so in this preliminary study we content
ourselves with working in the 3-jet CM.

We use two different
strategies to identify the neutralino decay jets.
The first, blind 3-jet reconstruction,
probably does worse than what could be done with more elaborately designed
cuts, and the second, optimized 3-jet reconstruction, probably errs on the
side of too optimistic results. Thus, the actual experimental result should
lie somewhere between these two extremes.

\subsubsection{Blind 3-jet reconstruction}
We term this analysis ``blind'' since we assume no \emph{a priori}
``divine knowledge'' of which jets are the correct ones etc. Thus,
we use a selection of cuts to
attempt to identify the neutralino decay jets
as precisely as possible, then reconstruct the presumed neutralino CM and
look for low-momentum particles there. Since our
cuts have not been
through a long, rigourous optimization procedure, it is likely that
we err on the low side of what could be done with better optimized cuts.

Note also that our cuts
rely to some extent on features specific to the SUSY scenario we
use (Snowmass point 1a), but they should produce acceptable results for any
scenario that shares its general features: a neutralino LSP and a rather
light sparticle spectrum with a relatively large gap between the
squarks (and the gluino) and the LSP.

With this type of spectrum, the squark/gluino decays release much kinetic
energy, giving rise to generally high-$p_\perp$
neutralinos. Since the $\schi^0_1$ mass itself is small
in comparison, its three decay jets always occupy a relatively small total
region of the detector. These two considerations
form the backbone of our cuts.
In decays like $\schi^0_1\to \c\s\b$, one could
furthermore require heavy-quark tagged jets, but we do not rely on
such extra information here.

To quantify, we select events with 8 or more reconstructed jets and in those
look for systems of 3 jets
which have a total $p_\perp^{\mathrm{3j}}>200$~GeV. The measure we use
for how far the three jets are from
each other in the detector is the maximal jet--jet $R$ distance:
\begin{equation}
\Delta R^{\mathrm{max}}_{\j\j}=\mathrm{Max}(\Delta R_{\j_1\j_2},
\Delta R_{\j_2\j_3},\Delta R_{\j_3\j_1})
\end{equation}
where $\j_i$ stands for the $i$'th jet in the candidate 3-jet system.
We impose a cut requiring $\Delta R^{\mathrm{max}}_{\j\j}<0.8$. Since we use a
cone size of $\Delta R=0.4$, this cut translates into
looking for 3 jets which overlap or just touch each other. The details of how
the energy in the overlapping regions is assigned to the each of the
overlapping jets depends on the particular jet algorithm used. However,
as long as double counting is avoided, we do not expect these details to
influence our analysis; ideally, the CM of the 3 jets should be invariant to
changes in the sharing of energy between its 3 constituent jets.

To further bring down the combinatorics, we require that none of the
2-jet pairs in the 3-jet configuration
have an invariant mass close to or larger than the neutralino mass. Assuming
that this is already known to some precision, we
reject candidate 3-jet systems where any pair of jets has a mass larger than
90 GeV. Of the remaining candidate 3-jet systems, we select those
which reconstruct to the neutralino mass $\pm 10$~GeV.

For each of these 3-jet systems
we search for candidate junction protons within the detector acceptance
described above. We check whether there are any (anti)protons
within $\Delta R_{\p3\j}<0.5$ of the 3-jet system momentum direction. Any
such protons are accepted as candidate junction protons. About 5\% of the
junction protons within the detector acceptance survive these cuts. The
corresponding number for the non-junction protons is 0.2\%.

\begin{figure}
\center
\includegraphics*[scale=0.7]{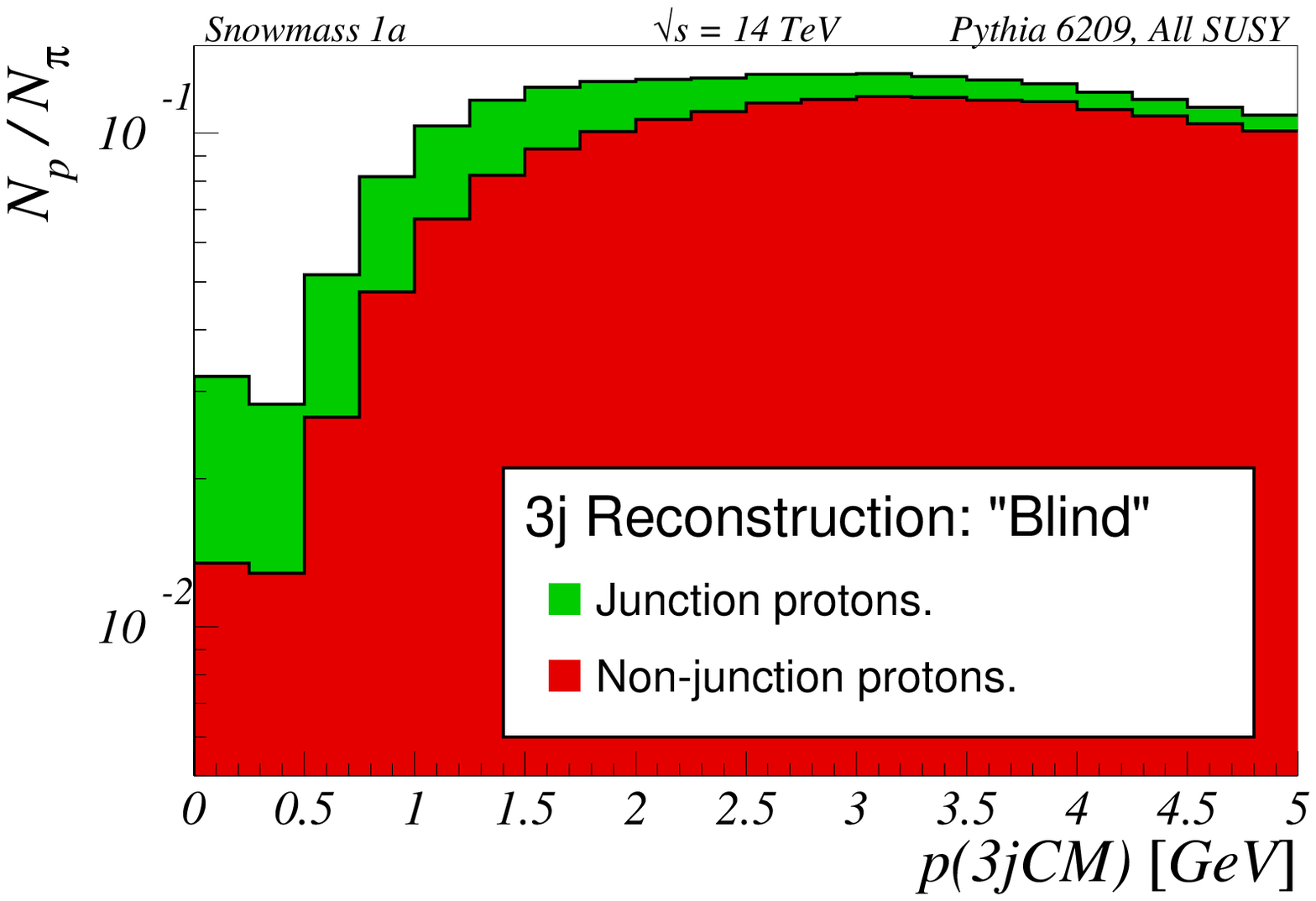}\vspace*{-5mm}
\captive{Reconstructed proton momentum spectrum in
the 3-jet CM frame for the ``blind'' neutralino decay jet identification.
\label{fig:problind}}
\end{figure}

The interesting quantity now is the momenta of these protons
in the 3-jet system CM frame, our first
approximation of the junction rest frame. As
has been argued above, we expect that the real junction protons will exhibit
a softer momentum spectrum in this frame than will be the case for
non-junction protons. Thus, we expect that
the enrichment of the sample by junction protons
should be largest for small proton momenta in the reconstructed neutralino CM.
Results are depicted in Fig.~\ref{fig:problind}.

The situation has improved
noticeably over the one in Fig.~\ref{fig:baryondet}, but
there is still a lot of non-junction protons at small $\pT$ which, as we
shall see in the next subsection, may be further reduced by reducing the
combinatorial background, i.e.\ by having a more precise identification of
the correct neutralino decay jets.

\subsubsection{Optimized 3-jet reconstruction}
We now assume that a nearly perfect neutralino decay jet
identification can be performed, by applying some selection of
cuts, the details of which are not interesting here. Rather, we
mimic the effects of such optimized cuts by using the event
generator information.

For each of the neutralino daughters, we select that jet which
lies closest in $\vec{p}_\perp$ and use only those jets for
reconstructing the presumed $\schi^0_1$ CM's. The combinatorics
are thus brought down to very low levels. One technicality
involved is how we deal with situations where one jet turns out
to be the closest in $\vec{p}_\perp$ to more than one of the
neutralino daughters. This situation can arise for instance if
those two daughters were sufficiently close together to be
clustered to a single jet, or if one of the daughters produced a
jet that lies outside the active calorimetry range.
 Such systems are not included in the analysis.

For each accepted 3-jet system we search for candidate junction protons and
charged pions in exactly the same manner as for the blind
analysis. Only 1\% of the junction protons within the detector acceptance
survive, still with a contamination of about 0.2\% of the non-junction
protons, but the junction proton distribution is now more
sharply peaked towards low momenta (in the 3-jet CM frame) and the
non-junction protons are found at higher momenta.

\begin{figure}
\center
\includegraphics*[scale=0.7]{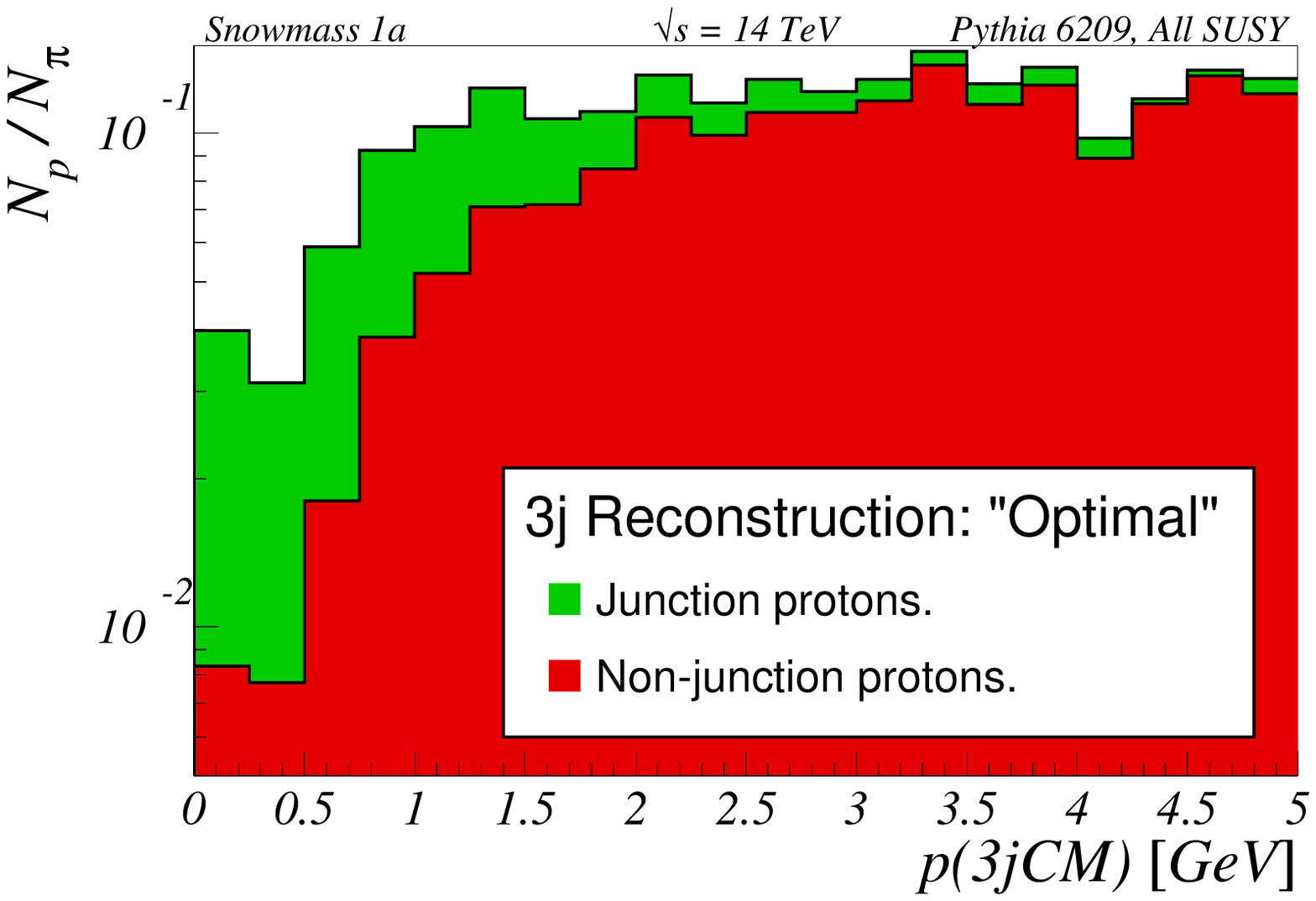}\vspace*{-5mm}
\captive{Reconstructed proton spectrum in the 3-jet CM frame
for the ``optimized''
  neutralino decay jet identification.
\label{fig:proopt}}
\end{figure}

The distribution of proton to pion ratios as a function of $p$ in
the 3-jet system CM frame is shown in Fig.~\ref{fig:proopt}.
A noticeable improvement of the ratio of junction protons to non-junction
protons can be seen in the region below $\sim 1$~GeV. For the blind analysis,
 cf.\ Fig.~\ref{fig:problind}, this ratio was about 2. The
 corresponding number for Fig.~\ref{fig:proopt} is closer to 4, approximately
 the same factor as for the isolated neutralino decay,
 Fig.~\ref{fig:baryonnumber}.

We thus conclude that even
 with all the surrounding activity in fully generated
pp collisions, it \emph{is} possible, with a good jet
 selection, to arrive at a proton sample of almost the same purity
 as for the case of isolated decay. It should be remembered, though, that we
 are not here talking about an \emph{excess} of protons of a factor of 4,
since the division into junction baryons and non-junction
 baryons is not an experimental observable. What is expected from ordinary
 $\q\qbar$ fragmentation is represented in Fig.~\ref{fig:baryonnumber} by
 circles and crosshairs, and we see that relative to \emph{this} number, the
 excess is only about a factor 1.5. We note that one does not have to rely on
 a Monte Carlo estimate of the proton to pion ratio. A study of reference
 background samples, of multijet events in configurations not significantly
 contaminated by SUSY events, will allow an absolute determination of the
 proton fraction from data. We therefore believe that even such a modest
 difference as a 50\% enrichment should be clearly visible, given enough
 statistics.

Note that the rate of baryon production in high-energy collisions
offers one puzzling outstanding problem in our understanding of QCD:
why are fewer kaons and protons produced in $\e\p$/$\gamma\p$ events than
in $\e^+\e^-$ ones? This problem has been with us since several years
\cite{tsjphysg}, and has been reconfirmed by recent HERA studies
\cite{patHERA}. One possible explanation is the `quiet string scenario'
\cite{tsjphysg}, namely that soft gluon emission around the
perturbative--nonperturbative border is less profuse in
$\e\p$/$\gamma\p$ events, where only part of the event feels a hard
scale, than in $\e^+\e^-$ ones, where the whole event derives from a
hard-scale process. The string would then be less `wrinkled' in the
former case than the latter, thus have a smaller effective string
tension, and thereby a lesser production of heavier particles. This is
very speculative, however, and having to give up jet universality would be
uncomfortable, to say the least. One possible test in Deeply Inelastic
Scattering would be a comparison of the current hemisphere in the Breit
frame with the rest of the event. It is also unknown whether
Tevatron collider events more resemble LEP or HERA ones in this respect.
In the quiet string scenario, one would expect particles associated with
the perturbative jets, i.e.\ the particles we have studied, to resemble
LEP, while the underlying event again could be quieter and therefore
also contain fewer baryons. The older UA5 measurements on
$\K^0_{\mathrm{S}}$ \cite{UA5} do not indicate any problem for the
standard string model \cite{Zijl}. Hopefully the issue will be settled
well before a study of observed potential BNV events is initiated.

\subsection{BNV two-body decays\label{subsec:stop}}
We now follow up the discussion in subsection \ref{subsec:altstring},
considering the process $\e^+\e^-\to\st_1(\to\dbar\sbar)\st_1^*(\to\d\s)$ for a
linear collider, TESLA or the NLC/JLC, 
using the mSUGRA parameters of Snowmass point 5 to represent a generic
``light stop'' scenario. 
Although the 220~GeV stop mass of this benchmark point 
is small enough to allow pair-production at 500~GeV, 
the amount of junction--junction annihilation depends strongly on the 
stop boost, meaning that
essentially no junctions would survive at CM energies close to
threshold in our preferred stringlength-minimization scenario. 
Thus, to increase the 2-junction rate, 
and also to get a higher total production cross section, we are lead 
to consider an 800~GeV option. Note that if the 
BNV couplings are small, we expect string breaks
between the stops to
occur \emph{before} decay. Such breaks would sever 
the connection in colour space between
the two junctions and thus effectively eliminate the possibility 
of obtaining 0-junction topologies. 

At the chosen CM energy, the stop boost 
$\gamma_{\st}=E_{\st}/m_{\st}=\ECM/2m_{\st}\sim$1.8,
making us expect that, in the case of large BNV couplings, 
about 40\% of the junction--junction systems 
survive to give rise to junction baryons, cf.\ Fig.~\ref{fig:2j0j}. The 
remaining 60\% of the events presumably represent a more or less
irreducible background to our search, since they contain no signal 
(junction) baryons and only differ from the signal events at the 
hadronization level, cf.\ Fig.~\ref{fig:stoprates2} and the discussion there. 
Of course, once BNV is established, also the 
rate of junction annihilation by itself could offer interesting
information on QCD.

Similarly to the LHC study, we assume that a clean sample of $\st_1\st_1^*$
events has been isolated, so that SM as well SUSY backgrounds can be
neglected. A further simplification we make is to disregard the effects of 
brems- and beam-strahlung in the initial state, mostly since the latter
depends sensitively on the beam parameters and we do not wish to become
too machine-specific at this stage.

So as not to be completely unrealistic in regard to detector
acceptance, we ignore all particles closer than $5^\circ$ to the beam pipe
and reconstruct only charged particles which have $\pT>0.5$~GeV. 
Beyond these
crude cuts, no effort is made to simulate detector effects.
For jet reconstruction we use Durham $\yD=10^{-2.5} \approx 0.003$. This
results in about 65\% 4-jet events, 20\% 5-jets, and 10\% 3-jets for the
process considered. Naturally, the events with more than 4 jets 
are those which involve emissions of hard gluons. This means 
partly that we are facing a larger combinatorics, and partly that
we are less certain in which direction the junctions went and hence in which
direction to look for the junction protons. For the 3-jet events, two of the
original four jets have ended up very close to each other. With a smaller
$\yD$ those jets might be resolved, but it is not clear that 
much could be gained by this. Thus, all but the 4-jet events are removed 
from the further analysis.
 
\begin{figure}
\vspace*{-.5cm}
\begin{center}
\begin{tabular}{c}
\includegraphics*[scale=0.7]{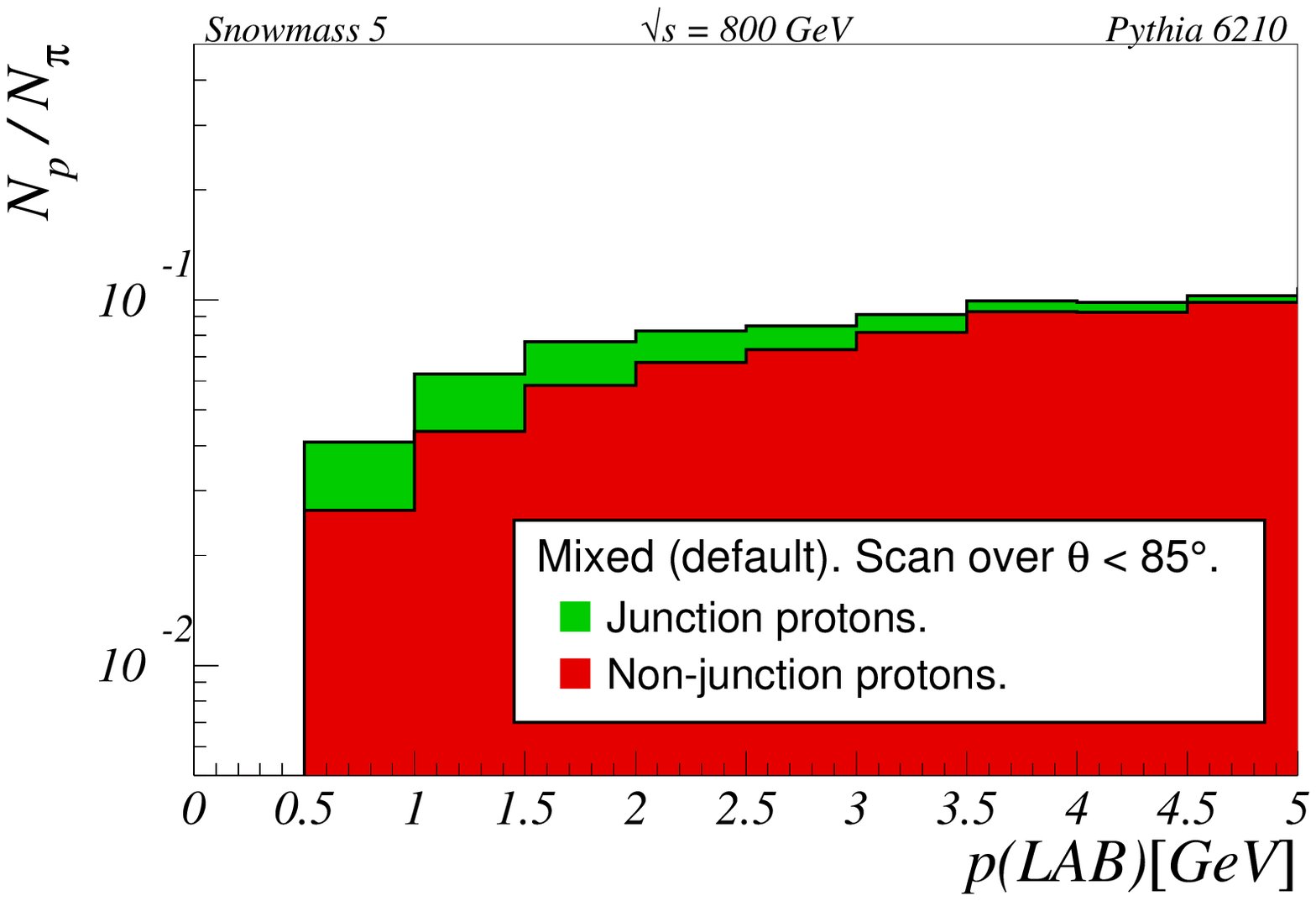}\vspace*{-.7cm}\\
(a)\vspace*{-.5cm}\\
\includegraphics*[scale=0.7]{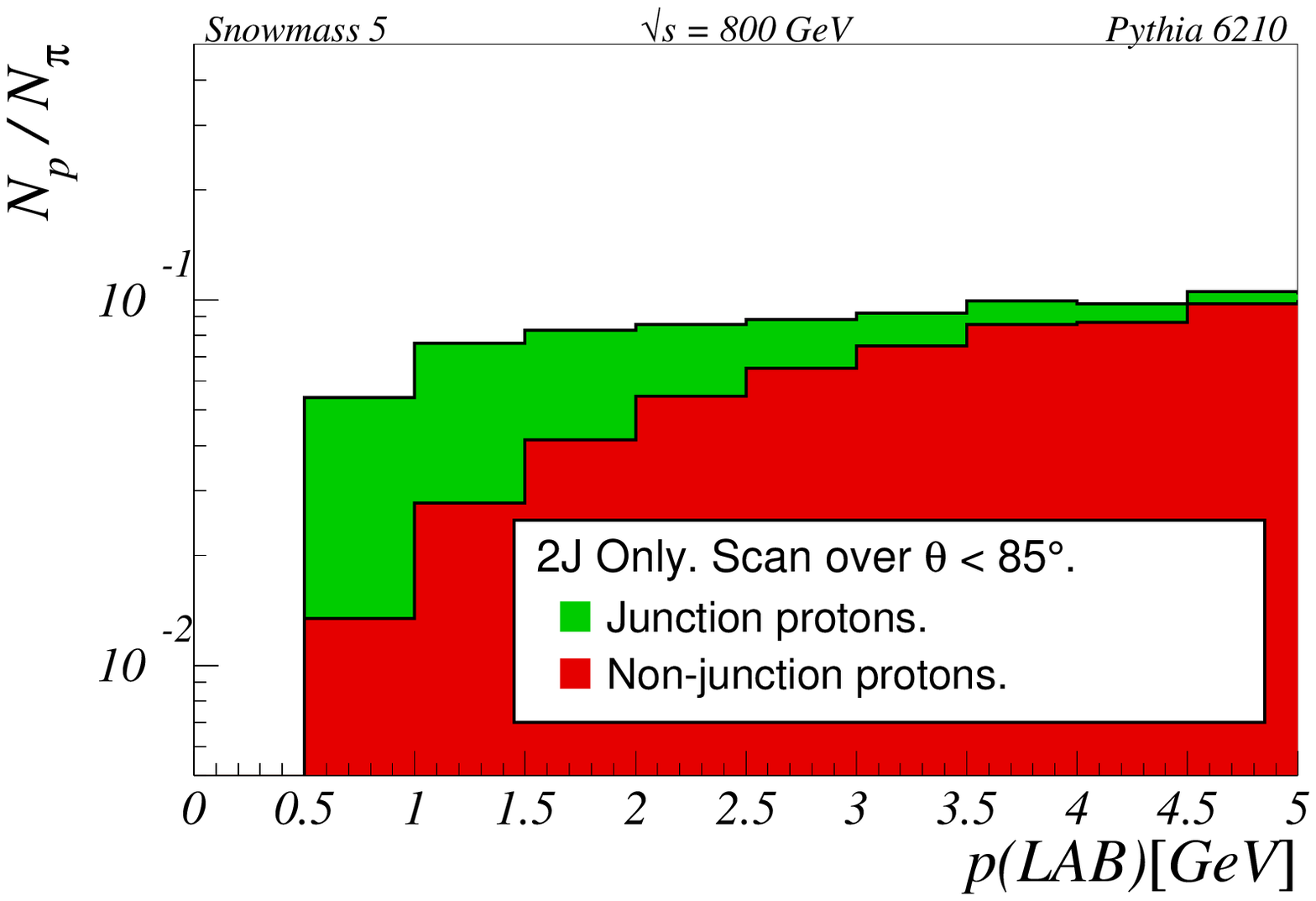}\vspace*{-.7cm}\\
(b)
\end{tabular}
\end{center}
\vspace*{-5mm}
\captive{%
Reconstructed momentum spectrum for the
ratio of protons to charged pions in a simple scan over the detector
volume: (a) mixed
2-junction and 0-junction topologies and (b) pure 2-junction topologies.
\label{fig:stopcount}}
\end{figure}

We begin by simply performing a proton-count within the detector acceptance,
normalizing to the charged pion multiplicity,
as we did for the LHC study. The result, shown in
Fig.~\ref{fig:stopcount}, is markedly 
better than for the LHC case, Fig.~\ref{fig:baryondet}, chiefly due to the
cleaner environment and that we now go down to $\pT=0.5$~GeV. 
Even though, in Fig.~\ref{fig:stopcount}a,
we have only a 40\% rate of 2-junction configurations, 
about 30\% of the protons below 1.5~GeV are junction protons. This should be
compared with 
Fig.~\ref{fig:stopcount}b, where we have turned off junction--junction
annihilation. 

Fortunately, it is again possible to purify the sample somewhat more. First,
we identify which of the 3 possible pair-by-pair combinations of the four jets
is most consistent with both pairs having the invariant mass of the $\st_1$,
by selecting the configuration with the smallest value of 
$\Delta M \equiv |m_{ij} - m_{\st_1}| + |m_{kl} - m_{\st_1}|$. 

\begin{figure}
\vspace*{-0.5cm}
\begin{center}
\includegraphics*[scale=0.7]{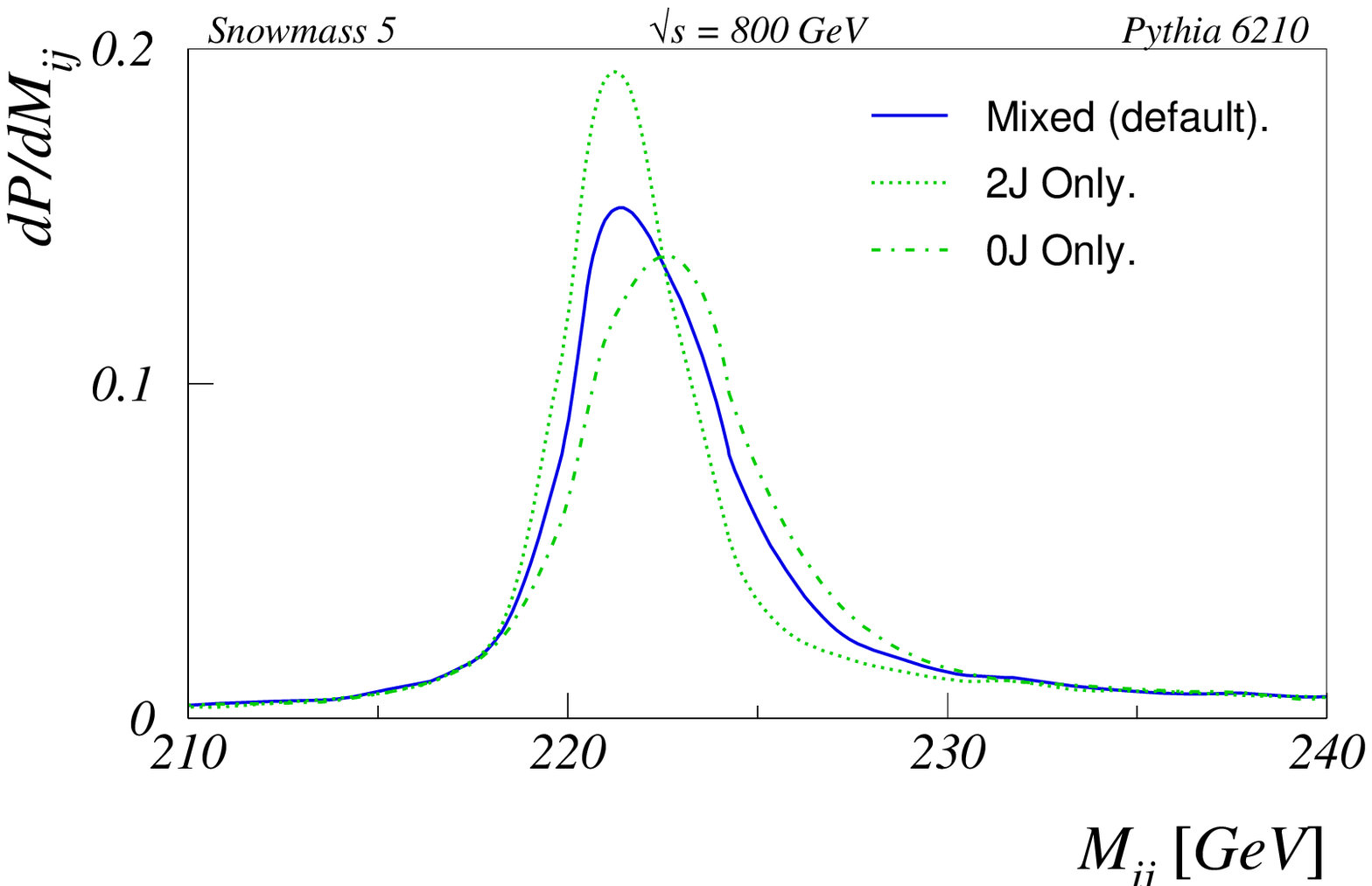}\vspace*{-0.8cm}
\end{center}
\captive{%
Reconstructed jet--jet mass spectrum. The distributions are normalized so as
to integrate to the 
4-jet rates.\label{fig:masspeaks}}
\end{figure}

The mass distributions of the pairs thus selected are shown in
Fig.~\ref{fig:masspeaks} for dynamically mixed (default) 2-junction and
0-junction topologies (solid curve), for 2-junction topologies in all events
(dotted curve), and for 0-junction topologies in all events (dot-dashed
curve). As before, note that the latter sample has 
a small contamination of 2-junction events, from perturbative breakups in the
$\st_1\st_1^*$ shower. 
All the distributions are shifted a few GeV towards larger masses, 
relative to the nominal $\st_1$ mass, but the 0-junction topologies 
are shifted a bit more and have a wider distribution compared to the 
2-junction ones. The shift of the peaks is due to a string effect: 
the fact that either one junction--junction string or two $\q\qbar$ 
strings are spanning the range \emph{between} the two jet systems
increases the hadronic multiplicity in this region. In the former 
topology, all the hadrons produced in a junction--junction string add 
mass to whichever stop they are clustered with. (The energy of these 
hadrons is taken from the kinetic energy of the stops, which are 
slowed down correspondingly.) In the latter topology, the two 
separately hadronizing $\q\qbar$ systems produce even more extra 
soft particles in between the $\st_1$ and $\st^*_1$ jets. Also note 
that this skews the reconstructed jet directions away from the original 
parton directions, towards larger opening angles, which is an alternative
way of expressing that the jet--jet pairs end up having invariant masses 
slightly above the actual resonance mass. In events with multiple gluon
emissions in the $\st_1 \st^*_1$ system --- few to begin with and further
reduced by our cuts --- the zig-zag pattern illustrated in
Fig.~\ref{fig:jjlen}b further tends to favour large string lengths and
thereby large mass shifts in the 0-junction configuration. It is important to
remember that the bulk of the particle production is associated with the four
(anti)quark jet directions, however, and what we discuss here are smaller
perturbations on this picture.

To summarize, we do not expect that a double peak, one from the 2-junction
topologies and another from the 0-junction ones, will be visible in the
jet--jet mass spectrum, but we do expect a small enrichment of 2-junction
topologies on the low side of the peak relative to the high side of
it. Therefore we impose a slightly asymmetric cut on the jet--jet masses,
$m_{\st_1}-10$~GeV~$<M_{ij}<m_{\st_1}+5$~GeV. Note that both jet--jet pairs
in an event are required to pass this cut, or the event is rejected, this
mainly to ensure that both sides of the event are well reconstructed.

Having thus reduced the combinatorics, it is a matter of simple kinematics to
show that the two daughters from each stop decay will be separated by more
than 60$^\circ$ at the 800 GeV CM energy.  Adding parton showers and
hadronization effects will not change this limit appreciably, and so we place
a cut requiring jet pair opening angles larger than 60$^\circ$. Larger
opening angles, from asymmetric decays and/or slowed-down stops, tend to
favour the 0-junction topologies, i.e.\ give less signal. Therefore, and
since we aim to look for junction protons \emph{between} the two daughter
jets, we require a maximal jet--jet opening angle of 120$^\circ$.

For the remaining jet--jet pairs, we may now safely presume that the junction
baryon, if there is one, is predominantly travelling in roughly the same
direction as the stop did, in between the two daughter jets.  To measure how
much ``in between'' two jets a particle is, we construct
\begin{equation}
\hat{\theta}_{\p\j_1\j_2} = \frac{1}{\sqrt{2}}
\sqrt{\left(\frac{2\theta_{\p\j_1}}{\theta_{\j_1\j_2}}-1\right)^2
+\left(\frac{2\theta_{\p\j_2}}{\theta_{\j_1\j_2}}-1\right)^2},
\end{equation}
where $\theta_{\p\j_i}$ is the angle between jet $i$ and the particle in
question, and $\theta_{\j_1\j_2}$ is the inter-jet angle. The measure is
constructed so as to have the value $\hat{\theta} = 0$ when a particle is
lying exactly between the two jets and $\hat{\theta} = 1$ when a particle is
exactly aligned with one or the other of the jet directions. Note that
$\theta_{\p\j_1}+\theta_{\p\j_1}\ge\theta_{\j_1\j_2}$, where the equality
holds for particles lying between the two jets and in the plane defined by
the jet directions.

\begin{figure}
\vspace*{-0.5cm}
\begin{center}
\includegraphics*[scale=0.7]{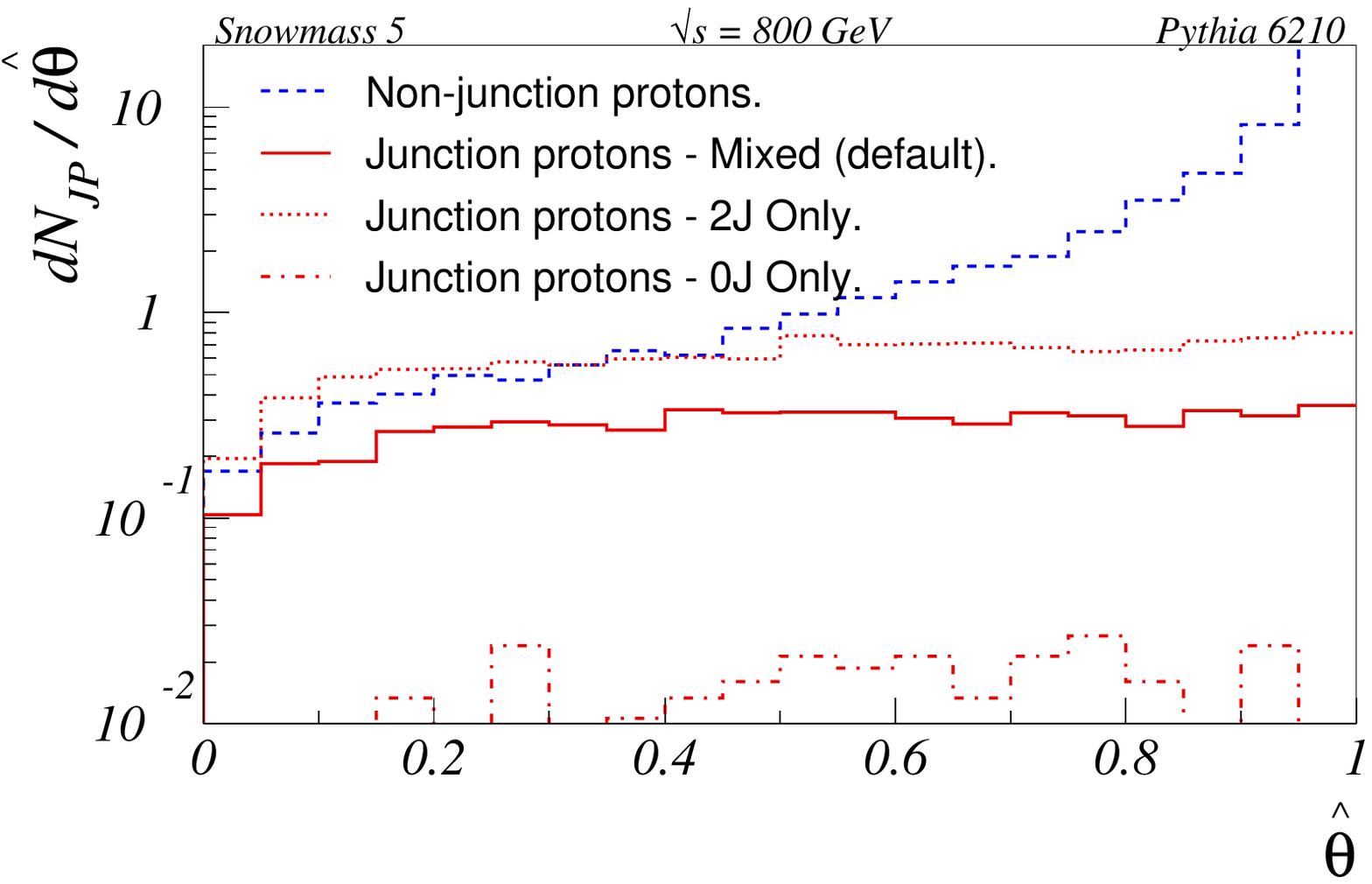}\vspace*{-0.8cm}
\end{center}
\captive{%
Distributions in $\hat{\theta}_{\p\j_1\j_2}$ of 
junction protons for dynamically mixed (default) 2-junction and 0-junction
topologies (solid lines), 2-junction only (dotted lines), and 
0-junction only (dash-dotted lines). The dashed histogram 
represents the non-junction protons. These are only shown for the default 
case, the non-default spectra being very similar. 
\label{fig:thpjj}}
\end{figure}

In Fig.~\ref{fig:thpjj} we show the distributions of junction protons and
ordinary protons in this variable for events passing the previous cuts.
Note that the junction proton distributions, even for the pure 2-junction
topologies, do not integrate to 2 over the
$\hat{\theta}$ range shown on the plot. This owes to the simple facts that
many junction protons are irretrievably lost by the
$\pT\ge0.5$~GeV cut we imposed above and that  
a non-vanishing fraction of them have $\hat{\theta}_{\p\j_1\j_2}>1$.  
The logarithmic scale on the plot is called for by the very 
strong peaking of the non-junction proton spectrum towards the jet axes. We
observe that the shape of the junction proton spectrum is rather flat,
independently of whether we allow junction--junction annihilation or not,
although the normalization of course changes. When junction--junction
annihilation is forced (dot-dashed curve), only those junctions which are
separated by $\g\to\q\qbar$ splittings in the cascades of the
$\st_1\st_1^*$ system survive after hadronization, hence extremely 
few junction protons are produced. We impose a cut requiring 
$\hat{\theta}_{\p\j_i\j_j}<0.5$, thereby 
cleaning out most of the jet-associated protons.

\begin{figure}
\vspace*{-0.5cm}
\begin{center}
\includegraphics*[scale=0.7]{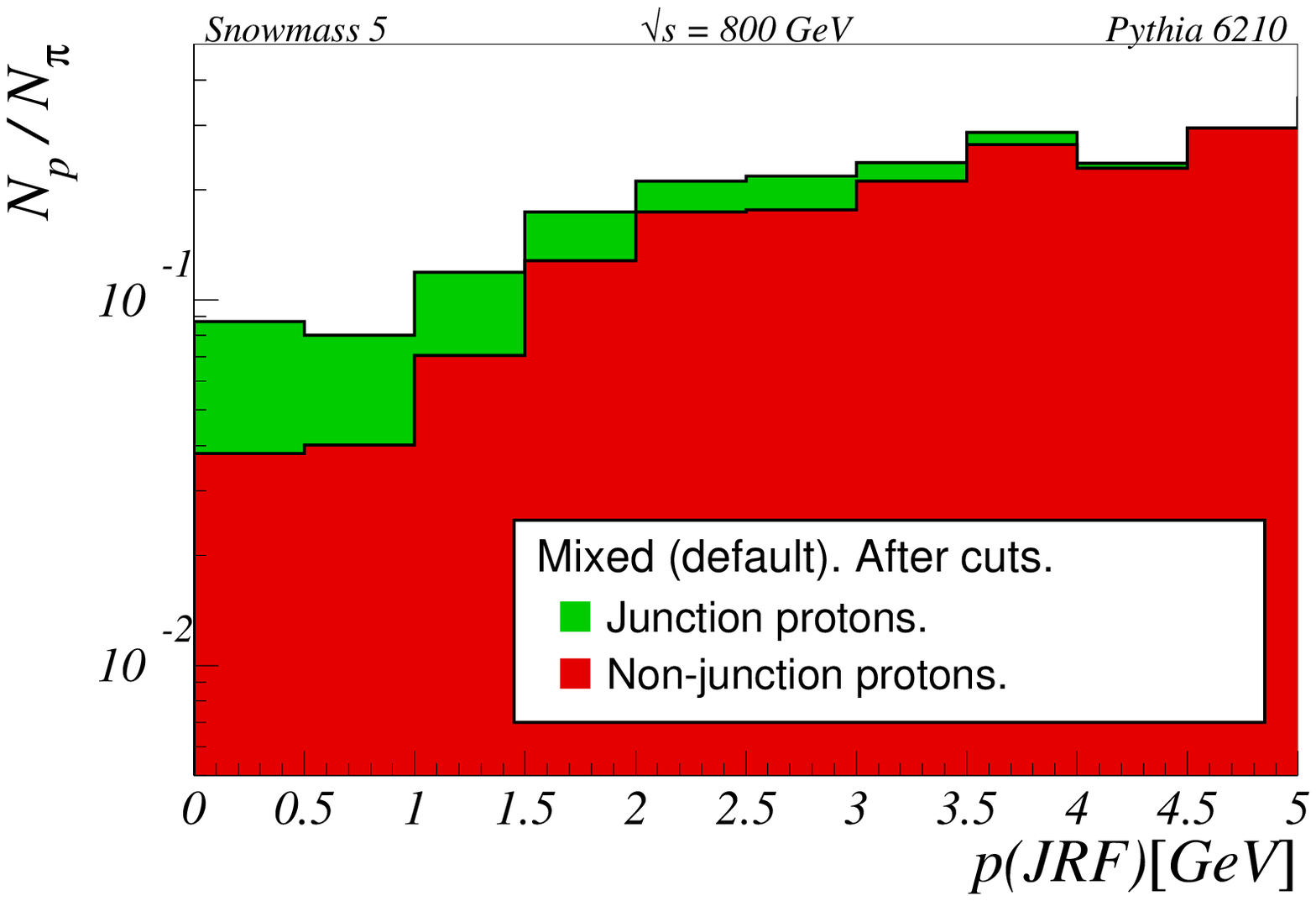}\vspace*{-.7cm}\\
(a)\\
\vspace*{-0.5cm}
\includegraphics*[scale=0.7]{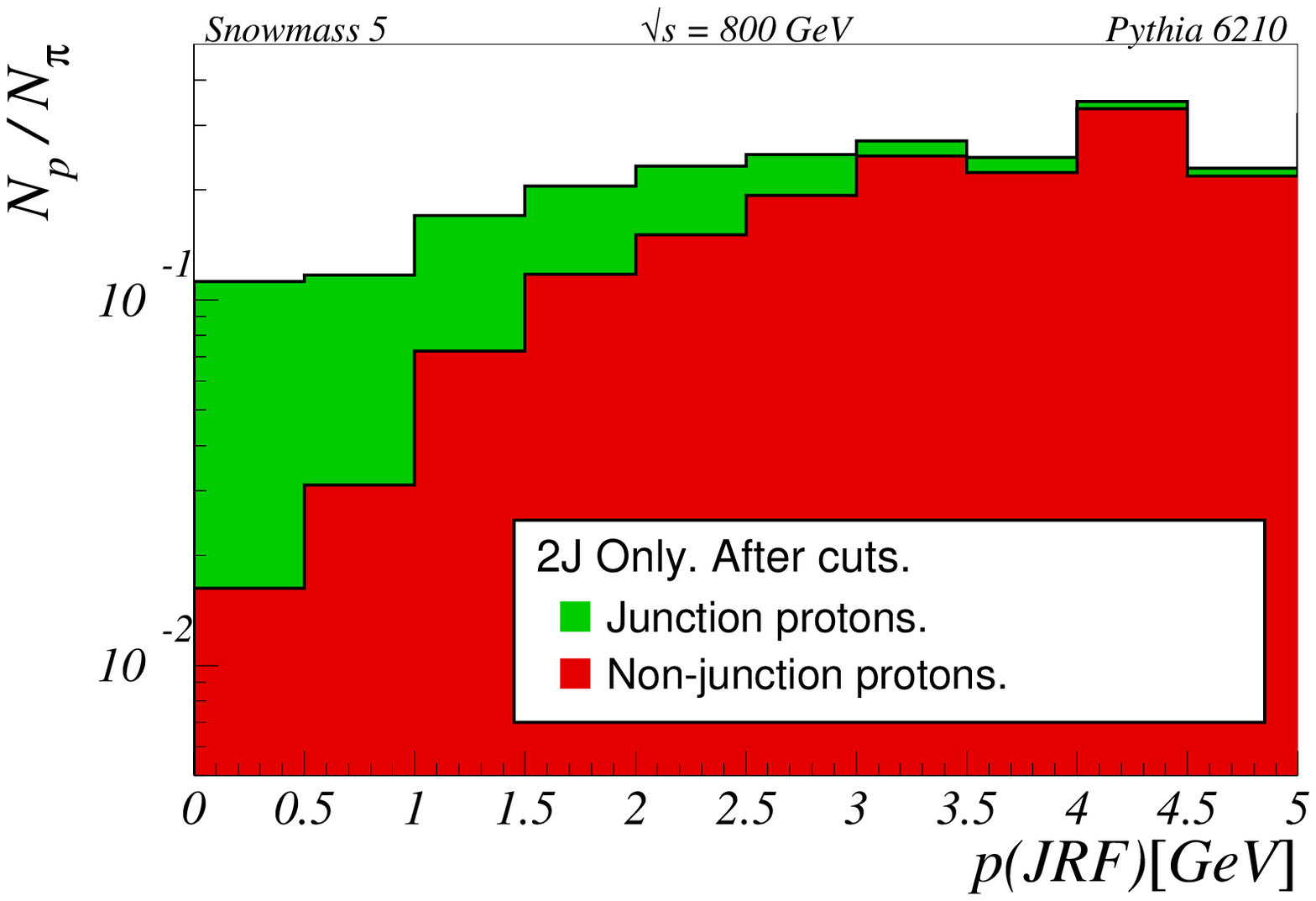}\vspace*{-.7cm}\\
(b)
\end{center}
\captive{%
Reconstructed momentum spectrum in the approximated junction rest frame 
for the ratio of protons to charged pions: (a) dynamically mixed
 (default) 2-junction and 0-junction topologies and (b) 2-junction topologies
 only. 
\label{fig:stopstopspec}}
\end{figure}

Furthermore, we may use the fact that the junction proton momentum spectrum
is most strongly peaked in the junction rest frame (JRF), cf.\
Fig.~\ref{fig:jbmom}. Analogously to what we did for the LHC study, we
therefore construct an approximate JRF for each jet pair, in each case 
using the linear combination of the momenta of the opposing pair as the 
``third leg''. Including the above-mentioned cuts, this results in the
multiplicity distribution Fig.~\ref{fig:stopstopspec}a, now with about 
60\% of the
protons below 1~GeV being junction protons. 
A rough estimate is that this
corresponds to about 50 junction protons per 100~fb$^{-1}$ of integrated
luminosity. With junction--junction annihilation turned off, the distribution
changes to that of Fig.~\ref{fig:stopstopspec}b, with about 80\% of the
protons below 1~GeV being junction protons, corresponding to more than 100
junction protons per 100~fb$^{-1}$. 

Thus, for the process and SUSY scenario
considered here, small BNV couplings substantially 
increase the possibility of directly detecting the extra (anti)baryons. 

Going further and establishing 
the absolute rate of junction protons per $\st_1\st_1^*$ event, with the BNV
branching fractions known, would obviously
amount to a direct measurement of the junction--junction annhilation rate. 
Such a measurement would be interesting from a QCD perspective, 
since it enables us to confront the value predicted by our model with an 
experimental result. We therefore here also briefly discuss an alternative 
way of measuring this quantity. 

\begin{figure}
\vspace*{-0.5cm}
\begin{center}
\includegraphics*[scale=0.7]{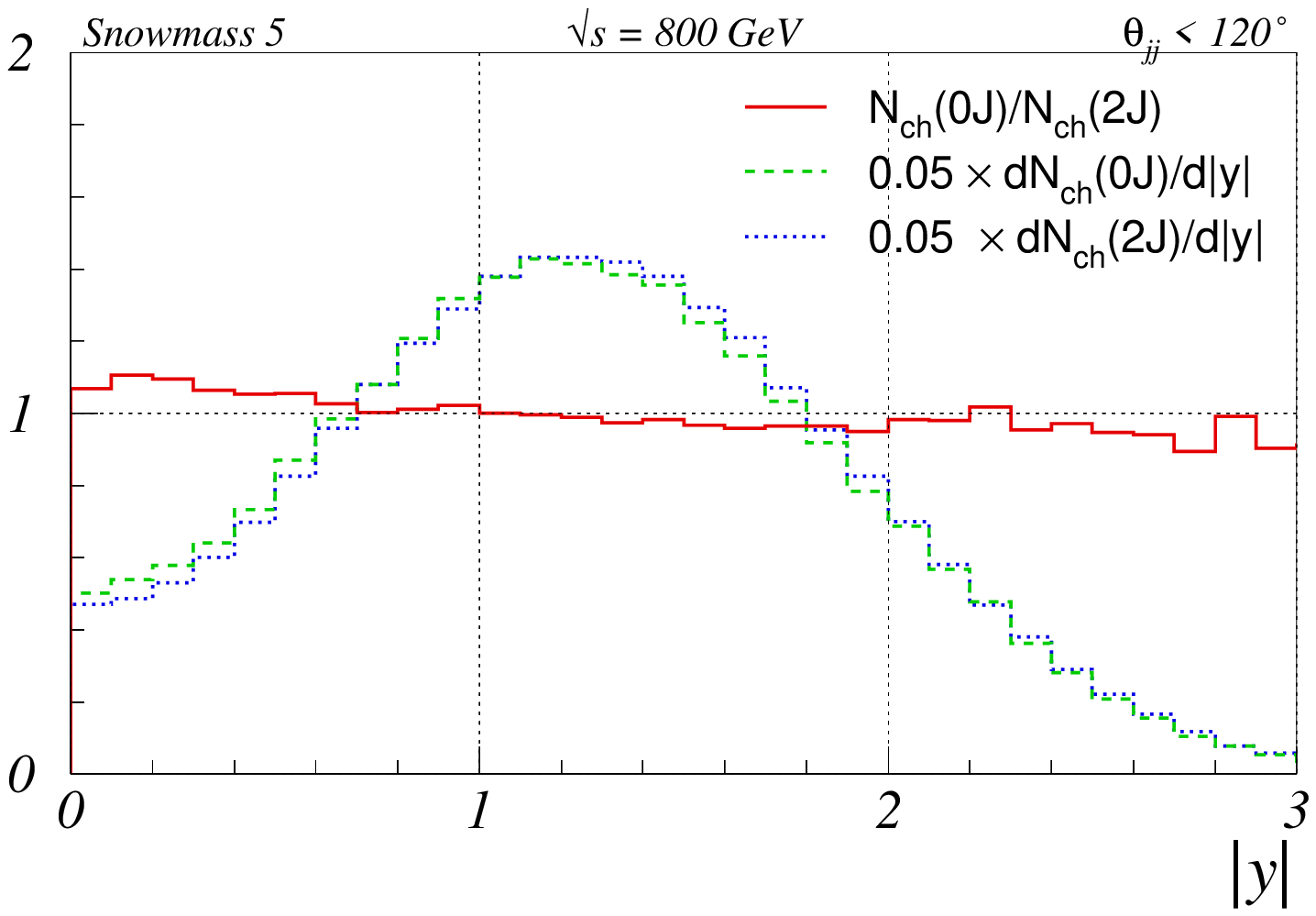}\vspace*{-.7cm}\\
(a)\\
\vspace*{-0.5cm}
\includegraphics*[scale=0.7]{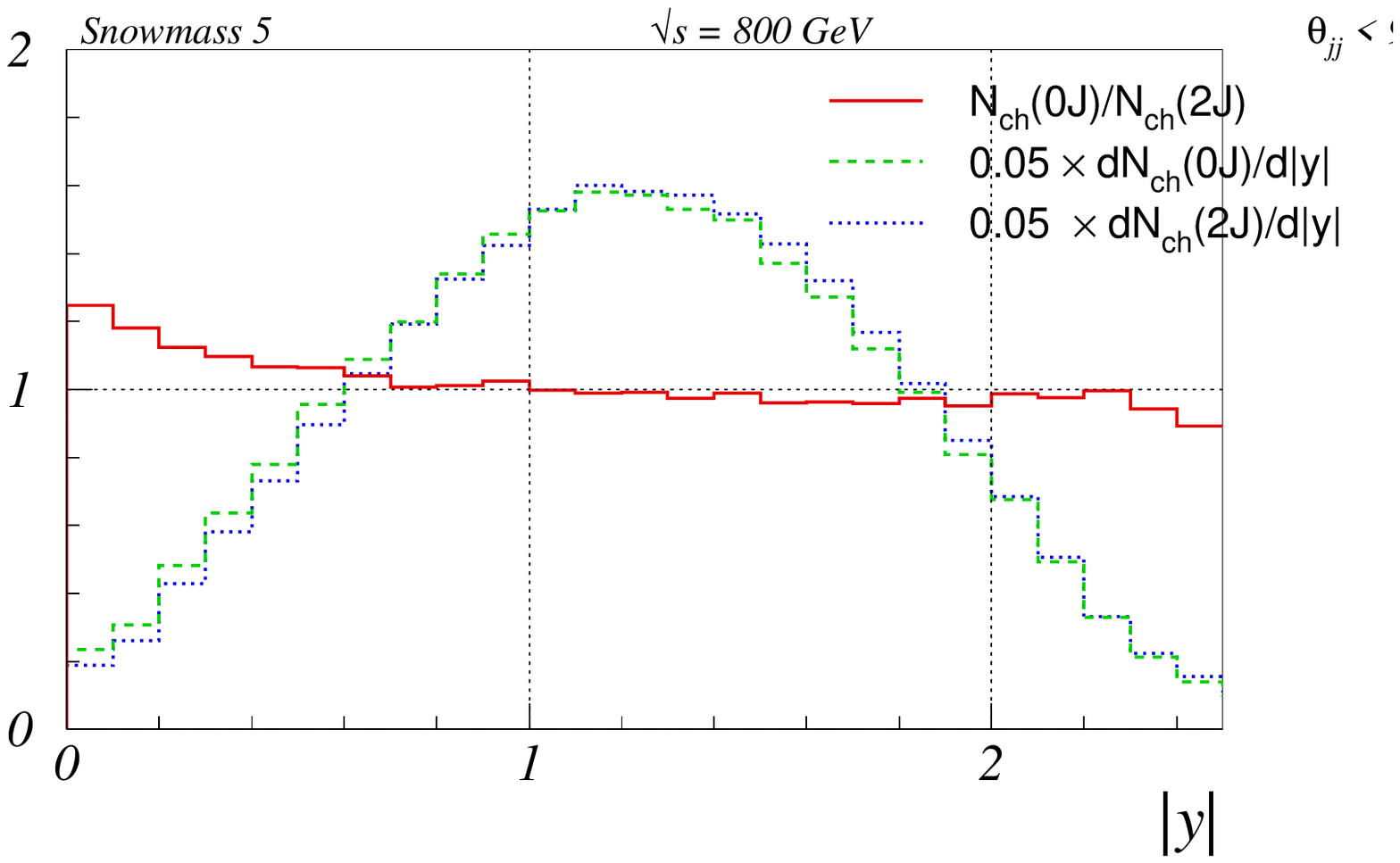}\vspace*{-.7cm}\\
(b)
\end{center}
\captive{%
Charged multiplicities of 0-junction topologies (dashed lines) and
2-junction topologies (dotted lines) 
as functions of rapidity along the stop-antistop thrust
axis for (a) $\theta_{\j\j}\le 120^\circ$ and (b) $\theta_{\j\j}\le
90^\circ$ (all other cuts are as for the junction proton search). The ratios
of the distributions are shown with solid lines.
\label{fig:stopmult}}
\end{figure}

In the 0-junction topology, with at least two strings spanning the rapidity
region between the stops, 
we here expect a higher average multiplicity 
than for the 2-junction case, in which only one string spans this
region. Na\"{\i}vely then, the charged multiplicity between the two jet
pairs should be at least twice as large for 0-junction topologies as compared
to 2-junction ones. However, this drastically oversimplifies the true
situation. 
Asymmetric decays of the stops and the addition of parton showers 
will in general smear the difference to a much lower factor, to the point
where it becomes questionable whether this method is at all feasible. 
To investigate, we select
events passing the above-mentioned cuts and plot the charged multiplicity as
a function of the absolute value of the 
rapidity with respect to the stop--antistop momentum axis. 
The resulting rapidity distributions,
shown in Fig.~\ref{fig:stopmult}a, only exhibit a negligible 
increase in the ratio $(\d N_{0\J}/\d |y|)/(\d N_{2\J}/\d |y|)$ in the region
$|y|\sim 0$. 

One reason for this is that the central region is not free from
jet activity. Specifically, if the decay products of one of the stops are
aligned with the original direction of motion of the stop, the backward jet
will end up in the ``wrong hemisphere''. The most extreme of these cases are
already removed by the requirement of jet--jet opening angles smaller than
120$^\circ$, as was imposed above, but this cut still leaves room for jet
activity in the mid-rapidity region. To clean out the worst of this
contamination,  
in Fig.~\ref{fig:stopmult}b, we have tightened 
the cut on maximal jet--jet opening angles to $\theta_{\j\j}\le 90^\circ$. One
notices that the average multiplicity in the central region drops, and that
the ratio between the multiplicities in the two topologies
increases in this region. 
Ultimately, to get as little contamination from jet tails as
possible in the central region, one should go as close as possible to the
kinematic limit, $\theta_{\j\j}=60^\circ$, where the rapiditity of all jets,
with respect to the stop-antistop momentum axis, is $|y|\sim 1.3$. The
necessary compromise thus lies between minimizing the 
jet contamination in the central region and obtaining 
sufficient statistics. For the present, we merely note that the central
multiplicity \emph{is} sensitive to the choice of string topology but that
it will presumably be hard to use it for a precise determination of 
the junction--junction annihilation rate.

\section{Summary}

The physics scenario studied here may not be the most likely one. First
we have to accept supersymmetry, then that $R$-parity is violated, and
finally that it is baryon rather than lepton number that is affected.
Nevertheless, it is one possibility among many that should be considered
in studies of the high-energy frontier, and one that has maybe received
less attention than others of equal likelihood to be correct.

In particular, it is important to know whether such a scenario introduces
new special demands on detectors. Obviously the whole normal artillery of
jet reconstruction and mass peaks need to be brought to bear on the issue,
in order to find a signal for new physics. This may already be enough to
favour some specific scenarios, such as BNV SUSY. However, given that the
unique aspect of BNV is the creation of an additional baryon or antibaryon,
a discovery of BNV SUSY would not be complete without a clear signal that
indeed baryon number is violated.

Since baryons are produced also in normal hadronization processes,
it will not be feasible to associate one specific baryon uniquely with
the BNV process, but only to establish an excess of baryon production.
This excess will not be large enough to help us, unless we know where to
look, and thereby can reduce the background noise level.

The string junction model we develop in this article offers us the tools
to do precisely that. The string concept itself is the most
successful approach we have to the description of hadronization at
high energies. The junction concept is also well-established in the
literature, to describe baryon colour topologies and confinement aspects.
It has never before (to the best of our knowledge) been developed and
implemented into a realistic hadronization framework, however.

Armed with this detailed model, we can predict where the baryon excess should
be found. In general, the junction baryon will have low momentum in the
rest frame of the process that generates the BNV. For the specific
case of neutralino or chargino BNV decay at LHC, we show how this
translates into explicit predictions. Typically we need to impose some
cut like $\pT > 1$~GeV to get away from the bulk of the underlying
event activity, but beyond that the main demand is for a good $\p/\pbar$
identification at rather low momenta, i.e.\ a few GeV. Failing this,
at least $\Lambda$ tagging should be well developed.

We believe this to be the first example where non-QCD physics interests
put precise demands on baryon identification. Thereby it offers an
interesting test case for further detector performance studies. The
specific distributions we show for LHC are only intended to ``whet the
appetite'', and are in no way blueprints for a realistic analysis.
Background studies will be essential, not only in the search of a
signal, but also to help calibrate the normal rate of baryon production.
Especially disconcerting is here the experimental indications of a
difference in the baryon fraction between different kinds of events,
e.g. $\e\p$ vs. $\e^+\e^-$.

Studies could also be performed for other machines. Unfortunately, it is
doubtful whether the Tevatron could produce enough BNV events to allow an
analysis along the lines we have in mind, given that no signs of SUSY have
been seen so far. In contrast, BNV neutralino or chargino decays
at a future linear $\e^+\e^-$ collider should be much cleaner and thereby
simpler to analyse --- one reason we have not considered it here. The
difficult questions at such a collider could rather come in the analysis
of BNV $\st\st^*$ events, where the threshold region production would allow
two different event topologies, one with an extra baryon--antibaryon pair
and one without. We find that the relative importance of these topologies may
not be easy to pin down experimentally.

It can be questioned how unique the proposed junction hadronization
framework is, given that it is not based on any experimental tests to
existing data. On the technical level, certainly a number of approximations
and simplifications have been introduced, but we have demonstrated that
the resulting uncertainties appear under control. More serious is the
issue whether the whole ansatz as such is what happens in Nature, and
here no guarantees can be given. What can be said its that there is
currently no alternative approach that looks anywhere near as credible.

One specific example is here offered by \textsc{Herwig}. This generator
maps a three-quark cluster onto a quark--diquark one, which then is
fragmented along a single axis. Since the baryon cluster mass typically
is very large --- in part a consequence of a low shower
activity in these events --- the fragmentation can deform the event
appreciably and, more significantly,  kick the baryon out to surprisingly
large momenta.

That is not to say there is no room for improvement in our
\textsc{Pythia} implementation. We did not (yet) include
the matrix-element information to give non-isotropic
three-body BNV decays, the BNV production processes are not
implemented at all, and our shower
description is not as sophisticated as e.g. for $\Z^0\to\q\qbar$ decays.
Such imperfections could influence estimates of the experimental
acceptance rate of BNV decays. Given all the other uncertainties at this
stage of the game, one should not exaggerate their importance, however.
Once SUSY is observed, with indications of BNV, further work would be
reasonable.

The development of a model for the hadronization of junction string
topologies is not only relevant for BNV SUSY. In principle we would
assume the same topology for the incoming baryons in high-energy
$\p\p$ or $\p\pbar$ collisions. So long as only one valence quark is
kicked out of a proton, the two remaining quarks drag the junction along
and thereby form an effective diquark. But in multiple interaction
approaches there is every possibility for two quarks to be kicked out,
in different directions, and with colours rearranged in the process.
Thereby the baryon number can start to drift in the event. This
scenario was advertised in \cite{Zijl}, but has not been studied
further till now. The recent increased interest in semihard physics at
hadron colliders \cite{multintdata}, which has confirmed the basic
validity of the \textsc{Pythia} framework but also pointed at problems,
here provides a stimulus for further developments. Hopefully,
refined models could allow us to understand the excess
of baryon number observed in the central region of events in a number
of experiments (\cite{barexcess} is a far from complete list).
Here junction scenarios have already been proposed as a possible
mechanism \cite{barexcessthy}, but without any solid modelling efforts.

A reason for us to begin with the SUSY BNV studies, rather than with pure
QCD ones, is that the latter inevitably will come to involve further
uncertainties, such as the details of the multiple interactions
scenario. Nevertheless, the hope is that continued QCD-related studies
may add support for the validity of the junction scenario proposed in
this article. Thereby the physics of junction fragmentation offers a
prime example of a topic where the exploration of very conventional and
very unconventional physics goes hand in hand.

\subsection*{Acknowledgements}

The authors would like to thank G\"osta Gustafson for helpful discussions
on the string length of junction topologies, and Frank Paige and Albert
de Roeck for useful information on the ATLAS and CMS proton identification
capabilities. We are also grateful to the NorduGrid project, for inviting
us to test their facilities. Many of our results were obtained with 
grid computing.

\end{document}